%% file: nb_px_paper.tex
\documentclass[3p]{elsarticle}
\usepackage[american]{babel}
\usepackage{amsmath}
\usepackage[version=3]{mhchem} 
\usepackage{mathtools}

\usepackage{xfrac}
\usepackage{lmodern}
\usepackage[hidelinks]{hyperref}
\usepackage{listings}
\usepackage{mcode}
\usepackage [autostyle, english = american]{csquotes}
\usepackage{booktabs,siunitx}
\usepackage{gensymb}
\usepackage[normalem]{ulem}

\usepackage{color}

\usepackage[colorinlistoftodos]{todonotes}

\usepackage[section]{placeins}
\usepackage{multirow}

\usepackage{graphicx}
\usepackage{dcolumn}
\usepackage{bm}

\usepackage{ecrc}

\volume{00}

\firstpage{1}

\journalname{Nuclear Instrum. and Methods in Phys. Res. B}

\runauth{A.S. Voyles et al.}

\jid{nimb}

\jnltitlelogo{Nucl Instrum Meth B}

\usepackage{amssymb}
\usepackage{amsthm}

\biboptions{sort&compress}

\usepackage[figuresright]{rotating}

\usepackage{fancyvrb}
\usepackage{color}
 
\definecolor{mygreen}{rgb}{0,0.6,0}
\definecolor{mygray}{rgb}{0.5,0.5,0.5}
\definecolor{mymauve}{rgb}{0.58,0,0.82}

\newcommand{\pp}[1]{\left( #1\right)}

\newcommand{\cmmnt}[1]{}

\newcommand{\etal}{\emph{et\,al.}}

\newcommand{\subfigimg}[4][,]{%
  \setbox1=\hbox{\noindent\includegraphics[#1]{#3}}
  \leavevmode\rlap{\usebox1}
  \rlap{\hspace*{#4pt}\raisebox{\dimexpr\ht1-2\baselineskip}{#2}}
  \phantom{\usebox1}
}

\usepackage{subfig}
\newcommand{\mmicro}{\si\micro} 

\usepackage{adjustbox}



\begin{document}

\begin{frontmatter}

\author[ucb]{Andrew S. Voyles \corref{cor1}}
\ead{andrew.voyles@berkeley.edu}

%
%

\author[lbl,ucb]{Lee A. Bernstein}

\author[lanl]{Eva R. Birnbaum}

\author[uwm]{Jonathan W. Engle}
\ead{jwengle@wisc.edu}

\author[iowa]{Stephen A. Graves}

\author[tlanl]{Toshihiko  Kawano}

\author[ucb]{Amanda M. Lewis}

\author[lanl]{Francois M. Nortier}

\address[ucb]{Department of Nuclear Engineering, University of California, Berkeley,  Berkeley, CA 94720, USA}
\address[lbl]{Nuclear Science Division, Lawrence Berkeley National Laboratory,  Berkeley, CA 94720, USA}
\address[uwm]{Department of Medical Physics, University of Wisconsin -- Madison,  Madison, WI 53705, USA}
\address[lanl]{Isotope Production Facility, Chemistry Division, Los Alamos National Laboratory,  Los Alamos, NM 87544, USA}
\address[iowa]{Department of Radiation Oncology, University of Iowa,  Iowa City, IA 52242, USA}
\address[tlanl]{Theoretical Division, Los Alamos National Laboratory,  Los Alamos, NM 87544, USA}

%
%
%

\title{Excitation functions for (p,x) reactions of niobium in the energy range of E\texorpdfstring{$_{\text{p}}$\,=\,40--90\,}{Ep = 40--90 }MeV}

\begin{abstract}

\input{./nb_abstract_text}

\end{abstract}





\end{frontmatter}




\input{./nb_body_text}

 \section{Acknowledgements}
 
 
 The authors would like to particularly acknowledge the assistance and support of  Michael Gallegos and Don Dry in the LANL C-NR Countroom, David Reass and Mike Connors at LANSCE-IPF, and the LANSCE Accelerator Operations staff. 
 
We gratefully acknowledge support for this work from the United States Department of Energy, Office of Science via the Isotope Development and Production for Research and Applications subprogram in the Office of Nuclear Physics. 
This work has been carried out  under the auspices of the U.S. Department of Energy by  Lawrence Berkeley National Laboratory and the U.S. Nuclear Data Program under contract \# DE-AC02-05CH11231.
This research was performed under appointment to the Rickover Fellowship Program in Nuclear Engineering, sponsored by the Naval Reactors Division of the U.S. Department of Energy.
Additional support has been provided by the U.S. Nuclear Regulatory Commission.

This research used the Savio computational cluster resource provided by the Berkeley Research Computing program at the University of California, Berkeley (supported by the UC Berkeley Chancellor, Vice Chancellor for Research, and Chief Information Officer).

\appendix

\input{./nb_appendix_text}

\pagebreak



\bibliographystyle{elsarticle-num}
\bibliography{../../library}







\end{document}

%% file: nb_abstract_text.tex
A stack  of thin Nb 
foils was irradiated with the 100\,MeV proton beam at  Los Alamos National Laboratory's Isotope Production Facility,  to investigate the \ce{^{93}Nb}(p,4n)\ce{^{90}Mo} nuclear reaction as a  monitor for intermediate energy proton experiments and to benchmark state-of-the-art reaction model codes.
A set of 38 measured  cross sections for  \ce{^{nat}Nb}(p,x) and  \ce{^{nat}Cu}(p,x) reactions between 40--90\,MeV, as well as 5 independent measurements of isomer branching ratios, are reported. 
These are useful in medical and basic science radionuclide productions at intermediate energies. 
The
\ce{^{nat}Cu}(p,x)\ce{^{56}Co}, \ce{^{nat}Cu}(p,x)\ce{^{62}Zn}, and \ce{^{nat}Cu}(p,x)\ce{^{65}Zn}  reactions were used to
determine proton fluence, and all activities were quantified using HPGe spectrometry.
Variance minimization techniques were employed to reduce systematic uncertainties in proton energy and fluence, improving the reliability of these measurements. 
The measured cross sections are shown to be in excellent agreement with  literature values, and have been measured with improved precision compared with previous measurements.
This work also reports the first measurement of the  \ce{^{nat}Nb}(p,x)\ce{^{82m}Rb} reaction, and of the independent cross sections for    
\ce{^{nat}Cu}(p,x)\ce{^{52\text{g}}Mn} and \ce{^{nat}Nb}(p,x)\ce{^{85\text{g}}Y} in the 40--90\,MeV region.
The effects of  \ce{^{nat}Si}(p,x)\ce{^{22,24}Na} contamination, arising from  silicone adhesive in the Kapton tape used to encapsulate the aluminum monitor foils, is also discussed as a cautionary note to future stacked-target cross section measurements.
\emph{A priori} predictions of the reaction modeling codes  CoH, EMPIRE, and TALYS are compared with experimentally measured values and used to explore the differences between codes for the
\ce{^{nat}Nb}(p,x) and  \ce{^{nat}Cu}(p,x) reactions.


%% file: nb_body_text.tex
\section{Introduction} \label{sec:intro}

%
%
%


Every year, approximately 17 million nuclear medicine procedures (both diagnostic and therapeutic) are performed in the U.S. alone 
\cite{Delbeke2011,NSACIsotopesSubcommittee2015}.
Most of the radionuclides currently used for these procedures are produced 
by low- (E \textless\ 30\,MeV / A) and intermediate-energy (30 \textless\ E \textless\ 200\,MeV / A) accelerators, 
e.g., \ce{^{11}C}, \ce{^{18}F}, \ce{^{68}Ga}, \ce{^{82}Rb}, and \ce{^{123}I}. 
These accelerators also produce 
non-medical radionuclides with commercial value, such as  \ce{^{22}Na}, \ce{^{73}As}, \ce{^{95m}Tc}, and  \ce{^{109}Cd} \cite{international2009iaea,schlyer2008cyclotron}. 
Novel applications are being explored for several radionuclides whose production methodologies are not established, but their production requires accurate, high-fidelity cross section data.
Candidate isotopes to meet these needs 
have been identified based on their chemical and radioactive decay properties \cite{NSACIsotopesSubcommittee2015,Qaim201731,bernstein2015nuclear}, and a series of campaigns are underway to perform targeted, high-priority measurements of thin-target cross sections and thick-target integral yields.
These studies will serve to facilitate the production of clinically relevant quantities of radioactivity.

Accurate cross section measurements using  activation methods benefit from well- characterized monitor reactions.  
Currently there is 
a paucity of such data at intermediate energies, and much of what exists have  high uncertainties (\textgreater15\%).
Indeed, the development of new monitor reaction standards and the improved evaluation of existing standards is one of the areas of greatest cross-cutting need for nuclear data \cite{bernstein2015nuclear}. 
New reactions can expand the available range of options for the monitoring of charged particle beams.
This work 
is an attempt to characterize a new monitor reaction for  proton beams in excess of 40\,MeV, for possible use at isotope production facilities such as the  Brookhaven Linac Isotope Producer (BLIP) at Brookhaven National Laboratory, 
the Isotope Production Facility (IPF) at  Los Alamos National Laboratory, or the Separated Sector Cyclotron at the 
iThemba Laboratory for Accelerator Based Sciences.
Desirable monitor reactions 
possess several hallmark characteristics,  
including intense, distinct gamma-rays, 
which can be used for unique identification during post-activation assay, and lifetimes long enough to enable removal after a reasonable length irradiation.    
Care should also be taken to avoid cases where two radionuclides which are produced by two different reactions on the same monitor foil lead to states in the same daughter nuclide.  
For example, \ce{^{48}V}  ($t_{1/2}$ = 15.97 d, $\epsilon=100\%$ to \ce{^{48}Ti}) and \ce{^{48}Sc}  ($t_{1/2}$ = 43.67 h, $\beta^-=100\%$ to \ce{^{48}Ti}) can both be formed via $^\text{nat}$Ti(p,x) reactions, yielding the same 983.52\,keV transition in \ce{^{48}Ti} \cite{Burrows2006}.
It is also of vital importance that the proposed monitor nucleus have well-characterized decay data.
This includes a precise and well-established half-life,  
and well-characterized decay gamma-ray intensities.
From a targetry  perspective, it is preferable to use a naturally mono-isotopic target that is readily available
and  chemically inert.
Targets which can be formed into a wide thickness range are convenient, as selection is subject to the context of an experiment, seeking to maximize thickness without overly perturbing the energy uncertainty of  measurements.
Lastly, and perhaps most importantly for high-energy monitor reaction applications, it is  of utmost importance to choose a reaction channel which cannot be populated via secondary particles incident upon the monitor target.
Typically, this is  mostly a concern for secondary neutrons produced through (z,xn) reactions, 
but any monitor reaction channel which can be populated by anything other than the primary beam should be avoided, as it is often 
difficult to accurately and unambiguously separate out the fraction of secondary particles contributing to the total activation.

One  reaction which satisfies these requirements is that of a new, intermediate-energy proton monitor reaction standard based on \ce{^{93}Nb}(p,4n)\ce{^{90}Mo}. 
Niobium is naturally mono-isotopic, readily  available commercially in high purity, is fairly chemically inert, and can easily be rolled down to foils as thin as 1 \mmicro m.  
\ce{^{90}Mo} also has 
a sufficiently long lifetime ($\epsilon=100\%, t_{1/2}=5.56 \pm 0.09$ h \cite{Browne1997})  
and seven strong, distinct gamma lines (notably its 122.370\,keV [$I_\gamma = 64 \pm 3\%$] and 257.34\,keV [$I_\gamma = 78 \pm 4\%$] lines) which can be used to uniquely and easily   quantify \ce{^{90}Mo} production. 
In addition,  \ce{^{90}Mo} is completely immune from (n,x) production on  \ce{^{93}Nb}, being produced only via the primary proton beam, and the \ce{^{90}Mo} decay lines can only be observed in its decay, as its daughter, \ce{^{90}Nb}, is also unstable and decays via $\epsilon$ to stable \ce{^{90}Zr}. 
 
The purpose of the present work is to  measure the production of the long-lived radionuclide \ce{^{90}Mo}  
via the $^\text{nat}$Nb(p,x) reaction. 
In addition to the $^\text{nat}$Nb(p,x)\ce{^{90}Mo} measurement, this experiment has also yielded measurements of 37 other (p,x) production cross sections between 40--90\,MeV  for a number of additional reaction products, including several emerging radionuclides with medical applications.
These include the non-standard positron emitters \ce{^{57}Ni}, \ce{^{64}Cu},  \ce{^{86}Y}, \ce{^{89}Zr},  \ce{^{90}Nb}, 
and the diagnostic agent \ce{^{82\text{m}}Rb}. 

%

In addition to providing a potentially highly-valuable beam monitor, the Nb(p,x) reactions offer an opportunity to study the angular momentum deposition via pre-equilibrium reactions and the spin distribution in g$_{9/2}$ subshell nuclei via the observation of isomer-to-ground state ratios.  
Measurements of isomer-to-ground state ratios have been used for over 20\,years to probe the spin distribution of excited nuclear states in the A\,$\approx$\,190 region \cite{PhysRevC.73.034613,PhysRevC.45.1171}.
These include the \ce{^{52\text{m}}Mn} ($t_{1/2}=21.1\pm0.2$\,m; J$^\pi=2^+$) to \ce{^{52\text{g}}Mn}  ($t_{1/2}=5.591\pm0.003$\,d; J$^\pi=6^+$), \ce{^{58\text{m}}Co} ($t_{1/2}=9.10\pm0.09$\,h; J$^\pi=5^+$) to \ce{^{58\text{g}}Co}  ($t_{1/2}=70.86\pm0.06$\,d; J$^\pi=2^+$),  \ce{^{85\text{m}}Y} ($t_{1/2}=4.86\pm0.13$\,h; J$^\pi=\sfrac{9}{2}^+$) to \ce{^{85\text{g}}Y}  ($t_{1/2}=2.68\pm0.05$\,h; J$^\pi=\sfrac{1}{2}^-$),  \ce{^{87\text{m}}Y} ($t_{1/2}=13.37\pm0.03$\,h; J$^\pi=\sfrac{9}{2}^+$) to \ce{^{87\text{g}}Y}  ($t_{1/2}=79.8\pm0.3$\,h; J$^\pi=\sfrac{1}{2}^-$),  and \ce{^{89\text{m}}Nb} ($t_{1/2}=66\pm2$\,m; J$^\pi=\sfrac{1}{2}^-$) to \ce{^{89\text{g}}Nb}  ($t_{1/2}=2.03\pm0.07$\,h; J$^\pi=\sfrac{9}{2}^+$)  ratios \cite{Dong2015,Nesaraja2010,Singh2014,Johnson2015,Singh2013}.

The measurements described in this paper involve the use of multiple monitor reactions in conjunction with statistical calculations and proton transport simulations to reduce systematic uncertainties in beam energy assignments, leading to some of the first and most precise measurements  for many of the excitation functions reported here. 
By expanding the available set of monitor reaction standards and well-characterized isotope production excitation functions, this work should help optimize medical isotope production modalities, making more options   available for modern medical imaging and cancer therapy.

\section{Experimental methods and materials}\label{sec:experiment}

The work described herein follows the  methods established by Graves \etal\ 
for monitor reaction characterization of beam energy and fluence in stacked target irradiations \cite{Graves2016}.

\subsection{Stacked-target design }\label{sec:target_design}

\begin{figure}
 \centering
 \includegraphics[width=0.5\linewidth,clip=true,trim=13cm 0cm 3cm 6cm]{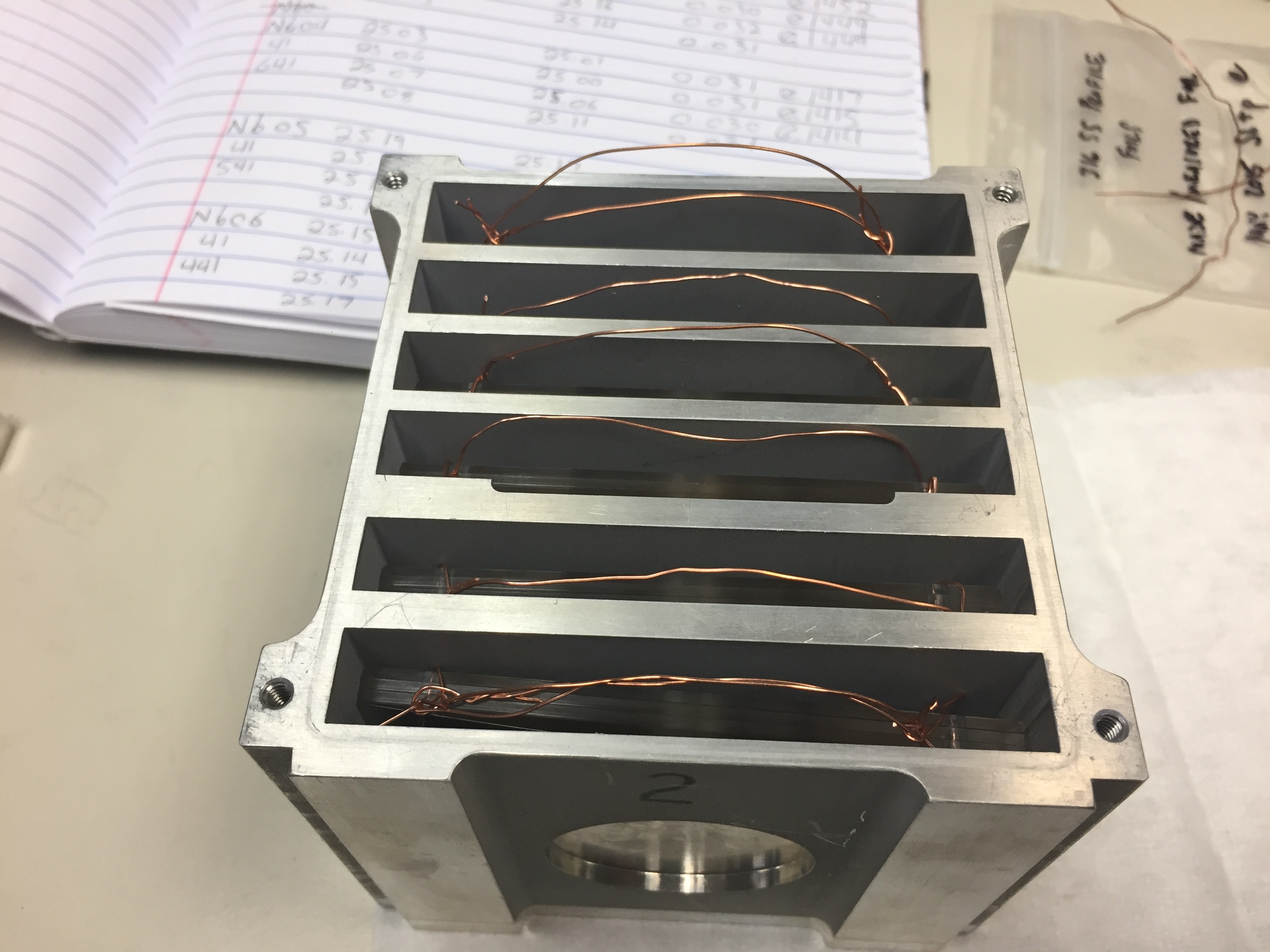}
 \caption{\label{fig:target_stack}Photograph of the assembled IPF target stack, before the stack's o-ring lid was sealed in place. The baling wire handles affixed to each bunch of Al+Cu+Nb foils are visible in each energy position, to facilitate removal of activated foils via  manipulators in the IPF hot cell. The circular Inconel beam entrance aperture is visible in the bottom center of the photograph.  }
\end{figure}

A stacked-target design was utilized for this work in order that the (p,x) cross sections for each reaction channel could be measured at multiple energy positions in a single irradiation \cite{Cumming1963}. 
A series of nominal 25 \mmicro m \ce{^{nat}Nb} foils (99.8\%, lot \#T23A035), 25 \mmicro m \ce{^{nat}Al} foils (99.999\%, lot \#M06C032), and 50 \mmicro m \ce{^{nat}Cu} foils (99.9999\%, lot \#N26B062) were used as targets (all from Alfa Aesar, Ward Hill, MA, 01835, USA).
Six foils of each metal were cut down to 2.5 $\times$ 2.5\,cm squares and characterized --- for each foil, length and width measurements were taken at four different locations using a digital caliper (Mitutoyo America Corp.), thickness measurements were taken at four different locations using a digital micrometer (Mitutoyo America Corp.), and four mass measurements were taken using an analytical balance after cleaning the foils with isopropyl alcohol.
Using these length, width, and mass readings, the areal density and its uncertainty (in mg/cm$^2$) for each foil was calculated.
The foils were tightly sealed into \enquote{packets} using two pieces of  3M 5413-Series Kapton polyimide film tape --- each piece of tape consists of 43.2 \mmicro m of a silicone adhesive (nominal 4.79\,mg/cm$^2$) on 25.4 \mmicro m of a polyimide backing (nominal 3.61\,mg/cm$^2$).
The sealed foils were mounted over the hollow center of a 1.575\,mm-thick plastic frame.
One \ce{^{nat}Al}, one \ce{^{nat}Cu}, and one \ce{^{nat}Nb} mounted foil were bundled together using baling wire for each energy position.
These foil packet bundles were lowered into the beamline by inserting them into a  water-cooled production target box.
The box, seen in \autoref{fig:target_stack}, is machined from 6061 aluminum alloy, has a thin (0.64\,mm) Inconel beam entrance window, and  contains 6 \enquote{energy positions} for targets, formed by  5 slabs of 6061 aluminum alloy (previously characterized) which serve as proton energy degraders  between energy positions.
After loading all targets in the stack, the lid of the target box is sealed in place, using an inset o-ring to create a water-tight seal, and the box is lowered through a hot cell into the beamline, where it sits electrically isolated.
The specifications of the target stack design for this work is presented in \autoref{tab:stack_table}.

This target stack was assembled and irradiated at the Isotope Production Facility (IPF) at the Los Alamos National Laboratory (LANL), using the LANSCE linear accelerator. 
The stack was irradiated for approximately 2\,h with a nominal current of 1\,mA, using a 50 \mmicro s pulse at a frequency of 2\,Hz, for an anticipated integral current of 205.9 nAh.
The beam current, measured using an inductive pickup, remained stable under these conditions for the duration of the irradiation, with the exception of approximately 70\,s of downtime, which occurred approximately 3\,min into irradiation.
The proton beam incident upon the stack's Inconel beam entrance window had an average energy of 100\,MeV determined via time-of-flight, with an approximately Gaussian energy distribution width of 0.1\,MeV --- this energy profile was used for all later analysis.
At the end of the irradiation, the target stack was withdrawn from the beamline into the IPF hot cell, where it was disassembled and the activated foils removed using robotic manipulators.
The activated foils were cleaned of all surface contamination, and transported to a counting lab for gamma spectrometry, which started approximately 6\,h following end-of-bombardment.

\begin{table}
\centering
\caption{Specifications of the  target stack design in the present work. The proton beam enters the stack upstream of the 249.8 \mmicro m SS profile monitor, and is transported through the stack in the order presented here. The 6061 aluminum degraders have a measured density of approximately 2.80 g/cm$^3$. Their areal densities were determined using the variance minimization techniques described  in this work  and the earlier paper by Graves \etal\ \cite{Graves2016}. At both the front and rear of the target stack's foils, a 316 stainless steel foil is inserted to serve as a beam profile monitor --- after end-of-bombardment (EoB), decay radiation emitted from these activated stainless steel foils were used to develop radiochromic film (Gafchromic EBT), revealing the spatial profile of the beam entering and exiting the stack.}
\label{tab:stack_table}
\small
\begin{tabular}{@{}llll@{}}
\toprule
Target layer       & \begin{tabular}[c]{@{}l@{}}Measured \\ thickness\end{tabular} & \begin{tabular}[c]{@{}l@{}}Measured areal\\density (mg/cm$^2$)\end{tabular} & \begin{tabular}[c]{@{}l@{}}Areal density \\ uncertainty (\%)\end{tabular} \\ \midrule
SS profile monitor & 249.8 \mmicro m         & 194.56                                      & 0.29                      \\
Al-1               & 25.0 \mmicro m          & 6.52                                        & 0.72                      \\
Cu-1               & 61.3 \mmicro m          & 53.74                                       & 0.15                      \\
Nb-1               & 30.0 \mmicro m          & 23.21                                       & 0.17                      \\
Al Degrader 01     & 4.96 mm           & -                                            & -                          \\
Al-2               & 25.5 \mmicro m          & 6.48                                        & 0.36                      \\
Cu-2               & 61.8 \mmicro m          & 53.85                                       & 0.17                      \\
Nb-2               & 30.8 \mmicro m          & 22.91                                       & 0.17                      \\
Al Degrader 02     & 4.55 mm           & -                                            & -                          \\
Al-3               & 25.8 \mmicro m          & 6.47                                        & 0.31                      \\
Cu-3               & 61.5 \mmicro m          & 53.98                                       & 0.11                      \\
Nb-3               & 31.0 \mmicro m          & 22.91                                       & 0.24                      \\
Al Degrader 03     & 3.52 mm           & -                                            & -                          \\
Al-4               & 26.3 \mmicro m          & 6.51                                        & 0.41                      \\
Cu-4               & 61.3 \mmicro m          & 53.46                                       & 0.22                      \\
Nb-4               & 30.8 \mmicro m          & 22.55                                       & 0.25                      \\
Al Degrader 04     & 3.47 mm           & -                                            & -                          \\
Al-5               & 26.5 \mmicro m          & 6.48                                        & 0.29                      \\
Cu-5               & 61.5 \mmicro m          & 53.57                                       & 0.11                      \\
Nb-5               & 30.8 \mmicro m          & 22.11                                       & 0.25                      \\
Al Degrader 05     & 3.46 mm           & -                                            & -                          \\
Al-6               & 26.3 \mmicro m          & 6.48                                        & 0.62                      \\
Cu-6               & 62.0 \mmicro m          & 53.84                                       & 0.32                      \\
Nb-6               & 31.3 \mmicro m          & 22.12                                       & 0.13                      \\
SS profile monitor & 124.4 \mmicro m         & 101.34                                      & 0.23                      \\ \bottomrule
\end{tabular}
\end{table}

\subsection{Measurement of induced activities}\label{sec:spectroscopy}

\begin{figure}
 \centering
 \includegraphics[width=6in]{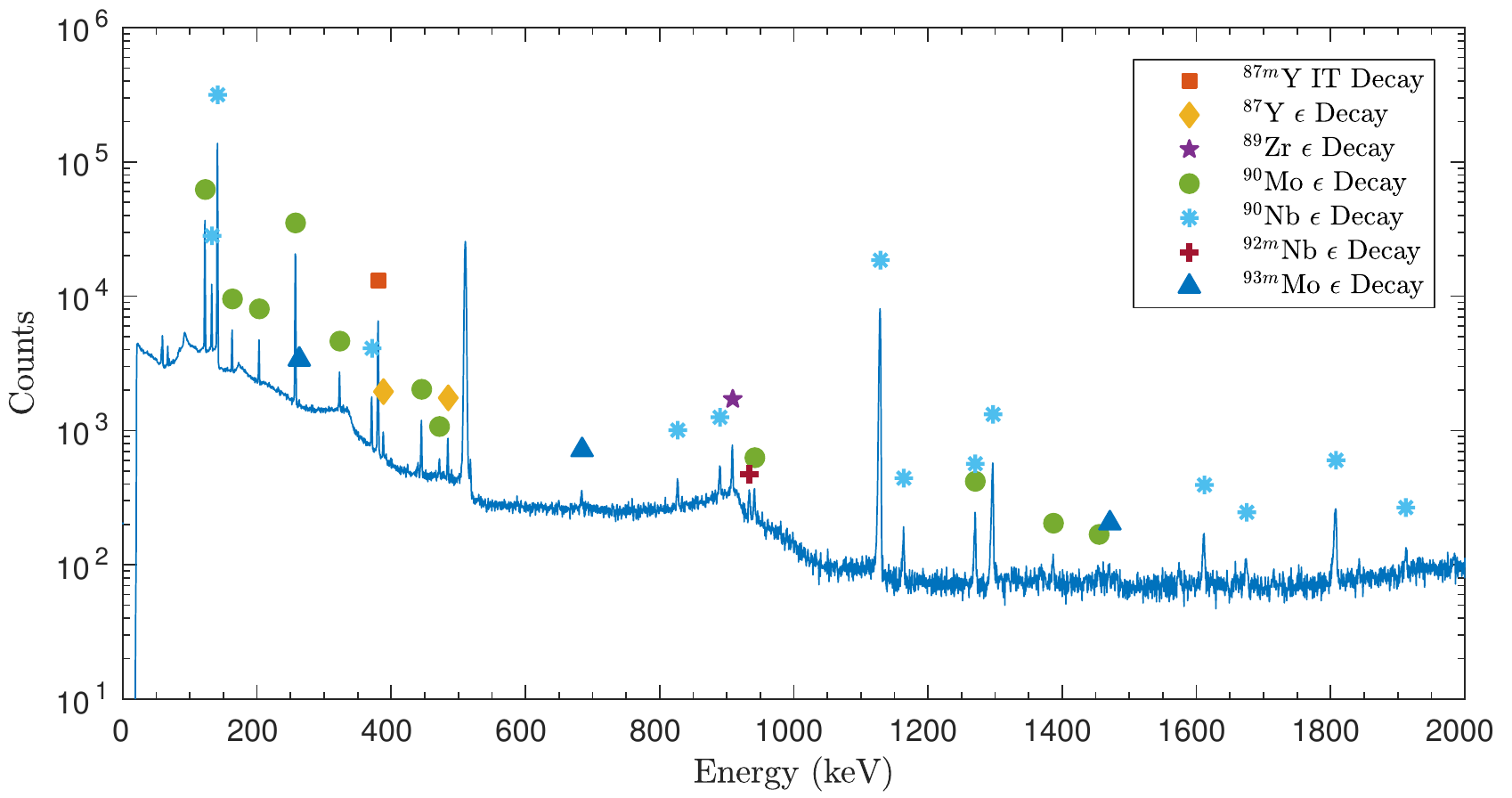}
 \caption{A gamma spectrum collected from an activated Nb foil at approximately 80 MeV. While the majority of observed reaction products are visible in this spectrum,  the \ce{^{90}Mo} decay lines, which form the basis of the \ce{^{93}Nb}(p,x)\ce{^{90}Mo} monitor reaction, are high in intensity and clearly isolated from surrounding peaks.}
 \label{fig:gspec}
\end{figure}

A single detector was used in this measurement, an ORTEC GEM Series (model \#GEM10P4-70)  High-Purity Germanium (HPGe) detector.
The detector is a mechanically-cooled coaxial p-type HPGe with a 1\,mm aluminum window, and a 49.2\,mm diameter, 27.9\,mm long crystal.
Samples were counted at fixed positions ranging 4.5--83.5\,cm (5\% maximum permissible dead-time) from the front face of the detector, with a series of standard calibration sources used to determine energy, efficiency, and pileup calibrations for each position.
The foils were counted  for a period of 2\,weeks following end-of-bombardment (EoB), to accurately quantify all induced activities,  with dead time never exceeding 5\%.
An example of one of the gamma-ray spectra collected in such a fashion is shown in \autoref{fig:gspec}.
For all spectra collected, net peak areas were fitted using the gamma spectrometry analysis code UNISAMPO \cite{Aarnio2001}, which has been shown to perform best in comparisons with other common analysis codes \cite{Jackman2014}.

Following  acquisition, the decaying product nuclei corresponding to each observed peak in the collected spectra were identified.
The calibrated detector efficiencies, along with gamma-ray intensities for each transition and  corrections for gamma-ray attenuation within each foil packet, were used to convert the net  counts in each fitted gamma-ray photopeak into an activity for the decay of the activation products.  
The nuclear decay data used in this work is tabulated in Tables \ref{tab:nudat_table_monitors} and \ref{tab:nudat_table_nb} of \ref{data}.
Data for photon attenuation coefficients were taken from the XCOM photon cross sections database  \cite{berger2011xcom}.
Decay gamma-rays from the product nuclei were measured at multiple points in time (up to 2\,weeks after EoB), and as nearly all of the product nuclei have multiple high-intensity gamma-rays, this provided independent activity measurements at each time point.
The total propagated uncertainty of the measured activity is the quadrature sum of the uncertainty in  fitted peak areas, uncertainty in detector efficiency calibration, and uncertainty in the gamma-ray branching ratio data.

Since many of the reaction products populated by energetic protons are more than one decay off of stability, many of these are produced not only  directly by reactions, but also indirectly by decay down a mass chain.
To this end, it is useful to differentiate between the types of cross sections reported in this work. 
For the first observable product nuclei in a mass chain, its (p,x) cross section will be reported as a cumulative cross section ($\sigma_c$), which is the sum of direct production of that nucleus, as well as decay of its  precursors and any other independent cross sections leading to that nucleus. 
Cumulative cross sections will be reported whenever it is impossible to use decay spectrometry to distinguish independent production of a nucleus from decay feeding.
For all remaining observed reaction products in the mass chain, and cases where no decay precursors exist, independent cross sections ($\sigma_i$) will be reported, allowing for determination of the independent production via subtraction  and facilitating comparison to reaction model calculations.

Corrections must be made for the decay of the various reaction products during the time between EoB and the spectrum acquisition, in order to calculate $A_0$, the initial activity at EoB, from the measured activities.
The use of  multiple gamma-rays at multiple points after EoB to calculate initial activities  for each observed product nucleus allows for a more accurate  determination of $A_0$ than simply basing its calculation off of a single gamma-ray observation.
For the case of cumulative cross sections, EoB activities were quantified by fitting the activities observed at multiple time points $t$ (since EoB) to the well-known radioactive decay law.
Nonlinear regression was used for this fitting process, minimizing on $\chi^2$\,/\,degree of freedom, so that not only would the uncertainty-weighted EoB activities be fitted, but that a 1-$\sigma$ confidence interval in $A_0$ could be reported as  well.
As with the gamma-ray intensities, all lifetimes used in this work are tabulated in Tables \ref{tab:nudat_table_monitors} and \ref{tab:nudat_table_nb} of \ref{data}.
In the case of independent cross sections, a similar process was followed, quantifying $A_i\pp{t=0} = A_{i,0}$, the EoB activity of nuclide $i$, by instead regressing to the solutions to the Bateman equation \cite{bateman1910solution,Cetnar2006}:
\begin{equation}
A_n\pp{t} = \lambda_n \sum_{i=1}^n \left[  N_{i,0} \times \pp{\prod_{j=i}^{n-1}\lambda_j} \times \pp{\sum_{j=i}^n \dfrac{e^{-\lambda_j t}}{\prod_{i\neq j}^n \pp{\lambda_i - \lambda_j}}  }   \right]
\end{equation}
where $j$ refers to a precursor nucleus populating a specific end-product.  
While higher-order terms were added if needed, typically for an isomeric state in a particular mass chain,  the second-order expansion ($n=2$) was often sufficient to quantify EoB activities in a mass chain, simplifying to:
\begin{equation}
A_2\pp{t} = \dfrac{A_{1,0}\lambda_2}{\lambda_1 - \lambda_2} \pp{e^{-\lambda_2 t} - e^{-\lambda_1 t}} + A_{2,0} e^{-\lambda_2 t}
\end{equation}
In these cases, the previously-quantified EoB activities from decay precursors ($A_{1,0}$, etc) would be substituted in, so that the feeding contributions from decay could be separated and an independent cross section reported.
After quantifying the cumulative EoB activities at the top of a mass chain and all subsequent independent EoB activities, these will be later used to report the various cross sections for all observed reaction products and isomeric states.


\subsection{Proton fluence determination}\label{sec:dosimetry}


%
%
%
In addition to the LANSCE-IPF beamline's direct beam current measurements, thin \ce{^{nat}Al} and \ce{^{nat}Cu} foils were included along with the \ce{^{nat}Nb} targets at each energy position, to provide more sensitive beam current monitors.
The IAEA-recommended \ce{^{nat}Al}(p,x)\ce{^{22}Na}, \ce{^{nat}Al}(p,x)\ce{^{24}Na}, \ce{^{nat}Cu}(p,x)\ce{^{56}Co}, \ce{^{nat}Cu}(p,x)\ce{^{62}Zn}, and \ce{^{nat}Cu}(p,x)\ce{^{65}Zn} monitor reactions were used for  proton fluence measurement \cite{gul2001charged}.
Due to the large energy degradation between the front and  back of the target stack, a non-trivial broadening of the proton energy distribution was expected for all monitor and target foils.
As a result, the integral form of the well-known activation equation was used to accurately determine proton fluence ($I \Delta t $) in each monitor foil:
\begin{equation}
I \Delta t = \dfrac{A_0 \Delta t}{\rho \Delta r \pp{1-e^{-\lambda \Delta t}} \int \sigma\pp{E} \dfrac{d\phi}{dE} dE}
\end{equation}
where $A_0$ is the EoB activity for the monitor reaction product, $I$ is the proton current, $\rho \Delta r$ is the foil's areal density, $\lambda$ is the monitor reaction product's decay constant, $\Delta t$ is the length of irradiation, $\sigma\pp{E}$ is the IAEA recommended cross section at energy $E$, and $\frac{d\phi}{dE}$ is the differential proton fluence.
Using this formalism, the quantified EoB activities for each monitor reaction may be converted into a measured proton fluence at each energy position.

The propagated uncertainty in proton fluence is calculated as the quadrature sum of the uncertainty in quantified EoB activity, uncertainty in the duration of irradiation (conservatively estimated at 60 s, to account for any transient changes in beam current), uncertainty in foil areal density, uncertainty in monitor product half-life (included, but normally negligible), uncertainty in IAEA recommended cross section, and uncertainty in differential proton fluence.
Of these, the first four contributions are all easily quantified in the preparation and execution of a stacked target irradiation;  the last two contributions prove to be more nuanced, however.
The uncertainty in proton fluence for irradiated monitor foils is derived from statistical uncertainty in the modeling of proton transport in the stack irradiation, discussed in \autoref{sec:proton_transport}.
The uncertainty in IAEA recommended cross section values must be estimated indirectly, as no uncertainty in the  recommended cross sections is provided in the current IAEA evaluation.
Fortunately, the recommended cross section values for each monitor reaction tend to closely match one of the   selected experimental source data sets used in their evaluation.
Since these data sets have listed uncertainties in the original manuscripts, uncertainties in  IAEA recommended cross section values have been estimated by the uncertainty in the data set most closely matching the  IAEA recommended  values.
For the monitor reactions employed in this work, these data sets are G. Steyn (1990) for  \ce{^{nat}Al}(p,x)\ce{^{22}Na} \cite{Steyn1990}, M. Uddin (2004) for \ce{^{nat}Al}(p,x)\ce{^{24}Na} \cite{Uddin2004}, and S. Mills (1992) for \ce{^{nat}Cu}(p,x)\ce{^{56}Co}, \ce{^{nat}Cu}(p,x)\ce{^{62}Zn}, and \ce{^{nat}Cu}(p,x)\ce{^{65}Zn} \cite{Mills1992}.

\subsection{Proton transport calculations}\label{sec:proton_transport}

Initial estimates of the proton beam energy in all foils were calculated using the Anderson \& Ziegler (A\&Z) stopping power formalism \cite{Andersen_Ziegler_1977,Ziegler1985,Ziegler1999}.
These estimates of average beam energy in each foil are useful for the preliminary stack design.
However, for final energy and fluence determinations, a more rigorous method of proton transport modeling is needed.
The Monte Carlo N-Particle transport code MCNP6.1 was used for simulation of the full 3-D target stack, including determination of the full proton energy distribution for each stack position   \cite{Goorley2012}.
MCNP6 provides a far more robust method of proton transport, as it is able to account for beam losses due to scattering and reactions, as well as production of secondary particles.
As it is a Monte Carlo-based code, the uncertainty in energy distribution scales inversely with the number of source protons simulated.  $10^8$ source protons were used for all simulations, which places the statistical uncertainty in proton energy distributions at less than 0.01\%.

The ability to model the full energy distribution in each target position is vital for stacked target irradiations, due to the progressively larger energy straggling towards the rear of the stack.
The initial proton beam has a finite energy spread (an approximately 0.1\,MeV Gaussian width at 100\,MeV), and since stopping power for charged particles is inversely proportional to their energy, the low-energy tail of the energy distribution is degraded more in each stack element than the high-energy tail.
This effect compounds  towards the rear of the stack, creating a significantly broadened low-energy tail, and a progressively larger net shift of the centroid to a lower energy. 
To account for this increasing energy uncertainty, a suitably representative energy must be established for  each foil in the target stack.
In this work, the flux-weighted average proton  energy in each foil, $\langle E \rangle$,  represents the energy centroid for protons in a target stack component, calculated using the energy distributions $\frac{d\phi}{dE}$ from MCNP6 modeling of proton transport:
\begin{equation}
\langle E \rangle = \dfrac{{\displaystyle\int E \dfrac{d\phi}{dE} dE}}{{\displaystyle\int \dfrac{d\phi}{dE} dE}}
\end{equation}
Likewise, to represent the energy uncertainty for each stack position, the full width at half maximum (FWHM) of the MCNP6-modeled energy distribution is chosen for each energy position reported.
While most experimental uncertainties are reported at the 1$\sigma$ level, the $2.355\sigma$ FWHM is used here to ensure at the 98\% confidence interval that this width includes  the \enquote{true} energy centroid value.

The \enquote{variance minimization} techniques  described by  Graves \etal\  have been employed here to further reduce the uncertainty in proton energy assignments     \cite{Graves2016}.
This method is based on the assumption that the independent measurements of proton fluence from the five monitor reactions used in this work should all be consistent at each energy position.
If the monitor reaction cross sections and MCNP6-modeled energy distributions are both accurate, disagreement in the  observed proton fluences is due to poorly characterized stopping power in simulations, or a systematic error in the 
areal densities of the stack components \cite{Graves2016,Marus2015}. 
This disagreement is minor at the front of the stack, and gets progressively worse as the beam is degraded, due to the compounded effect of systematic uncertainties in stack areal densities.

\begin{figure}
 \centering
 \includegraphics[width=0.5\linewidth]{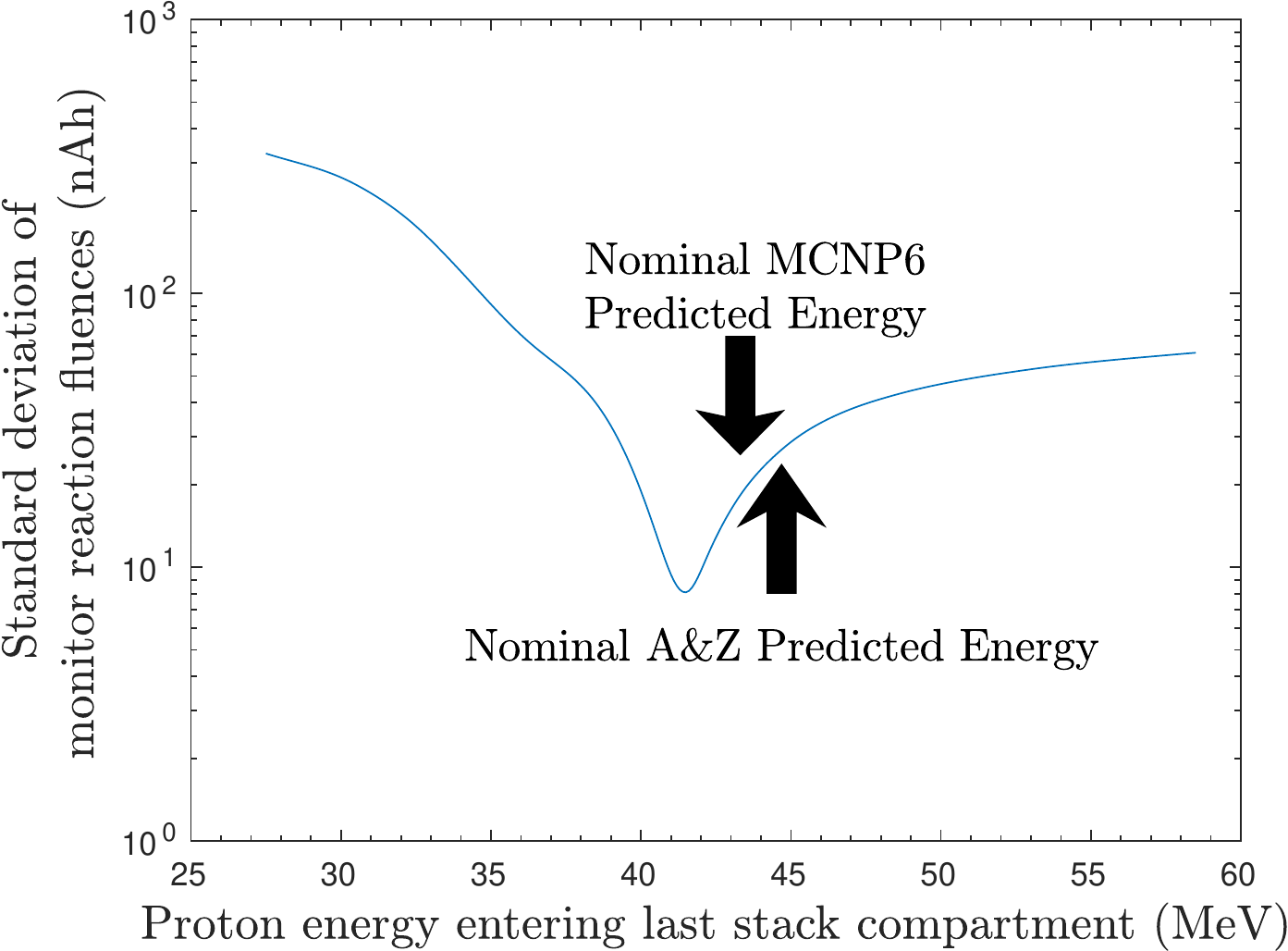}

 \caption{Result of the variance minimization performed by adjusting the degrader density in MCNP6 simulations of the target stack.  A flux-weighted average proton energy of 41.34 MeV entering the last energy position creates a clear minimum in observed reaction fluence variance, corresponding to an areal density 2.52\% greater than nominal. The variance minimum occurring at a lower incident energy than nominal MCNP6 and A\&Z calculations indicates that there exists an additional systematic beam degradation not accounted for in modeling of proton transport in the stack design.}
 \label{fig:variation_curve}
\end{figure}

Due to the  
significantly greater areal density of the thick 6061 aluminum degraders as compared to the other stack elements (nominal 3--5\,mg/cm$^2$, relative to nominal 1000--1400\,mg/cm$^2$),
the areal density of each of the 6061 aluminum degraders  were varied uniformly in MCNP6 simulations  by a factor of up to $\pm$25\% of nominal values, to find the effective density which minimized variance in the measured proton fluence at the lowest energy position (Al-6, Cu-6).
This lowest energy position was chosen as a minimization candidate, as it is most sensitive to systematic uncertainties in stack design.
The results of this minimization technique, shown in \autoref{fig:variation_curve}, indicate a clear minimum in proton fluence variance for flux-weighted average 41.34\,MeV protons entering the last energy position.
This is approximately 2\,MeV lower than the nominal MCNP6 simulations, and approximately 3\,MeV lower than nominal A\&Z calculations, both of which used the nominal 2.80\,g/cm$^3$ measured density of the 6061 aluminum degraders.
This energy corresponds to a 6061 aluminum areal density of 2.52\% greater than nominal measurements, and serves as a lump correction for other minor systematic uncertainties in stack design, including stack areal densities and incident beam energy.

\begin{figure}
    \centering
    \subfloat{
        \centering
        \hspace{-5pt}\subfigimg[width=0.5\textwidth]{a)}{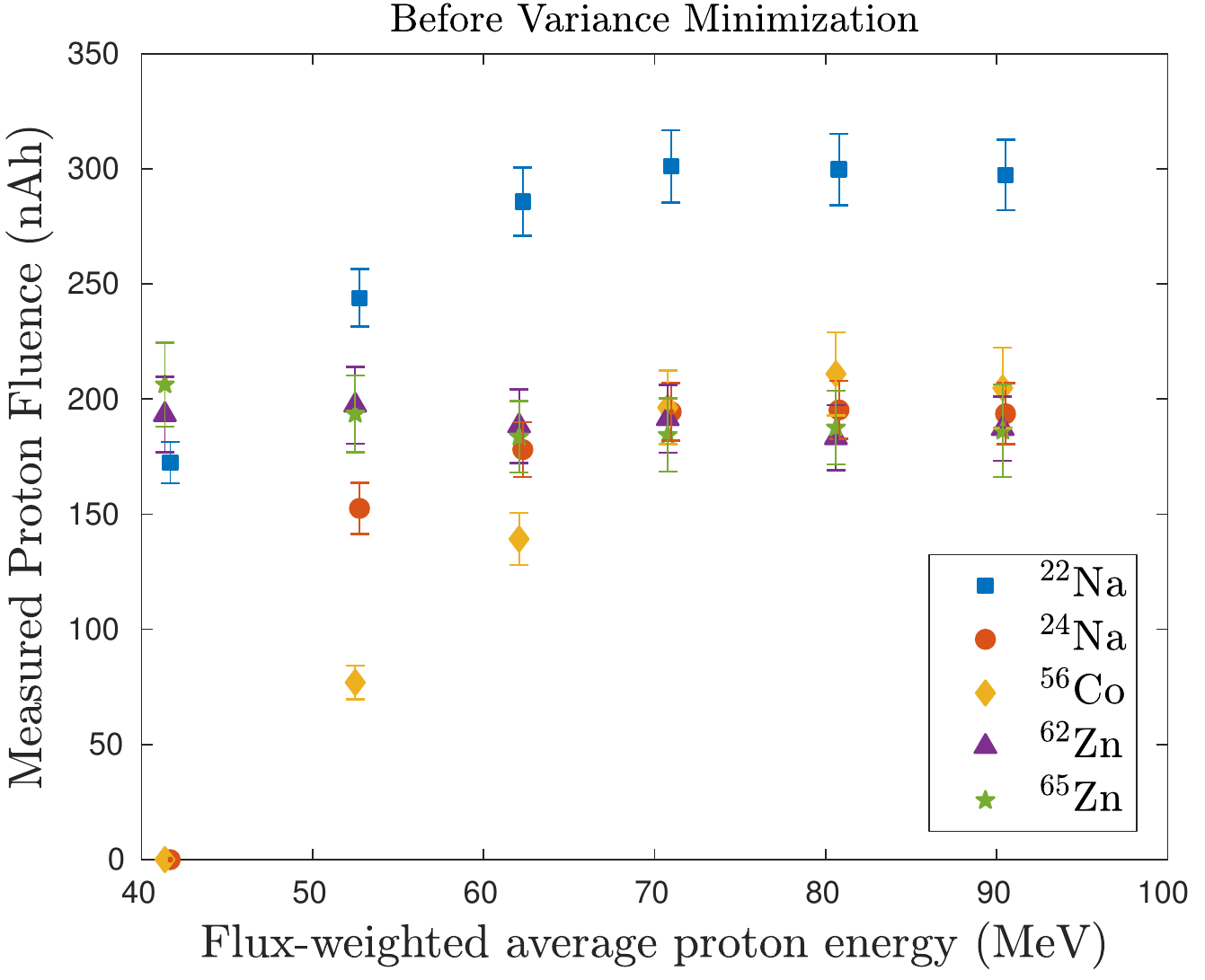}{50}
         \label{fig:before_minimization}
   \hspace{-5pt}}%
     \subfloat{
        \centering
        \subfigimg[width=0.5\textwidth]{b)}{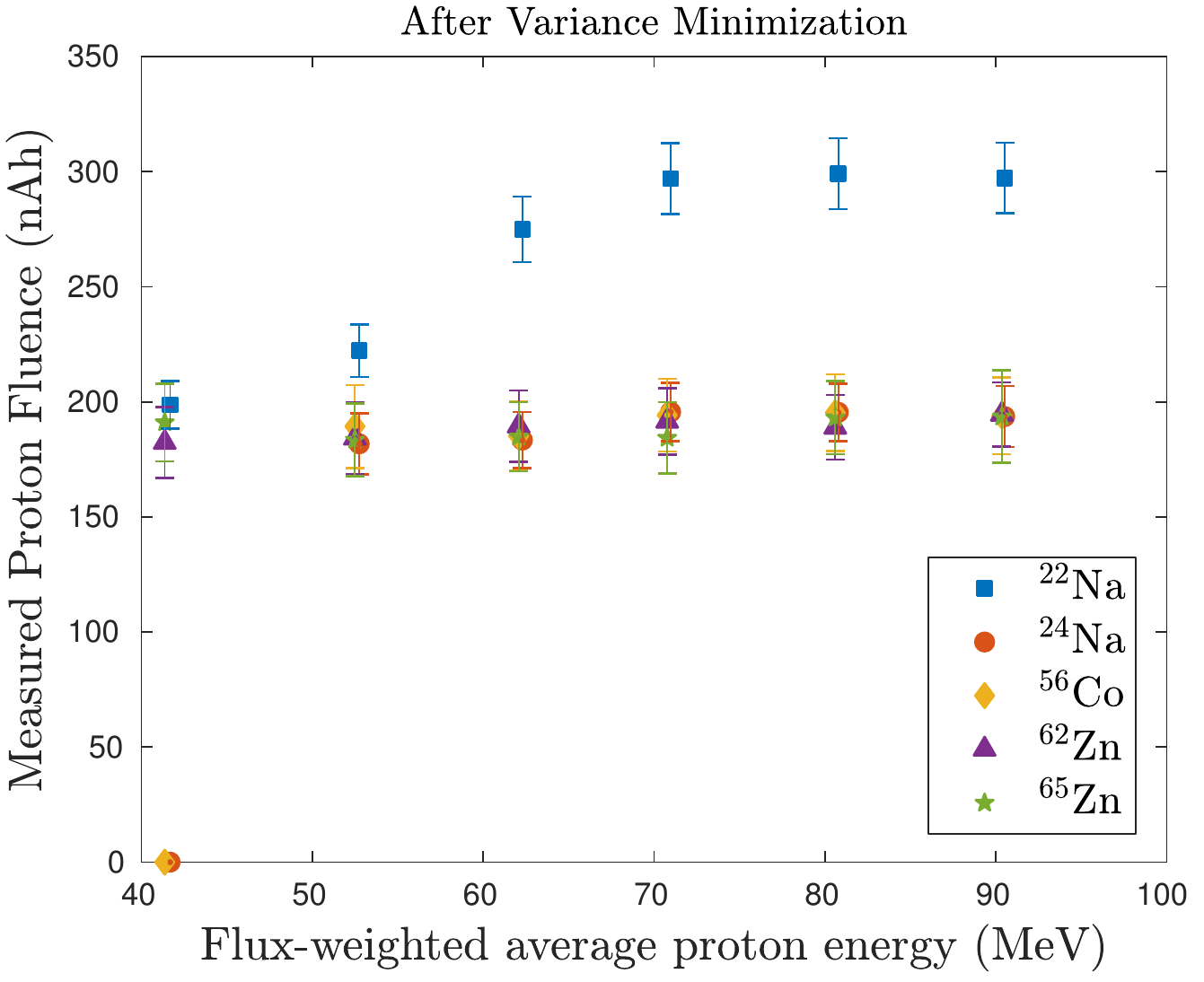}{50}
         \label{fig:after_minimization}
   \hspace{-5pt}}%
    \caption{Results of variance minimization through enhancement of the effective areal density of the 6061 aluminum degraders by 2.52\%. A noticeable reduction of variance in measured proton fluence is seen,  particularly at the  rear stack positions. Following minimization, additional apparent fluence is observed in the  \ce{^{nat}Al}(p,x)\ce{^{22}Na} and \ce{^{nat}Al}(p,x)\ce{^{24}Na} monitor channels, due to contamination from \ce{^{nat}Si}(p,x)\ce{^{22,24}Na} on the silicone adhesive used for sealing foil packets.}
     \label{fig:variance_mins}
\end{figure}

\begin{figure}
    \centering
    \subfloat{
        \centering
        \hspace{-5pt}\subfigimg[width=0.5\linewidth]{a)}{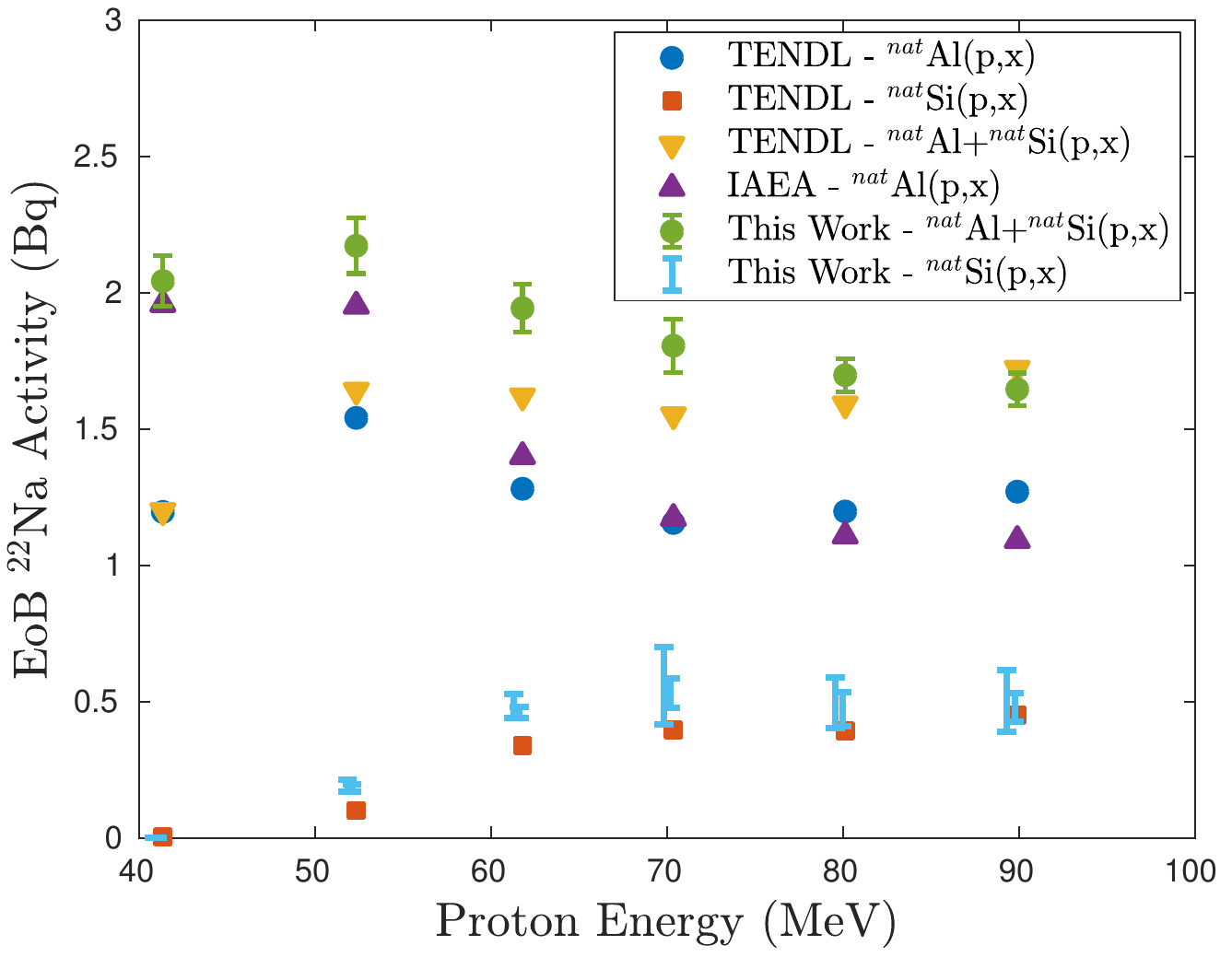}{50}
   \hspace{-5pt}}%
     \subfloat{
        \centering
        \subfigimg[width=0.5\linewidth]{b)}{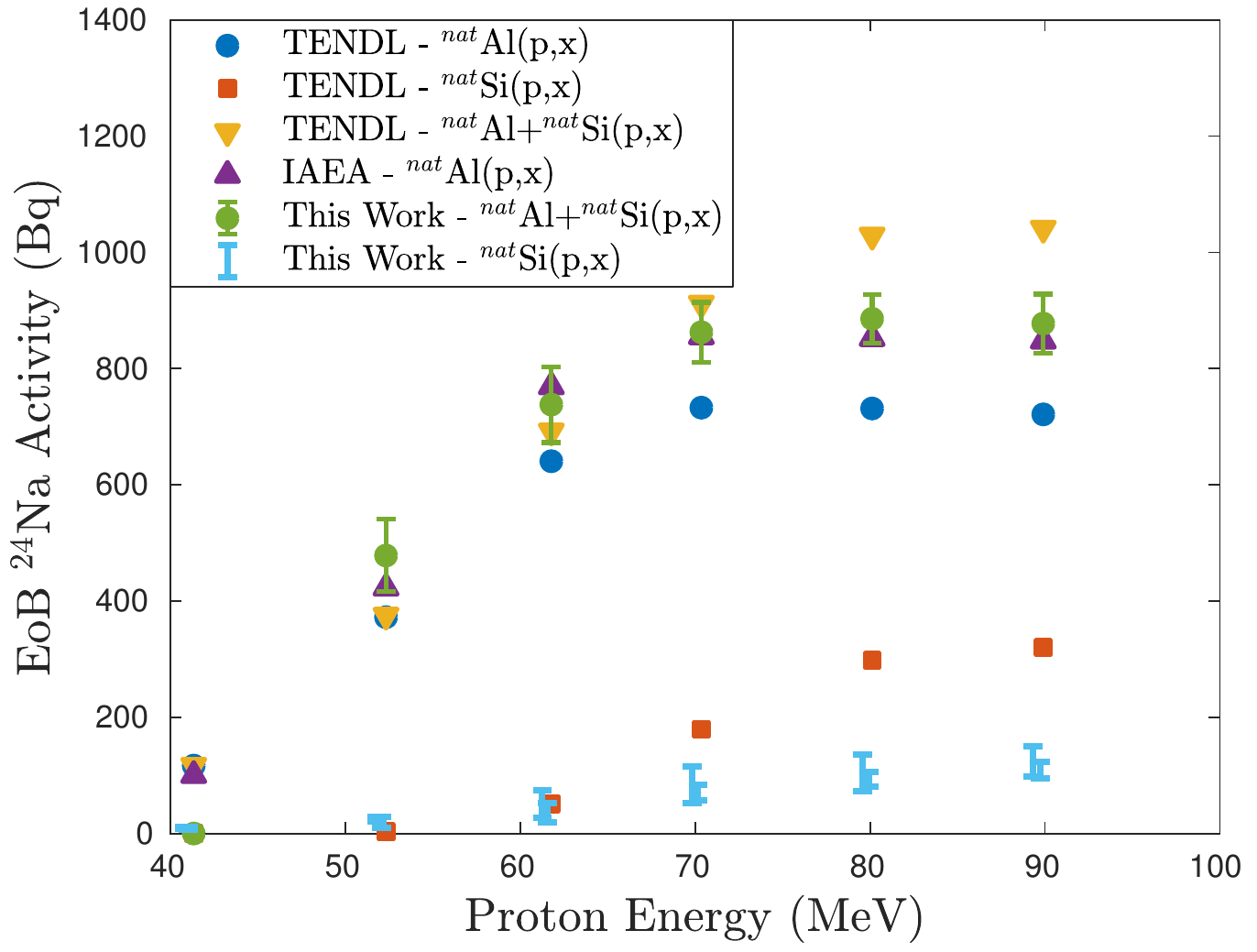}{180}
         \label{fig:Na24_activity_compare}
   \hspace{-5pt}}%
    \caption{Estimates of EoB \ce{^{nat}Al}(p,x)\ce{^{22,24}Na} and \ce{^{nat}Si}(p,x)\ce{^{22,24}Na} activities using TENDL-2015 cross sections, in comparison with the IAEA recommended \ce{^{nat}Al}(p,x)\ce{^{22,24}Na} cross sections. At low energies, experimentally observed apparent \ce{^{22,24}Na} activities in each Al foil packet are consistent with IAEA recommendations, but diverge at higher energies as the \ce{^{nat}Si}(p,x)\ce{^{22}Na} exit channels begin to open up. \ce{^{22,24}Na} activities consistent with TENDL-2015 estimates are observed in each Nb and Cu foil packet as well, confirming that contamination may be attributed to activation of silicone adhesives.}
     \label{fig:Na_activity_compare}
\end{figure}

The impact of this variance minimization is  clearly  seen in   \autoref{fig:variance_mins}.
As expected, the 2.52\% increase in 6061 aluminum areal density has an almost negligible impact on the higher-energy positions, but causes a progressively larger downshift  in proton energies at the later energy positions.
In addition, as one moves to the rear  positions, the disagreement in the independent proton fluence measurements is reduced.
It is worth noting that the proton fluence measured by the \ce{^{nat}Al}(p,x)\ce{^{22}Na} monitor reaction (threshold 21.0\,MeV) is consistently higher in magnitude than all other monitor channels, with an increasing disparity at higher energies.
This disparity is due to silicon in the Kapton tape (comprised of a silicone adhesive layer on a polyimide backing) used for sealing the foil packets,  making up approximately 10\% of the silicone on a stoichiometric basis.
The \ce{^{22}Na} and \ce{^{24}Na} monitor channels can also be populated off of natural silicon  (92.2\%\,\ce{^{28}Si}), predominantly via \ce{^{28}Si}(p,$\alpha$2pn)\ce{^{22}Na} (threshold 35.3\,MeV) and \ce{^{28}Si}(p,4pn)\ce{^{24}Na} (threshold 44.6\,MeV).
\ce{^{29}Si} and \ce{^{30}Si} are also potential targets for (p,x)\ce{^{22,24}Na}, albeit with higher energetic thresholds and smaller cross sections.
The attribution of excess Al(p,x)\ce{^{22,24}Na} activity to the silicone adhesive is supported by the observation of \ce{^{22}Na} and \ce{^{24}Na} activities in all Cu and Nb foil positions.

\ce{^{nat}Si}(p,$\alpha$2pn) 
is competitive with the \ce{^{nat}Al}(p,x) 
production route,
seen when comparing the total measured activities of \ce{^{22,24}Na}  
in each Al foil packet, 
relative to the expected EoB activities for each reaction channel 
(\autoref{fig:Na_activity_compare}).
Since no evaluated  cross section data exists in this energy region  for  \ce{^{28}Si}(p,x)\ce{^{22}Na}  (and only minimal \ce{^{nat}Si} data exists),  the TENDL-2015 library is used to estimate the expected relative EoB activities for \ce{^{nat}Al}(p,x)\ce{^{22,24}Na} and \ce{^{nat}Si}(p,x)\ce{^{22,24}Na}, 
relative to IAEA recommended \ce{^{nat}Al}(p,x)\ce{^{22,24}Na} cross sections.
Several observations are immediately obvious.
At lower energies, the magnitude of \ce{^{nat}Al}(p,x)\ce{^{22}Na} is large compared to \ce{^{nat}Si}(p,x)\ce{^{22}Na}, which is why the \ce{^{nat}Al}(p,x)\ce{^{22}Na} monitor agrees in fluence at the 40 (and almost at the 50) MeV position.  
At higher energies, the apparent \ce{^{nat}Al}(p,x)\ce{^{22}Na} activity begins to diverge from the IAEA expected activities as     \ce{^{nat}Si}(p,x)\ce{^{22}Na} production begins to open up,  which accounts for the nearly 50\% apparent excess fluence in \ce{^{22}Na} between 60--90\,MeV.
For  \ce{^{24}Na} production, we see  similar behavior, with only a minor increase in apparent \ce{^{24}Na} activity,  since the observed \ce{^{nat}Si}(p,x)\ce{^{24}Na} yield remains consistently low in magnitude.
The observed \ce{^{24}Na} activities also follow the shape of the TENDL-2015 \ce{^{nat}Si}(p,x)\ce{^{24}Na} yields, albeit smaller in magnitude at the higher energy positions.

\begin{figure}
    \centering
    \subfloat{
        \centering
        \hspace{-5pt}\subfigimg[width=0.5\textwidth]{a)}{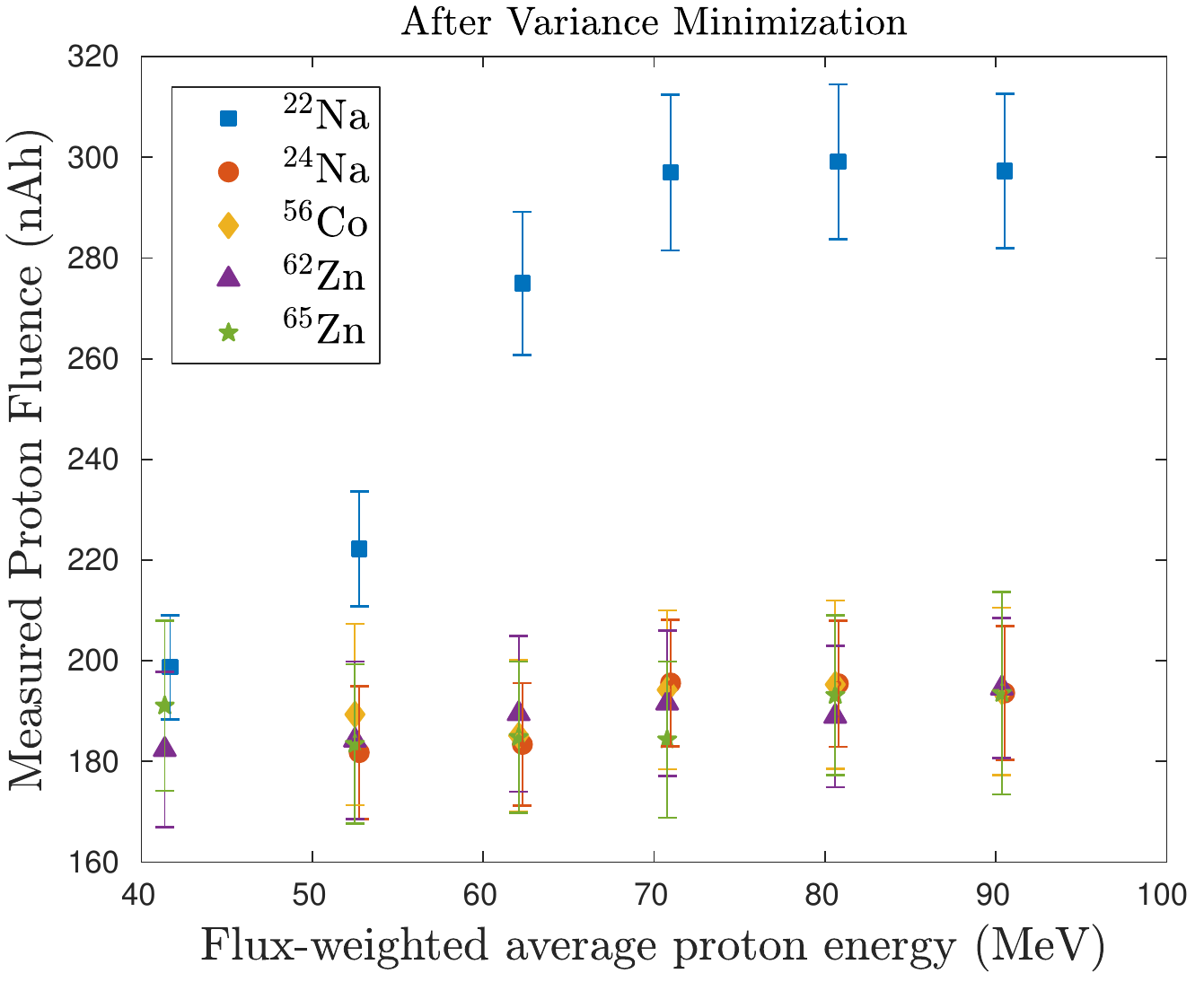}{80}
         \label{fig:before_subtraction}
   \hspace{-5pt}}%
     \subfloat{
        \centering
        \subfigimg[width=0.5\textwidth]{b)}{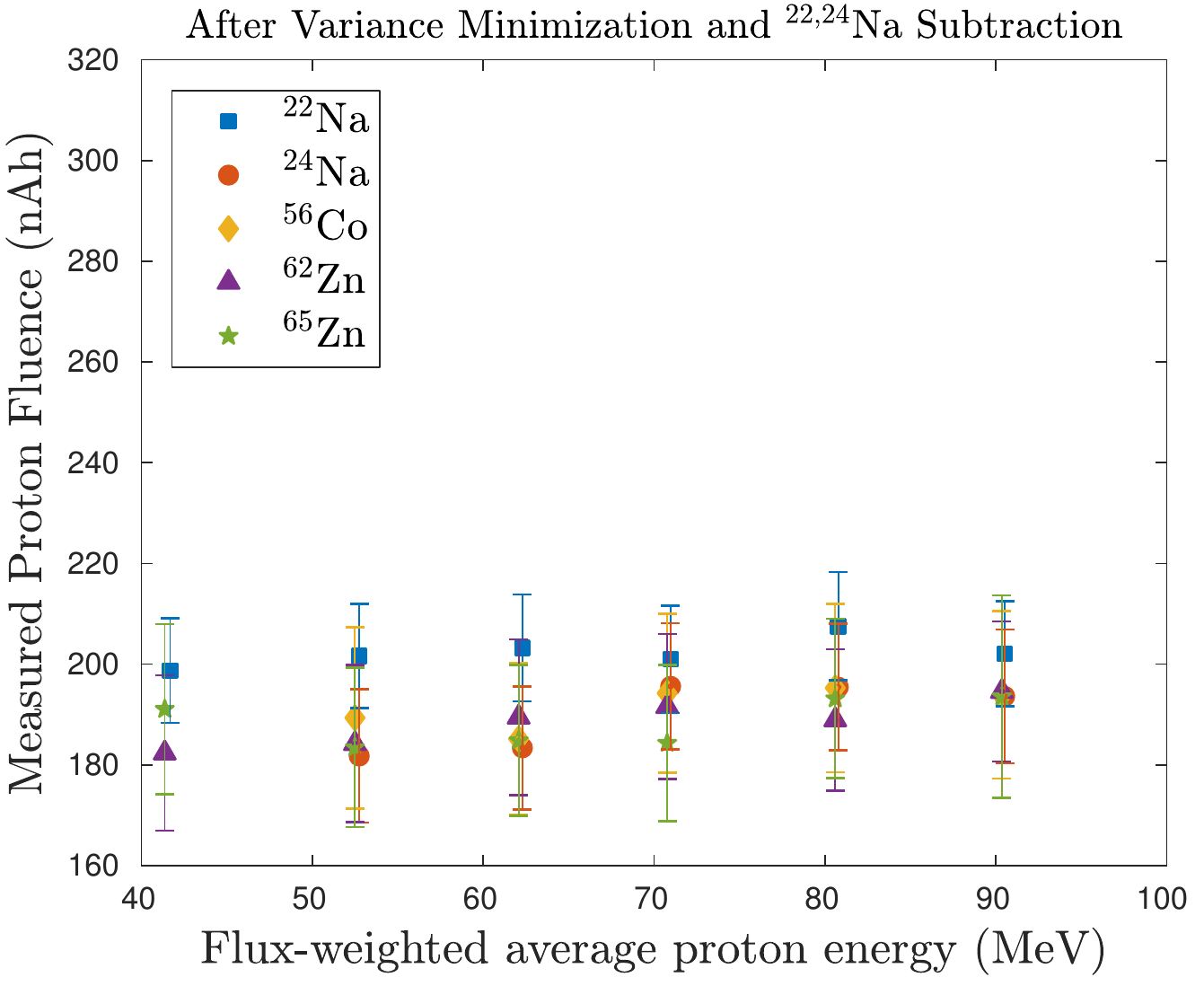}{80}
         \label{fig:after_subtraction}
   \hspace{-5pt}}%
    \caption{The \enquote{extra fluence} observed in the  \ce{^{nat}Al}(p,x)\ce{^{22}Na} and \ce{^{nat}Al}(p,x)\ce{^{24}Na} monitor channels is caused by contamination from \ce{^{nat}Si}(p,x)\ce{^{22,24}Na} on the silicone adhesive used for sealing foil packets. Following subtraction of \ce{^{22,24}Na} activities observed in the silicone adhesive of Nb and Cu foils in the same energy \enquote{compartment}, the consistency of the \ce{^{nat}Al}(p,x)\ce{^{22}Na} monitor reaction improves  dramatically.  By excluding these contaminated channels, the remaining 3 independent monitor reactions serve to minimize uncertainty in stack energy assignments and incident fluence.}
     \label{fig:na_subtraction}
\end{figure}

There are several important  conclusions to be drawn from this simple estimate using the TENDL   \ce{^{nat}Si}(p,x)\ce{^{22,24}Na} yields.
The observation of the \ce{^{22,24}Na} activities in Cu and Nb foils  represents an indirect measurement of the \ce{^{nat}Si}(p,x)\ce{^{22,24}Na} cross sections, but  will not be reported due to 
uncertainties in the areal density of the Si in the adhesive.
However, if we assume a 10\% Si stoichiometric basis and an areal density of 4.79\,mg/cm$^2$ (based on bulk density),
we can subtract out the measured \ce{^{22,24}Na} activity at each Nb and Cu foil position (correcting for the minor difference in proton energy between adjacent foils) from the apparent \ce{^{22,24}Na}  activities observed in each Al foil packet, in order to obtain the \enquote{true} or uncontaminated fluence via the Al monitor reactions, shown  
in \autoref{fig:na_subtraction}.
Following subtraction, the \ce{^{22,24}Na} fluences become more consistent with other monitor reaction channels, 
though  \ce{^{22}Na} fluence remains 3--6\% higher than the weighted mean of the remaining monitor reaction channels.
While the dramatic improvement in monitor reaction consistency builds confidence, in the interest of surety and because they are consistent, only the \ce{^{nat}Cu}(p,x)\ce{^{56}Co}, \ce{^{nat}Cu}(p,x)\ce{^{62}Zn}, and \ce{^{nat}Cu}(p,x)\ce{^{65}Zn} monitor reaction channels will be used for fluence determination for the reported cross sections.
This serves as a pointed example of the importance of selecting monitor reaction products inaccessible through channels aside from the primary reaction (\ce{^{nat}Al}(p,x)\ce{^{22,24}Na}, in this case ), as noted previously.


\begin{figure}
 \centering
 \includegraphics[width=0.5\linewidth]{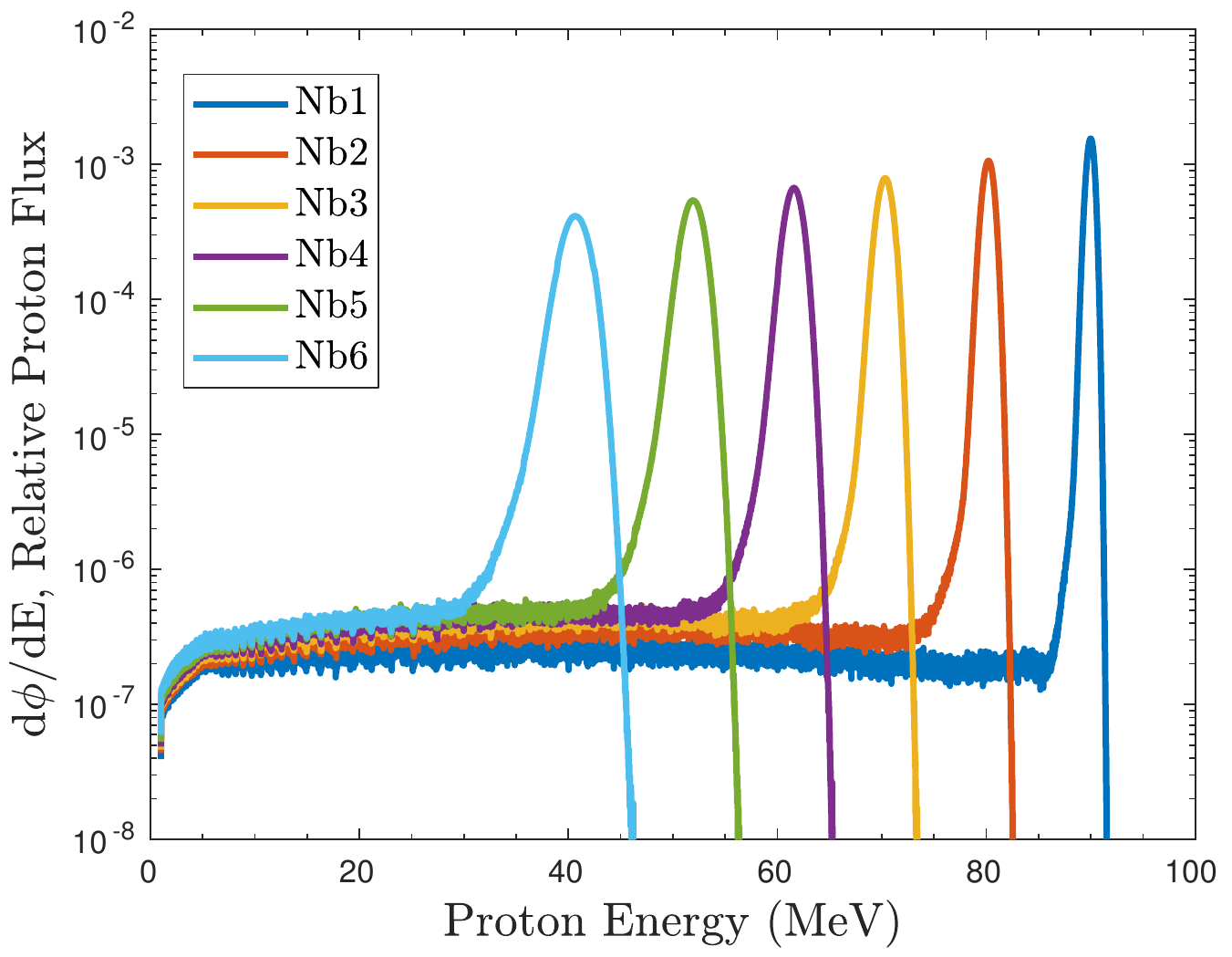}
 \caption{Final variance minimized incident proton energy distributions for the Nb foils, as simulated in MCNP6. The distribution tallies in each foil are all normalized to be per source proton, which was $10^8$ in all simulations. As the beam is degraded, proton energy distributions become visibly broadened due to straggling, and drop in magnitude due to scattering losses.}
 \label{fig:Nb_ptallies}
\end{figure}

Using this variance minimized degrader density, the final incident proton  energy distributions $\frac{d\phi}{dE}$ from MCNP6 simulation are shown for the six irradiated Nb foils in \autoref{fig:Nb_ptallies}. 
As expected, the energy distribution becomes increasingly more broadened at the lower energy positions, as a result of the beam energy degradation.
In addition, as the beam becomes more degraded, the magnitude of the peak of each energy distribution (as well as the integral of each distribution) is reduced, as beam fluence is lost due to scattering, and the peak-to-low-energy-tail ratio increases as more  secondary protons are produced upstream.
As with the monitor foils, these distributions were used to calculate the  energy centroid  (as the  flux-weighted average proton  energy) and  uncertainty (as the FWHM of the distribution) for the final proton energy assignment of each Nb foil.

\begin{figure}
 \centering
 \includegraphics[width=0.5\linewidth]{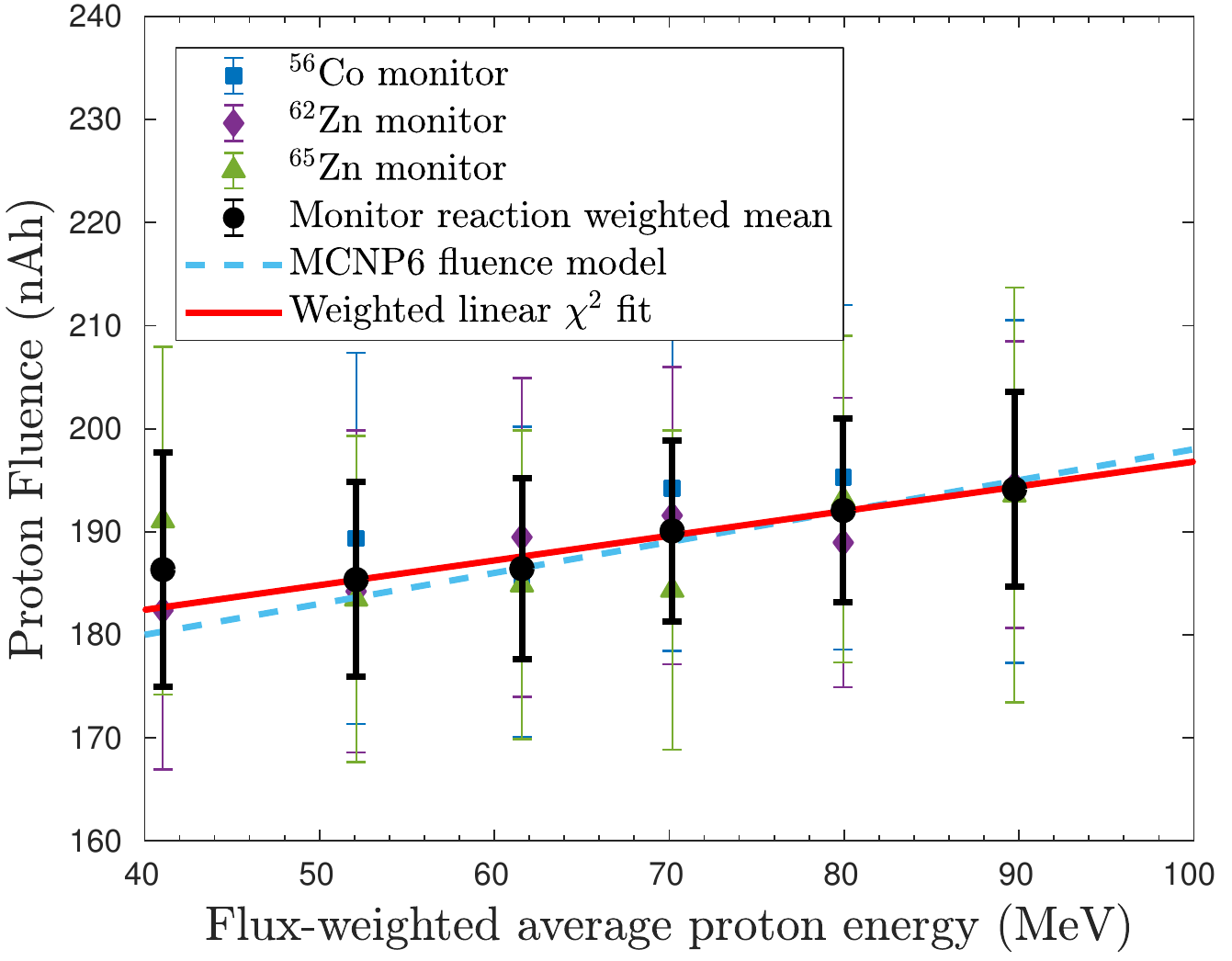}
 \caption{Final uncertainty-weighted mean proton fluences throughout the target stack, based on the variance-minimized observed fluence from the the  \ce{^{nat}Cu}(p,x)\ce{^{56}Co}, \ce{^{nat}Cu}(p,x)\ce{^{62}Zn}, and \ce{^{nat}Cu}(p,x)\ce{^{65}Zn} monitor reactions. 
  The fluence  drops by approximately 7.2--8.9\% from the incident fluence of 196.9--198.8 nAh over the length of the target stack, based on fluence loss models from MCNP6 simulations and an empirical fit to  fluence measurements.}  
 \label{fig:fluence_plot}
\end{figure}

An enhanced version of the final \ce{^{nat}Cu}(p,x)\ce{^{56}Co}, \ce{^{nat}Cu}(p,x)\ce{^{62}Zn}, and \ce{^{nat}Cu}(p,x)\ce{^{65}Zn} monitor reaction fluences is shown in \autoref{fig:fluence_plot}.
Without the reliable use of the  \ce{^{nat}Al}(p,x)\ce{^{22}Na} and \ce{^{nat}Al}(p,x)\ce{^{24}Na} monitor channels, local interpolation cannot be used for fluence assignment to the Nb foils, and global interpolation is reliant upon a validated model for fluence loss.
The uncertainty-weighted mean  for the three \ce{^{nat}Cu}(p,x) monitor channels was calculated at each energy position, to determine the final fluence assignments for the Nb and Cu foils.
Uncertainty in proton fluence  is likewise calculated by error propagation of the fluence values  at each energy position.
These weighted-mean fluences are  plotted  in \autoref{fig:fluence_plot}, along with the estimated fluence according to both  MCNP6 transport 
and an uncertainty-weighted linear $\chi^2$ fit to the individual monitor channel fluence measurements.
Both models reproduce the observed fluence data consistently within uncertainty, with the MCNP6 model predicting a slightly greater fluence loss throughout the stack.
These models are used purely to provide an extrapolation from the 90\,MeV energy position back to the \enquote{front} of the stack at 100\,MeV, to compare with the nominal fluence measured by  IPF upstream current monitors.

\subsection{Calculation of measured cross sections}\label{sec:calcs_sec}


Using the quantified EoB activities along with the variance-minimized proton fluence, it is possible to calculate the final cross sections for the various observed Nb(p,x) reactions.
While thin ($\approx$ 22\,mg/cm$^2$) Nb foils were irradiated to minimize the energy width of these cross section measurements, it is important to note that all cross sections reported here are flux-averaged  
over the energy distribution subtended by each foil, as seen in \autoref{fig:Nb_ptallies}.
For both the cumulative and independent activities quantified, cross sections were calculated as:
\begin{equation}
\sigma = \dfrac{A_0 }{\rho \Delta r I \pp{1-e^{-\lambda \Delta t}} }
\end{equation}
where $A_0$ is the EoB activity for the monitor reaction product, $I$ is the proton current, $\rho \Delta r$ is the foil's areal density, $\lambda$ is the monitor reaction product's decay constant, and $\Delta t$ is the length of irradiation.
The beam current, measured using an inductive pickup, remained stable for the duration of the irradiation, with the exception of approximately 70\,s of downtime,  occurring approximately 3\,min into irradiation.
The propagated uncertainty in cross section is calculated as the quadrature sum of the uncertainty in quantified EoB activity (which includes uncertainty in detector efficiencies), uncertainty in the duration of irradiation (conservatively estimated at 60\,s, to account for any transient changes in beam current), uncertainty in foil areal density, uncertainty in monitor product half-life (included, but normally negligible),  and uncertainty in proton current (quantified by error propagation of the monitor reaction fluence values  at each energy position, as seen in \autoref{fig:fluence_plot}).


%

\section{Results}

After irradiation, all foils were confirmed to still be sealed inside their Kapton packets, verifying that no activation products were lost due to packet failure and dispersal.
In addition, each activated foil had a small \enquote{blister} under the Kapton tape layer, caused by a combination of thermal swelling and the formation of short-lived beta activities.
This blister   shows the location where the primary proton beam was incident upon the foil.
The \ce{^{nat}Cu}(p,x)\ce{^{56}Co}, \ce{^{nat}Cu}(p,x)\ce{^{62}Zn}, and \ce{^{nat}Cu}(p,x)\ce{^{65}Zn} monitor reactions were used to determine the uncertainty-weighted mean fluence at each energy position (seen in \autoref{fig:fluence_plot}).
A fluence of 198.8$\pm$6.7\,nAh was calculated to be incident upon the target stack using the MCNP6 fluence model, and a  fluence of 196.9$\pm$11.3\,nAh using the linear fit model, both of which are consistent with the nominal fluence of 205.9\,nAh based on IPF upstream current monitors.
As fluence loss in the target box's entrance window scales with $\sigma_{\mathrm{tot}}\rho\Delta r$, it is expected that an extrapolation back to the stack entrance will underestimate the nominal fluence incident upon the box.
This incident fluence dropped by approximately 8.9\% to  180.9$\pm$5.4\,nAh (and by 7.2\% to  182.7$\pm$13.5\,nAh using the linear fit model) over the length of the target stack, which is consistent with similar measurements at IPF in the past \cite{Graves2016}.
This loss of fluence is due to a combination of 
(p,x) reactions throughout the target stack, as well as large-angle deflections (primarily in the aluminum degraders) from scattering of the beam.

Using the final proton fluence at each energy position, cross sections for  \ce{^{51}Cr},  \ce{^{52g}Mn}, \ce{^{52m}Mn}, \ce{^{54}Mn}, \ce{^{55}Co}, \ce{^{56}Ni}, \ce{^{57}Ni}, \ce{^{57}Co},  \ce{^{58g}Co}, \ce{^{58m}Co}, \ce{^{59}Fe}, \ce{^{60}Co}, \ce{^{61}Cu}, and \ce{^{64}Cu} were extracted for (p,x) reactions  on \ce{^{nat}Cu} foils in the 40--90\,MeV region, as recorded in \autoref{tab:cu_rp_table}.
For  (p,x) reactions on \ce{^{nat}Nb} foils, the (p,x) cross sections for \ce{^{82m}Rb}, \ce{^{83}Sr}, \ce{^{85g}Y}, \ce{^{85m}Y}, \ce{^{86}Zr}, \ce{^{86}Y}, \ce{^{87}Zr}, \ce{^{87g}Y}, \ce{^{87m}Y}, \ce{^{88}Zr}, \ce{^{88}Y}, \ce{^{89g}Nb}, \ce{^{89m}Nb}, \ce{^{89}Zr}, \ce{^{90}Mo}, \ce{^{90}Nb}, \ce{^{91m}Nb}, \ce{^{92m}Nb}, and \ce{^{93m}Mo} were extracted, as recorded in \autoref{tab:nb_rp_table}.
In addition, as there exist a number of isomers with radioactive ground states in these mass regions,  independent measurements of isomer-to-ground-state branching ratios for \ce{^{52m/g}Mn},\ce{^{58m/g}Co},\ce{^{85m/g}Y},\ce{^{87m/g}Y}, and \ce{^{89m/g}Nb} were  extracted and are recorded in \autoref{tab:ibr_table}.
Comparisons  of the measured cross sections and isomer branching ratios with literature data (retrieved from EXFOR \cite{Otuka2014272}) are seen in the figures of \ref{fit_figures} and \ref{ibr_figures}.
The propagated uncertainty in these cross sections varies widely based on the reaction product in question, with the major components  arising from uncertainty in EoB activity ($\pm$3--7\%), proton fluence ($\pm$4--6\%), and foil areal density ($\pm$0.1--0.6\%).

\begin{table}
\centering
\caption{Measured cross sections for the various \ce{^{nat}Cu}(p,x) reaction products observed in this work. Cumulative cross sections are designated as $\sigma_c$, independent cross sections are designated as $\sigma_i$.}
\label{tab:cu_rp_table}
\small
\resizebox{\textwidth}{!}{%
\begin{tabular}{@{}lllllll@{}}
\toprule
                            & \multicolumn{6}{c}{Production cross section (mb)}                                                                                                         \\ \cmidrule(l){2-7} 
E$_\text{p}$ (MeV)          & $89.74^{+0.48}_{-0.43}$ & $79.95^{+0.67}_{-0.64}$ & $70.17^{+0.91}_{-0.85}$ & $61.58^{+1.03}_{-0.98}$ & $52.10^{+1.25}_{-1.20}$ & $41.05^{+1.62}_{-1.54}$ \\ \midrule
\ce{^{51}Cr}\,($\sigma_c$)  & $0.919\pm0.079$         & $0.373\pm0.023$         & $0.450\pm0.028$         & $0.303\pm0.016$         & --\cmmnt{\hrulefill}    & --\cmmnt{\hrulefill}    \\
\ce{^{52}Mn}\,($\sigma_c$)  & $1.70\pm0.11$           & $0.570\pm0.031$         & $0.0407\pm0.0022$       & $0.00526\pm0.00057$     & --\cmmnt{\hrulefill}    & --\cmmnt{\hrulefill}    \\
\ce{^{52g}Mn}\,($\sigma_i$) & $0.673\pm0.043$         & $0.239\pm0.018$         & $0.0164\pm0.0023$       & $0.000986\pm0.000053$   & --\cmmnt{\hrulefill}    & --\cmmnt{\hrulefill}    \\
\ce{^{52m}Mn}\,($\sigma_c$) & $1.023\pm0.091$         & $0.331\pm0.030$         & $0.0244\pm0.0036$       & $0.00427\pm0.00052$     & --\cmmnt{\hrulefill}    & --\cmmnt{\hrulefill}    \\
\ce{^{54}Mn}\,($\sigma_i$)  & $5.87\pm0.37$           & $3.77\pm0.21$           & $4.14\pm0.22$           & $4.84\pm0.26$           & $1.680\pm0.091$         & --\cmmnt{\hrulefill}    \\
\ce{^{55}Co}\,($\sigma_c$)  & $1.71\pm0.11$           & $1.015\pm0.058$         & $0.193\pm0.012$         & $0.0299\pm0.0028$       & $0.00235\pm0.00022$     & --\cmmnt{\hrulefill}    \\
\ce{^{56}Ni}\,($\sigma_c$)  & $0.0806\pm0.0051$       & $0.1005\pm0.0055$       & $0.0906\pm0.0046$       & $0.0304\pm0.0016$       & --\cmmnt{\hrulefill}    & --\cmmnt{\hrulefill}    \\
\ce{^{57}Ni}\,($\sigma_c$)  & $1.465\pm0.093$         & $1.202\pm0.065$         & $1.400\pm0.071$         & $2.13\pm0.11$           & $1.565\pm0.083$         & $0.0262\pm0.0015$       \\
\ce{^{57}Co}\,($\sigma_i$)  & $40.1\pm2.5$          & $35.6\pm1.9$          & $35.8\pm1.8$          & $48.5\pm2.5$            & $47.7\pm2.5$            & $3.21\pm0.18$           \\
\ce{^{58}Co}\,($\sigma_c$)  & $57.7\pm4.5$            & $55.0\pm4.7$            & $42.7\pm3.4$            & $33.7\pm2.8$            & $39.0\pm3.8$            & $62.3\pm4.6$            \\
\ce{^{58g}Co}\,($\sigma_i$) & $14.0\pm2.5$            & $10.8\pm2.1$            & $6.1\pm1.6$             & $7.8\pm1.4$             & $7.1\pm1.7$             & $1.12\pm0.32$           \\
\ce{^{58m}Co}\,($\sigma_i$) & $43.6\pm3.7$            & $44.2\pm4.3$            & $36.6\pm3.0$            & $25.8\pm2.5$            & $31.9\pm3.3$            & $61.1\pm4.6$            \\
\ce{^{59}Fe}\,($\sigma_c$)  & $0.865\pm0.057$         & $0.837\pm0.046$         & $0.749\pm0.039$         & $0.616\pm0.034$         & $0.209\pm0.014$         & --\cmmnt{\hrulefill}    \\
\ce{^{60}Co}\,($\sigma_c$)  & $13.23\pm0.87$          & $13.47\pm0.78$          & $11.14\pm0.94$          & $11.44\pm0.80$          & $9.30\pm0.87$           & $6.6\pm1.1$             \\
\ce{^{61}Cu}\,($\sigma_c$)  & $50.5\pm3.3$            & $56.1\pm3.2$            & $65.1\pm3.6$            & $72.2\pm4.0$            & $80.6\pm4.7$            & $157.1\pm8.6$           \\
\ce{^{64}Cu}\,($\sigma_i$)  & $38.7\pm2.7$            & $42.8\pm2.4$            & $45.5\pm2.7$            & $50.2\pm2.8$            & $55.7\pm3.0$            & $63.3\pm3.6$                 \\ \bottomrule
\end{tabular}
}
\end{table}

\begin{table}
\centering
\caption{Measured cross sections for the various \ce{^{nat}Nb}(p,x) reaction products observed in this work. Cumulative cross sections are designated as $\sigma_c$, independent cross sections are designated as $\sigma_i$.}
\label{tab:nb_rp_table}
\small
\begin{tabular}{@{}lllllll@{}}
\toprule
                            & \multicolumn{6}{c}{Production cross section (mb)}                                                                                                         \\ \cmidrule(l){2-7} 
E$_\text{p}$ (MeV)          & $89.37^{+0.47}_{-0.45}$ & $79.55^{+0.68}_{-0.64}$ & $69.70^{+0.90}_{-0.85}$ & $61.07^{+1.05}_{-0.98}$ & $51.51^{+1.25}_{-1.21}$ & $40.34^{+1.58}_{-1.55}$ \\ \midrule
\ce{^{82m}Rb}\,($\sigma_c$) & $2.48\pm0.22$           & --\cmmnt{\hrulefill}    & --\cmmnt{\hrulefill}    & --\cmmnt{\hrulefill}    & --\cmmnt{\hrulefill}    & --\cmmnt{\hrulefill}    \\
\ce{^{83}Sr}\,($\sigma_c$)  & $4.02\pm0.61$           & $4.78\pm0.42$           & $3.49\pm0.36$           & --\cmmnt{\hrulefill}    & --\cmmnt{\hrulefill}    & --\cmmnt{\hrulefill}    \\
\ce{^{85}Y}\,($\sigma_c$)   & $13.78\pm0.55$          & $7.52\pm0.51$           & $2.11\pm0.14$           & --\cmmnt{\hrulefill}    & --\cmmnt{\hrulefill}    & --\cmmnt{\hrulefill}    \\
\ce{^{85g}Y}\,($\sigma_i$)  & $2.37\pm0.11$           & $2.08\pm0.17$           & $0.557\pm0.037$         & --\cmmnt{\hrulefill}    & --\cmmnt{\hrulefill}    & --\cmmnt{\hrulefill}    \\
\ce{^{85m}Y}\,($\sigma_i$)  & $11.41\pm0.54$          & $5.44\pm0.48$           & $1.55\pm0.13$           & --\cmmnt{\hrulefill}    & --\cmmnt{\hrulefill}    & --\cmmnt{\hrulefill}    \\
\ce{^{86}Zr}\,($\sigma_c$)  & $12.68\pm0.68$          & $18.21\pm0.93$          & $19.28\pm0.97$          & $6.16\pm0.32$           & --\cmmnt{\hrulefill}    & --\cmmnt{\hrulefill}    \\
\ce{^{86}Y}\,($\sigma_i$)   & $33.4\pm1.8$            & $41.6\pm2.2$            & $39.9\pm2.1$            & $13.56\pm0.72$          & --\cmmnt{\hrulefill}    & --\cmmnt{\hrulefill}    \\
\ce{^{87}Zr}\,($\sigma_c$)  & $47.4\pm7.3$            & $28.0\pm2.8$            & $32.2\pm2.9$            & $49.8\pm5.0$            & $38.2\pm3.7$            & $1.12\pm0.17$           \\
\ce{^{87}Y}\,($\sigma_i$)   & $110.0\pm7.2$           & $54.7\pm2.8$            & $61.0\pm2.9$            & $90.0\pm4.9$            & $67.2\pm3.6$            & $2.91\pm0.17$           \\
\ce{^{87g}Y}\,($\sigma_i$)  & $28.0\pm5.8$            & $7.4\pm1.3$             & $6.55\pm0.64$           & $5.8\pm2.2$             & $2.63\pm0.47$           & $0.942\pm0.073$         \\
\ce{^{87m}Y}\,($\sigma_i$)  & $82.0\pm4.3$            & $47.3\pm2.5$            & $54.4\pm2.8$            & $84.2\pm4.4$            & $64.6\pm3.6$            & $1.97\pm0.15$           \\
\ce{^{88}Zr}\,($\sigma_c$)  & $159.1\pm7.8$           & $144.6\pm6.8$           & $62.4\pm3.1$            & $21.2\pm1.0$            & $33.6\pm1.8$            & $65.3\pm4.0$            \\
\ce{^{88}Y}\,($\sigma_i$)   & $17.2\pm1.1$            & $13.27\pm0.86$          & $7.98\pm0.72$           & $2.91\pm0.25$           & $9.2\pm1.4$             & $9.88\pm0.69$           \\
\ce{^{89}Nb}\,($\sigma_c$)  & --\cmmnt{\hrulefill}    & --\cmmnt{\hrulefill}    & $179\pm14$              & $214.4\pm9.8$           & --\cmmnt{\hrulefill}    & --\cmmnt{\hrulefill}    \\
\ce{^{89g}Nb}\,($\sigma_i$) & --\cmmnt{\hrulefill}    & --\cmmnt{\hrulefill}    & $145\pm14$              & $186.4\pm9.6$           & --\cmmnt{\hrulefill}    & --\cmmnt{\hrulefill}    \\
\ce{^{89m}Nb}\,($\sigma_i$) & --\cmmnt{\hrulefill}    & --\cmmnt{\hrulefill}    & $34.7\pm2.6$            & $28.0\pm2.0$            & --\cmmnt{\hrulefill}    & --\cmmnt{\hrulefill}    \\
\ce{^{89}Zr}\,($\sigma_i$)  & $211\pm11$              & $243\pm13$              & $294\pm15$              & $257\pm13$              & $55.4\pm3.0$            & $15.5\pm1.0$            \\
\ce{^{90}Mo}\,($\sigma_i$)  & $21.3\pm1.1$            & $26.4\pm1.3$            & $34.5\pm1.6$            & $61.9\pm3.1$            & $122.0\pm6.1$           & $24.2\pm1.5$            \\
\ce{^{90}Nb}\,($\sigma_i$)  & $158.3\pm8.1$           & $174.9\pm8.5$           & $209.3\pm9.9$           & $272\pm14$              & $369\pm19$              & $163.9\pm9.8$           \\
\ce{^{91m}Nb}\,($\sigma_c$) & --\cmmnt{\hrulefill}    & --\cmmnt{\hrulefill}    & --\cmmnt{\hrulefill}    & --\cmmnt{\hrulefill}    & --\cmmnt{\hrulefill}    & $66.5\pm5.8$            \\
\ce{^{92m}Nb}\,($\sigma_i$) & $43.7\pm2.4$            & $47.3\pm2.4$            & $49.8\pm2.6$            & $52.9\pm2.8$            & $55.3\pm3.1$            & $59.9\pm3.9$            \\
\ce{^{93m}Mo}\,($\sigma_i$) & $0.97\pm0.20$           & $1.29\pm0.15$           & $1.62\pm0.24$           & $1.85\pm0.15$           & $1.86\pm0.14$           & $2.00\pm0.15$               \\ \bottomrule
\end{tabular}
\end{table}

\begin{table}
\centering
\caption{Measured isomer-to-ground-state branching ratios for the various \ce{^{nat}Nb}(p,x) and \ce{^{nat}Cu}(p,x) reaction products observed in this work.}
\label{tab:ibr_table}
\small
\resizebox{\textwidth}{!}{%
\begin{tabular}{@{}lllllll@{}}
\toprule
                               & \multicolumn{6}{c}{Isomer branching ratio}                                                                                                                \\ \cmidrule(l){2-7} 
E$_\text{p}$ (MeV)             & $89.74^{+0.48}_{-0.43}$ & $79.95^{+0.67}_{-0.64}$ & $70.17^{+0.91}_{-0.85}$ & $61.58^{+1.03}_{-0.98}$ & $52.10^{+1.25}_{-1.20}$ & $41.05^{+1.62}_{-1.54}$ \\ \midrule
\ce{^{nat}Cu}(p,x)\ce{^{52}Mn} & $0.603\pm0.066$         & $0.581\pm0.062$         & $0.598\pm0.095$         & $0.81\pm0.13$           & --\cmmnt{\hrulefill}    & --\cmmnt{\hrulefill}    \\
\ce{^{nat}Cu}(p,x)\ce{^{58}Co} & $0.757\pm0.088$         & $0.80\pm0.10$         & $0.858\pm0.099$         & $0.767\pm0.097$         & $0.82\pm0.12$           & $0.98\pm0.10$            \vspace{1em}     \\ 
E$_\text{p}$ (MeV)             & $89.37^{+0.47}_{-0.45}$ & $79.55^{+0.68}_{-0.64}$ & $69.70^{+0.90}_{-0.85}$ & $61.07^{+1.05}_{-0.98}$ & $51.51^{+1.25}_{-1.21}$ & $40.34^{+1.58}_{-1.55}$ \\ \midrule
\ce{^{nat}Nb}(p,x)\ce{^{85}Y}  & $0.828\pm0.051$         & $0.724\pm0.080$         & $0.736\pm0.080$         & --\cmmnt{\hrulefill}    & --\cmmnt{\hrulefill}    & --\cmmnt{\hrulefill}    \\
\ce{^{nat}Nb}(p,x)\ce{^{87}Y}  & $0.746\pm0.063$         & $0.865\pm0.063$         & $0.893\pm0.063$         & $0.936\pm0.070$         & $0.961\pm0.075$         & $0.676\pm0.065$         \\
\ce{^{nat}Nb}(p,x)\ce{^{89}Nb} & --\cmmnt{\hrulefill}    & --\cmmnt{\hrulefill}    & $0.193\pm0.021$         & $0.130\pm0.011$         & --\cmmnt{\hrulefill}    & --\cmmnt{\hrulefill}    \\ \bottomrule
\end{tabular}
}
\end{table}

These results have several notable features.
The various \ce{^{nat}Cu}(p,x) cross sections measured here are in excellent agreement with the body of measurements in the literature,  but have been measured nearly exclusively with the highest precision to date.
Similarly, the various \ce{^{nat}Nb}(p,x) cross sections measured here are in excellent agreement with literature data, which is far more sparse in the 40--90\,MeV region than for \ce{^{nat}Cu}(p,x) ---  fewer than three existing measurements have been performed for the majority of the reactions presented here.
Indeed,  the \ce{^{nat}Nb}(p,x)\ce{^{83}Sr}, \ce{^{nat}Nb}(p,x)\ce{^{85}Y}, \ce{^{nat}Nb}(p,x)\ce{^{89}Nb}, \ce{^{nat}Nb}(p,x)\ce{^{90}Mo}, \ce{^{nat}Nb}(p,x)\ce{^{91m}Nb}, and \ce{^{nat}Nb}(p,x)\ce{^{98m}Mo} reactions each possess no more than a total of three data points in this energy region.
Not only do the \ce{^{nat}Nb}(p,x) measurements in this work fill in the sparse data in this energy region, but they have been measured with the highest precision relative to existing literature data.

This work presents the first measurements of several observables in 
this mass region, including the \ce{^{nat}Nb}(p,x)\ce{^{82m}Rb} reaction in the 40--90\,MeV region, 
the independent cross section for       \ce{^{nat}Cu}(p,x)\ce{^{52\text{g}}Mn}, and the \ce{^{52\text{m}}Mn} ($2^+$) / \ce{^{52\text{g}}Mn}  ($6^+$)  isomer branching ratio via \ce{^{nat}Cu}(p,x).  
The cumulative cross sections from these data are also consistent with existing measurements of the cumulative \ce{^{nat}Cu}(p,x)\ce{^{52}Mn} cross section.
Similarly, this work offers the first measurement of the independent cross sections for \ce{^{nat}Nb}(p,x)\ce{^{85\text{g}}Y},  as well as the first measurement of the     \ce{^{85\text{m}}Y} ($\sfrac{9}{2}^+$) / \ce{^{85\text{g}}Y}  ($\sfrac{1}{2}^-$) isomer branching ratio via \ce{^{nat}Nb}(p,x).

Notably, this work is the most well-characterized measurement of the \ce{^{nat}Nb}(p,x)\ce{^{90}Mo} reaction below 100\,MeV to date, with cross sections measured  at the 4--6\% uncertainty level.
This is important, as it presents the first step towards characterizing this reaction for use as a proton monitor reaction standard below 100\,MeV.
\ce{^{nat}Nb}(p,x)\ce{^{90}Mo} can only be populated through the (p,4n) reaction channel, so no corrections for (n,x) contamination channels or decay down the A=90 isobar are needed.
\ce{^{90}Mo}  possesses seven strong, distinct gamma lines which can easily  be used for its identification and quantification.
Finally, the production of \ce{^{90}Mo}  in the 40--90\,MeV region is quite strong, with a peak cross section of approximately 120 mb.
Combining the reaction yield and gamma abundance, the use of approximately 23\,mg/cm$^2$ Nb targets easily provided sufficient counting statistics for activity quantification in the 40--90\,MeV region.
This result presents the first step towards the use of \ce{^{90}Mo} as a clean and precise   charged particle monitor reaction standard in irradiations up to approximately 24\,h in duration.

In addition to the $^\text{nat}$Nb(p,x)\ce{^{90}Mo} measurement, this experiment has also yielded measurements of  a number of additional  emerging radionuclides with medical applications.
These include the non-standard positron emitters 
\ce{^{57}Ni} \cite{PMID:7632762,zweit1996medium,Graves2016,Rosch2014}, 
\ce{^{64}Cu} \cite{Lewis2003,Bandari2014,mp500671j,Szelecsenyi1993,Aslam2009,Hilgers2003,Szelecsenyi2005,Voyles2017},  \ce{^{86}Y} \cite{Valdovinos2017,Nickles2003,Qaim2008,QaimSyedM2011,Rosch1993,doi:10.1139/v67-193,levkovski1991cross,Johnson2015,Singh2013,Kiselev1974,Kandil2009}, 
\ce{^{89}Zr}  \cite{Verel2003,Dijkers2009,Dijkers2010,PhysRevC.38.1624,Omara2009},  
\ce{^{90}Nb} \cite{Busse2002,Radchenko2012},  
and the Auger-therapy agent \ce{^{82\text{m}}Rb} \cite{Kovacs1991,Titarenko2011}. 
Production of these radionuclides offers no major advantages over established pathways, with the generally lower yields and radioisotopic purities failing to justify the convenience of natural targets  via   \ce{^{nat}Cu}(p,x) and  \ce{^{nat}Nb}(p,x). 
The one possible exception to this trend is the non-standard positron emitter \ce{^{57}Ni} ($t_{1/2}=35.60\pm0.06$ h, $\epsilon$=100\% to \ce{^{57}Co} \cite{Bhat1998}) --- the \ce{^{57}Ni}/\ce{^{56}Ni} ratio of production rates is approximately 290 at 61.58\,MeV, and varies from 45--75 at the 70--90\,MeV positions.
This \ce{^{nat}Cu}(p,x) route offers both higher yield and higher radioisotopic purity over the established  \ce{^{nat}Co}(p,3n) pathway, which suffers from approximately fivefold greater  \ce{^{56}Ni} contamination \cite{MICHEL1997153,Ditrói2013}.

We wish to urge caution in future stacked-target activation experiments by avoiding the use of silicone adhesive-based tapes for foil containment, especially when paired with the use of Al monitor foils.
Acrylic-based tape options are commercially available, and are immune from (p,x) production of \ce{^{22,24}Na} activities, due to being of too low-Z for these reaction channels to be possible.
Even with subtraction of \ce{^{22,24}Na} activities though irradiating a Kapton tape \enquote{blank} or similar, we observe the Al monitor channels to measure consistently higher proton fluence than via Cu monitor channels, by 5--8\%. 
If Al monitors are used in conjunction with silicone-based tapes, even with subtraction of excess \ce{^{22}Na} activities, a systematically enhanced fluence may be determined, leading to cross sections reported with inaccurately diminished magnitude.
Furthermore, since data for monitor reactions are often self-referencing, the propagated impact of this systematic enhancement in fluence may have far-reaching consequences for both medical isotope production, as well as for  the evaluated nuclear data libraries, which use these proton activation experiments as input.

%


As mentioned before, cumulative cross sections are reported here for the first observable product nuclei in a mass chain, or whenever it is impossible to use decay spectrometry to distinguish direct production of a nucleus from decay feeding.
For all remaining observed reaction products in the mass chain, and cases where no decay precursors exist, independent cross sections are reported, allowing for determination of the direct production via subtraction.  
This, in turn, offers the opportunity to 
gauge the predictive capabilities of modern nuclear models used in the reaction evaluation process.  
The reaction channels with independent cross sections were compared to calculations with the reaction modeling codes EMPIRE, TALYS, and CoH, run with the default settings. 
The default optical models and E1 gamma strength function models for each code are presented in Table \ref{tab:defaults}. 
The large energy range covered by many of the exit channels, which extends significantly beyond the range of pure compound nuclear/evaporation, allows the data to be used to study the differences between these modeling codes in the  pre-equilibrium regime.

The default level density in both CoH and TALYS is the Gilbert-Cameron model, which uses a Constant Temperature model below a critical energy and Fermi Gas model above it.
The default level density in EMPIRE is the Enhanced Generalized Superfluid Model (EGSM) which uses the Generalized Superfluid model below a critical energy, and Fermi Gas model above it \cite{Capote2009}.
The EGSM densities are normalized to $D_0$ and the discrete levels, but in such a way that only the level density below the neutron separation energy is  effected by the discrete levels chosen for the normalization.  
All three codes use a two-exciton phenomenological model to calculate the pre-equilibrium cross section, but the specific implementation differs between the codes.

\begin{table}
 \caption{Default settings for the reactions codes}
 \label{tab:defaults}
 \resizebox{\textwidth}{!}{%
\begin{tabular}{ c c c c}

 \underline{Code Version} & \underline{Proton/Neutron Optical Model} & \underline{Alpha Optical Model}  & \underline{E1 $\gamma$SF Model}  \\ 
 EMPIRE-3.2.3\cite{Herman2007}       & Koning-Delaroche\cite{Koning2003} & Avrigeanu(2009)\cite{Avrigeanu2009}       & Modified Lorentzian\cite{belgya2006handbook}  \\  
 TALYS-1.8\cite{Koning2012}          & Koning-Delaroche & Specific folded potential\cite{Koning2012}      & Brink-Axel Lorentzian\cite{Koning2012}   \\
  CoH-3.5.1\cite{kawano2003coh,KAWANO2010} & Koning-Delaroche     & Avrigeanu(1994)\cite{Avrigeanu1994} & Generalized Lorentzian\cite{kawano2003coh,KAWANO2010}
  
\end{tabular}
}
\end{table}

Given the large number of exit channels in this data set, we will limit our discussion to  cross sections for the production of a specific residual nucleus with experimental data through the full rise and fall of the peak, 
and at least 1\% of the total reaction cross section.
Exit channel cross sections that do not exhibit the full rise and fall of the peak, which is identified as 
being dominated by the formation of a compound nucleus, do not provide enough information to analyze the calculations.
Residual nuclei like \ce{^{88}Zr} that can be produced by multiple reaction channels, such as (p,$\alpha$2n) and by (p,2p4n)  are also not discussed in depth. 
We exclude reactions with cross sections with peak values less than 1\% of the total reaction cross section 
because their behavior is extremely sensitive  to  more dominant  channels.
The three residual nuclei that meet all of the above criteria for which there is an independent measurement of the residual production cross section are \ce{^{86}Y}, \ce{^{90}Mo}, and \ce{^{90}Nb}.

\begin{figure}
 \centering
 \includegraphics[width=0.5\linewidth]{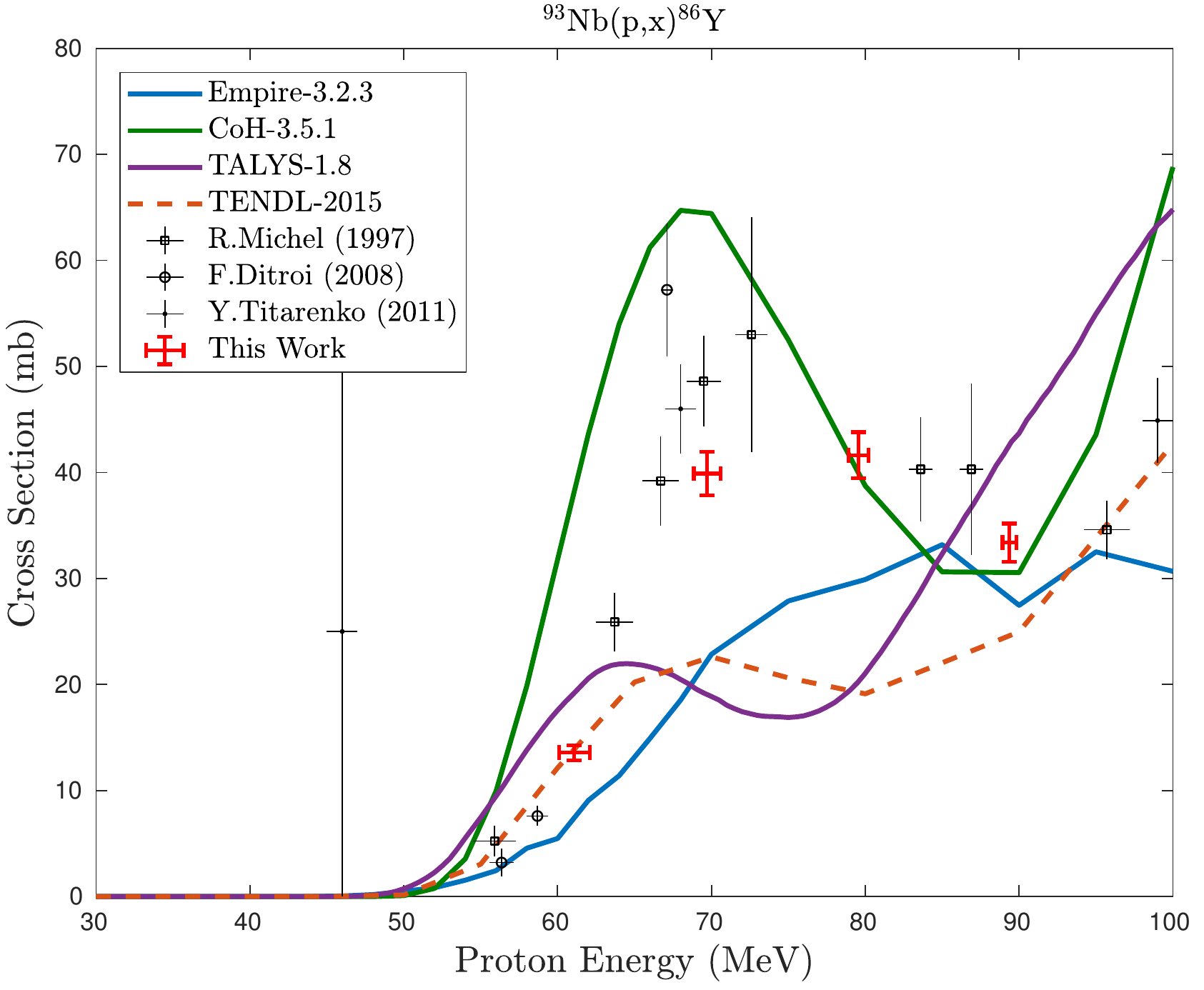}
 \caption{Measured \ce{^{93}Nb}(p,x)\ce{^{86}Y} cross section, with the \ce{^{93}Nb}(p,$\alpha$p3n)\ce{^{86}Y} reaction channel visibly peaking at approximately \mbox{70 MeV}.}
 \label{fig:86Y}
\end{figure}

\begin{figure}
 \centering
 \includegraphics[width=0.5\linewidth]{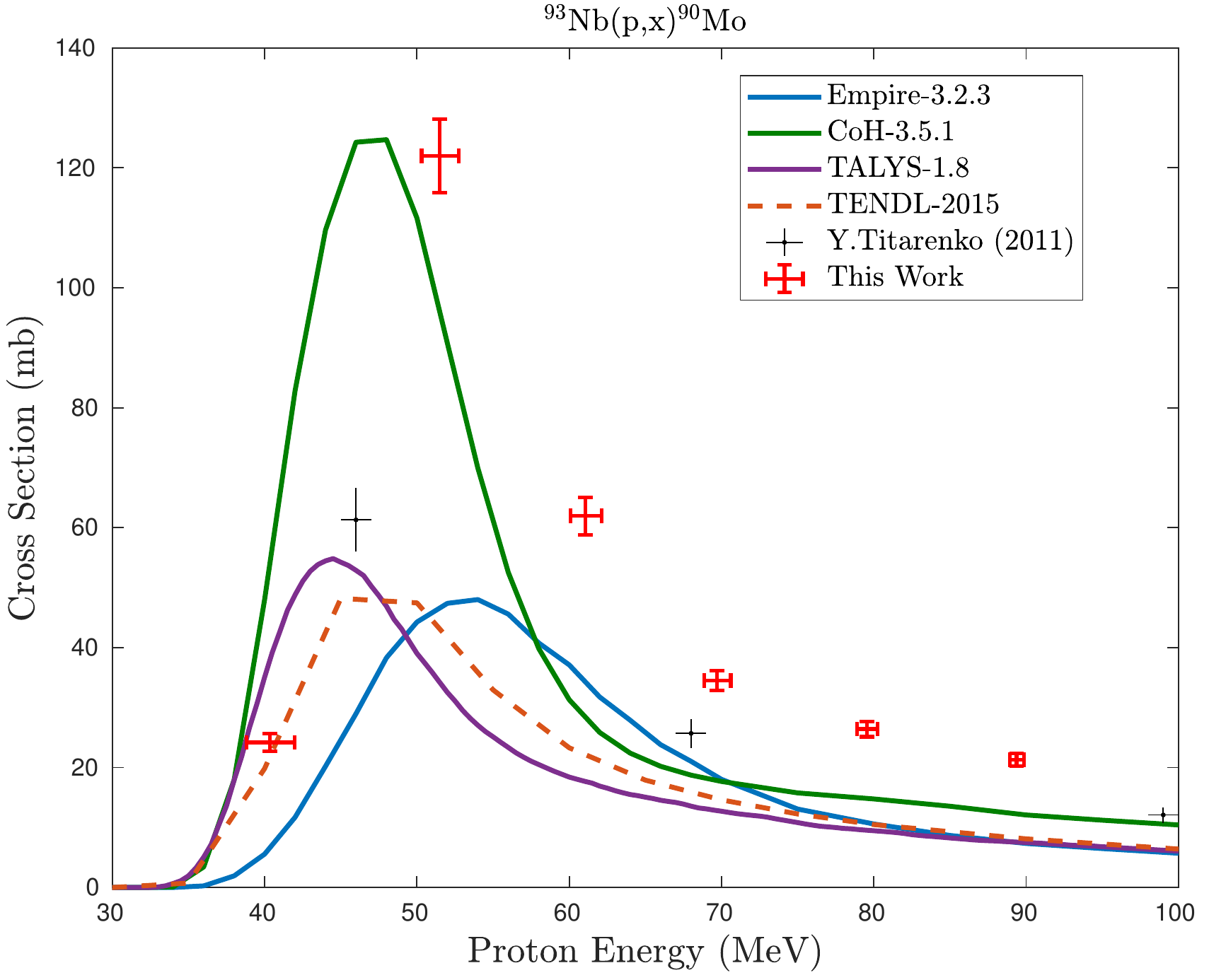}
 \caption{Measured \ce{^{93}Nb}(p,x)\ce{^{90}Mo} cross section, with the \ce{^{93}Nb}(p,4n)\ce{^{90}Mo} reaction channel visibly peaking at approximately \mbox{50 MeV}.}
 \label{fig:90Mo}
\end{figure}

\begin{figure}
 \centering
 \includegraphics[width=0.5\linewidth]{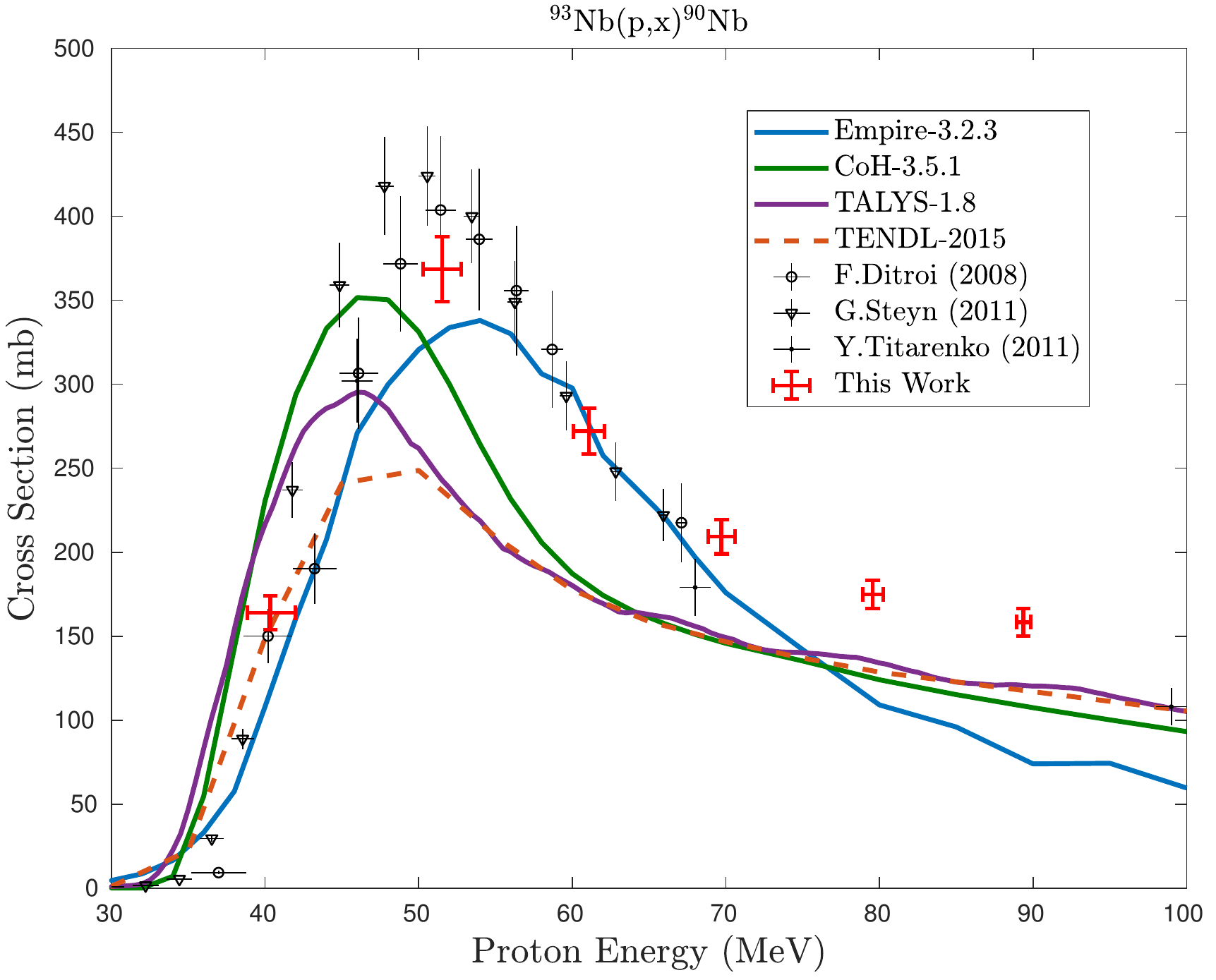}
 \caption{Measured \ce{^{93}Nb}(p,x)\ce{^{90}Nb} cross section, with the \ce{^{93}Nb}(p,p3n)\ce{^{90}Nb} reaction channel visibly peaking at approximately \mbox{50 MeV}.}
 \label{fig:90Nb}
\end{figure}

The \ce{^{93}Nb}(p,$\alpha$p3n)\ce{^{86}Y} reaction channel, which peaks at approximately \mbox{70\,MeV}, is well within the compound regime for the entire energy region of this experiment (\autoref{fig:86Y}).
The data collected on this residual is consistent with the one other data set available, taken in 1997 by Michel \emph{et al.} \cite{MICHEL1997153}. 
The \ce{^{93}Nb}(p,4n)\ce{^{90}Mo} and \ce{^{93}Nb}(p,p3n)\ce{^{90}Nb} channels both peak early in the energy region, around 50\,MeV, and the data clearly show the full rise, peak, and fall of the compound cross section (\autoref{fig:90Mo} \& \ref{fig:90Nb}). 
In both of these channels, this data  is consistent with the data by Titarenko \emph{et al.} in 2011 \cite{Titarenko2011}.

The \ce{^{90}Nb} production cross section exhibits a persistent pre-equilibrium \enquote{tail} that keeps the channel open  well after the compound cross section has fallen away. 
TALYS, TENDL, and CoH seem to have the correct shape for this pre-equilibrium cross section, with magnitudes that are just slightly too low.
EMPIRE, however, does not level off  as much as the data and the other codes are seen to, and misses the high-energy data points.

In all three channels, the TALYS, TENDL, and CoH calculations rise, peak, and fall at lower energies than the data, while EMPIRE calculates the peak to occur at higher energies.
For \ce{^{90}Mo}, the EMPIRE peak is representative of the data.
For \ce{^{86}Y} and \ce{^{90}Nb}, the peak is missed by all three  of the codes.

The magnitudes of the TALYS and TENDL calculations are consistently too low in the three channels studied here. 
For \ce{^{86}Y}, CoH and EMPIRE also predict smaller cross sections than the data would suggest, which may be influenced by incorrect modeling of other, stronger, channels.
The magnitude of the peak in the CoH calculation for \ce{^{90}Mo}  is consistent  with the data, while EMPIRE predicts a cross section that is approximately the same magnitude as that of TALYS.
\ce{^{90}Nb} is one of the strongest measured channels, approximately 10\% of the total reaction cross section, and the values from the three codes are all consistent, but  too small, in magnitude.



 

 \section{Conclusions}

We present here a set of measurements of 38 cross sections for the \ce{^{nat}Nb}(p,x) and  \ce{^{nat}Cu}(p,x) reactions between 40--90\,MeV, as well as  independent measurements of five isomer branching ratios.
Nearly all cross sections have been reported with higher precision than previous measurements.
We report the first measurements of the  \ce{^{nat}Nb}(p,x)\ce{^{82m}Rb} reaction, as well as the first measurement of the independent cross sections for    \ce{^{nat}Cu}(p,x)\ce{^{52\text{m}}Mn}, \ce{^{nat}Cu}(p,x)\ce{^{52\text{g}}Mn}, and \ce{^{nat}Nb}(p,x)\ce{^{85\text{g}}Y} in the 40--90\,MeV region.
We advise that future activation experiments avoid the use of silicone-based adhesives, particularly in conjunction with aluminum monitor foils, to avoid reporting an enhanced fluence due to \ce{^{22,24}Na} contamination.
We also use these measurements to illustrate the deficiencies in the current state of  reaction modeling for 40--90\,MeV \ce{^{nat}Nb}(p,x) and  \ce{^{nat}Cu}(p,x) reactions.
Finally, this work provides another example of the usefulness of the recently-described variance minimization techniques for reducing energy uncertainties in stacked target charged particle irradiation experiments.

%% file: nb_appendix_text.tex
\section{Decay data} \label{data}
%
The   lifetimes and gamma-ray branching ratios  listed in these tables were used for all calculations of measured cross sections reported in this work, and have been taken from the most recent edition of  Nuclear Data Sheets for each  mass chain  \cite{Basunia2015,Firestone2007,Wang2017,Dong2015,Dong2014,JUNDE2008787,Junde2011,Bhat1998,Nesaraja2010,BAGLIN2002,Browne2013,Zuber20151,NICHOLS2012973,Singh2007,Browne2010,Tuli2003,McCutchan2015,Singh2014,NEGRET20151,Johnson2015,McCutchan2014,Singh2013,Browne1997,Baglin2013,Baglin2012,Baglin2011}.



\begin{table}[ht]
\centering
\caption{Decay data for gamma-rays observed in \ce{^{nat}Al}(p,x) and \ce{^{nat}Cu}(p,x).}
\label{tab:nudat_table_monitors}
\small
\begin{tabular}{@{}llll@{}}
\toprule
Nuclide & Half-life & E$_\gamma$ (keV) & I$_\gamma$ (\%)\\
\midrule
\ce{^{22}Na} & 2.6018(22) y & 1274.537 & 99.940(14)\\
 
\ce{^{24}Na} & 14.997(12) h & 1368.626 & 99.9936(15)\\
 
\ce{^{51}Cr} & 27.704(3) d & 320.0824 & 9.910(10)\\
 
\ce{^{52m}Mn} & 21.1(2) m & 1434.0600 & 98.2(5)\\
 
\ce{^{52}Mn} & 5.591(3) d & 744.233 & 90.0(12)\\
 
 & 5.591(3) d & 935.544 & 94.5(13)\\
 
 & 5.591(3) d & 1246.278 & 4.21(7)\\
 
 & 5.591(3) d & 1434.092 & 100.0(14)\\
 
\ce{^{54}Mn} & 312.20(20) & 834.848 & 99.9760(10)\\
 
\ce{^{55}Co} & 17.53(3) h & 477.2 & 20.2(17)\\
 
 & 17.53(3) h & 931.1 & 75.0(35)\\
 
 & 17.53(3) h & 1316.6 & 7.1(3)\\
 
 & 17.53(3) h & 1408.5 & 16.9(8)\\
 
\ce{^{56}Ni} & 6.075(10) d & 158.38 & 98.8(10)\\
 
 & 6.075(10) d & 269.50 & 36.5(8)\\
 
 & 6.075(10) d & 480.44 & 36.5(8)\\
 
 & 6.075(10) d & 749.95 & 49.5(12)\\
 
 & 6.075(10) d & 811.85 & 86.0(9)\\
 
 & 6.075(10) d & 1561.80 & 14.0(6)\\
 
\ce{^{56}Co} & 77.236(26) d & 846.770 & 99.9399(2)\\
 
 
 & 77.236(26) d & 1037.843 & 14.05(4)\\
 
 & 77.236(26) d & 1238.288 & 66.46(12)\\
 
 & 77.236(26) d & 1360.212 & 4.283(12)\\
 
 & 77.236(26) d & 1771.357 & 15.41(6)\\
 
\ce{^{57}Ni} & 35.60(6) h & 127.164 & 16.7(5)\\
 
 & 35.60(6) h & 1377.63 & 81.7(24)\\
 
 & 35.60(6) h & 1757.55 & 5.75(20)\\
 
 & 35.60(6) h & 1919.52 & 12.3(4)\\
 
\ce{^{57}Co} & 271.74(6) d & 122.06065 & 85.60(17)\\
 
 & 271.74(6) d & 136.47356 & 10.68(8)\\
 
\ce{^{58}Co} & 70.86(6) d & 810.7593 & 99.450(10)\\
 
 & 70.86(6) d & 863.951 & 0.686(10)\\
 
 
\ce{^{59}Fe} & 44.495(9) d & 1099.245 & 56.5(18)\\
 
 & 44.495(9) d & 1291.590 & 43.2(14)\\
 
\ce{^{60}Co} & 5.2714(5) y & 1173.228 & 99.85(3)\\
 
 & 5.2714(5) y & 1332.492 & 99.9826(6)\\
 
\ce{^{61}Cu} & 3.339(8) h & 282.956 & 12.2(2.2)\\
 
 & 3.339(8) h & 373.050 & 2.1(4)\\
 
 
 & 3.339(8) h & 656.008 & 10.8(20)\\
 
 & 3.339(8) h & 1185.234 & 3.7(7)\\
 
\ce{^{62}Zn} & 9.193(15) h & 243.36 & 2.52(23)\\
 
 & 9.193(15) h & 246.95 & 1.90(18)\\
 
 & 9.193(15) h & 260.43 & 1.35(13)\\
 
 
 
 & 9.193(15) h & 394.03 & 2.24(17)\\
 
 & 9.193(15) h & 548.35 & 15.3(14)\\
 
 & 9.193(15) h & 596.56 & 26.0(20)\\
 
 
\ce{^{64}Cu} & 12.701(2) h & 1345.77 & 0.475(11)\\
 
\ce{^{65}Zn} & 243.93(9) d & 1115.539 & 50.04(10)\\
\bottomrule
\end{tabular}
\end{table}

\begin{table}[ht]
\centering
\caption{Decay data for gamma-rays observed in \ce{^{nat}Nb}(p,x).}
\label{tab:nudat_table_nb}
\small
\begin{tabular}{@{}llll@{}}
\toprule
Nuclide & Half-life & E$_\gamma$ (keV) & I$_\gamma$ (\%)\\
\midrule
\ce{^{82m}Rb} & 6.472(6) h & 554.35 & 62.4(9)\\
 
 & 6.472(6) h & 619.11 & 37.98(9)\\
 
 
 & 6.472(6) h & 776.52 & 84.39(21)\\
 
 & 6.472(6) h & 1044.08 & 32.07(8)\\
 
 
\ce{^{83}Sr} & 32.41(3) h & 418.37 & 4.2(3)\\
 
 & 32.41(3) h & 762.65 & 26.7(22)\\
 
\ce{^{85m}Y} & 4.86(13) h & 231.7 & 22.8(22)\\
 
\ce{^{85}Y} & 2.68(5) h & 231.65 & 84(9)\\
 
 & 2.68(5) h & 913.89 & 9.0(9)\\
 
\ce{^{86}Zr} & 16.5(1) h & 242.8 & 95.84(2)\\
 
 & 16.5(1) h & 612.0 & 5.8(3)\\
 
\ce{^{86}Y}  & 14.74(2) h & 443.13 & 16.9(5)\\

 
 
 
 
 
 & 14.74(2) h & 627.72 & 32.6(1)\\
 
 
 
 
 
 
 & 14.74(2) h & 1076.63 & 82.5(4)\\
 
 & 14.74(2) h & 1153.05 & 30.5(9)\\
 
%
 
 
 & 14.74(2) h & 1854.38 & 17.2(5)\\
 
 & 14.74(2) h & 1920.72 & 20.8(7)\\
 
\ce{^{87}Zr} & 1.68(1) h & 380.79 & 62.79(10)\\
 
 & 1.68(1) h & 1227.0 & 2.80(4)\\
 
\ce{^{87m}Y} & 13.37(1) h & 380.79 & 78.05(8)\\
 
\ce{^{87}Y} & 79.8(3) h & 388.5276 & 82.2(7)\\
 
 & 79.8(3) h & 484.805 & 89.8(9)\\
 
\ce{^{88}Zr} & 83.4(3) d & 392.87 & 97.29(14)\\
 
\ce{^{88}Y} & 106.627(21) d & 898.042 & 93.7(3)\\
 
 & 106.627(21) d & 1836.063 & 99.2(3)\\
 
\ce{^{89m}Nb} & 66(2) m & 588.0 & 95.57(13)\\
 
\ce{^{89}Nb} & 2.03(7) h & 1511.4 & 1.9(4)\\
 
 & 2.03(7) h & 1627.2 & 3.5(7)\\
 
 & 2.03(7) h & 1833.4 & 3.3(7)\\
 
\ce{^{89}Zr} & 78.41(12) h & 909.15 & 99.04(3)\\
 
 & 78.41(12) h & 1713.0 & 0.745(13)\\
 
 
\ce{^{90}Mo} & 5.56(9) h & 122.370 & 64(3)\\
 
 & 5.56(9) h & 162.93 & 6.0(6)\\
 
 & 5.56(9) h & 203.13 & 6.4(6)\\
 
 & 5.56(9) h & 257.34 & 78(4)\\
 
 & 5.56(9) h & 323.20 & 6.3(6)\\
 
 & 5.56(9) h & 472.2 & 1.42(16)\\
 
 & 5.56(9) h & 941.5 & 5.5(7)\\
 
\ce{^{90}Nb} & 14.6(5) h & 132.716 & 4.13(4)\\
 
 & 14.6(5) h & 141.178 & 66.8(7)\\
 
 
 & 14.6(5) h & 1611.76 & 2.38(7)\\
 
 
\ce{^{91m}Nb} & 60.86(22) d & 104.62 & 0.574(1)\\
 
 & 60.86(22) d & 1204.67 & 2.0(3)\\
 
\ce{^{92m}Nb} & 10.15(2) d & 912.6 & 1.78(10)\\
 
 & 10.15(2) d & 934.44 & 99.15(4)\\
 
\ce{^{93m}Mo} & 6.85(7) d & 263.049 & 57.4(11)\\
 
 & 6.85(7) d & 684.693 & 99.9(8)\\
 
 & 6.85(7) d & 1477.138 & 99.1(11)\\
\bottomrule
\end{tabular}
\end{table}

\section{Measured excitation functions} \label{fit_figures}

Figures of the cross sections measured in this work are presented here, in comparison with literature data \cite{Albouy1963,PhysRev.162.1055,PhysRevC.6.1235,Grutter1982,Greenwood1984,Aleksandrov1987,levkovski1991cross,Mills1992,MICHEL1997153,Fassbender1997,Ido2002,sisterson2002selected,YashimaH2003,A2006,Ditroi2008,Ditroi2009,steyn2011excitation,Titarenko2011,Shahid2015,Garrido2016,Graves2016}, the TENDL-2015 data library \cite{Koning2012}, and the reaction modeling codes CoH-3.5.1, EMPIRE-3.2.3, and TALYS-1.8 \cite{KAWANO2010,Herman2007,Koning2012}.




\begin{figure*}
    \sloppy
    \centering
    \subfloat{
        \centering
        \subfigimg[width=0.496\textwidth]{}{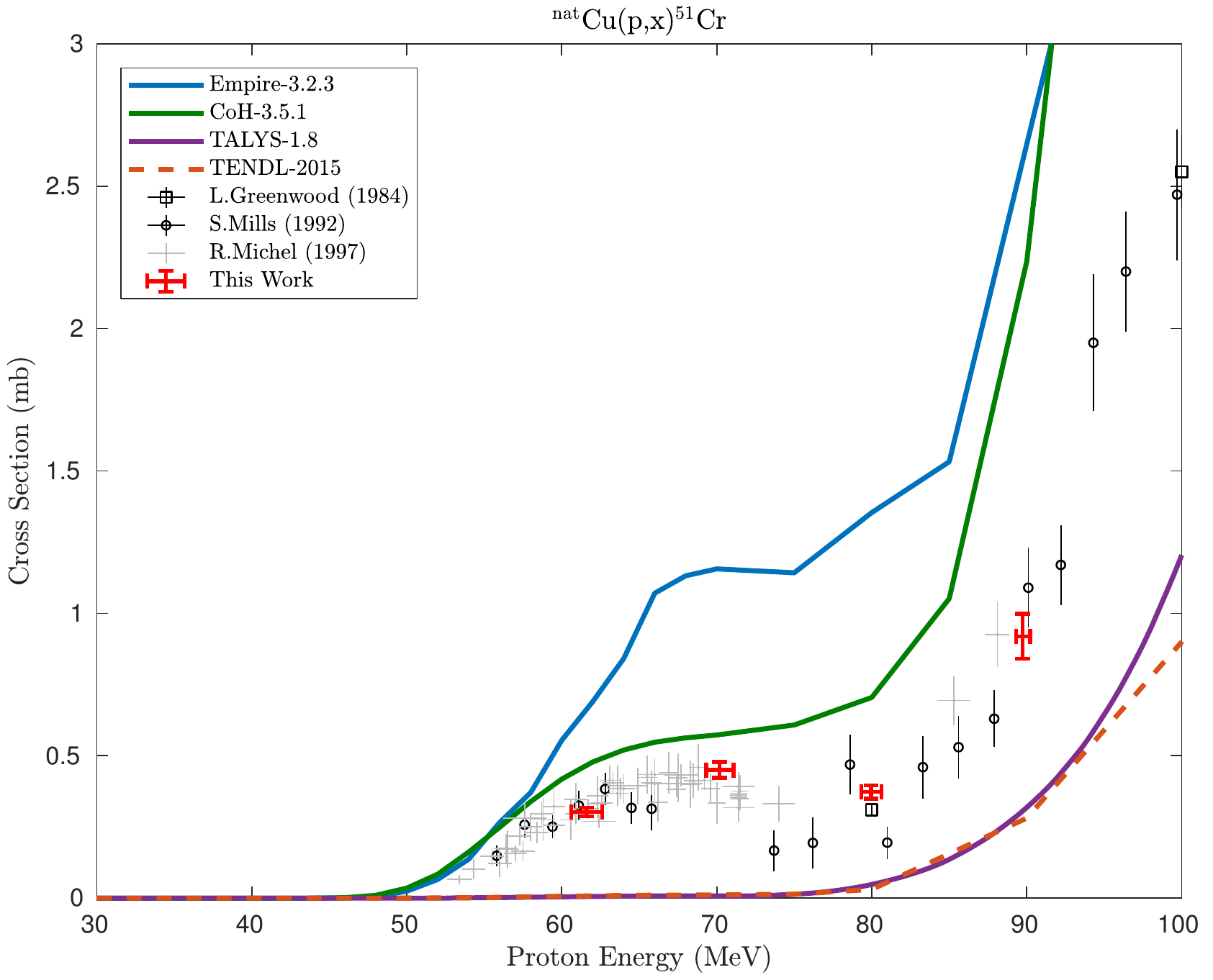}{50}
        \subfigimg[width=0.496\textwidth]{}{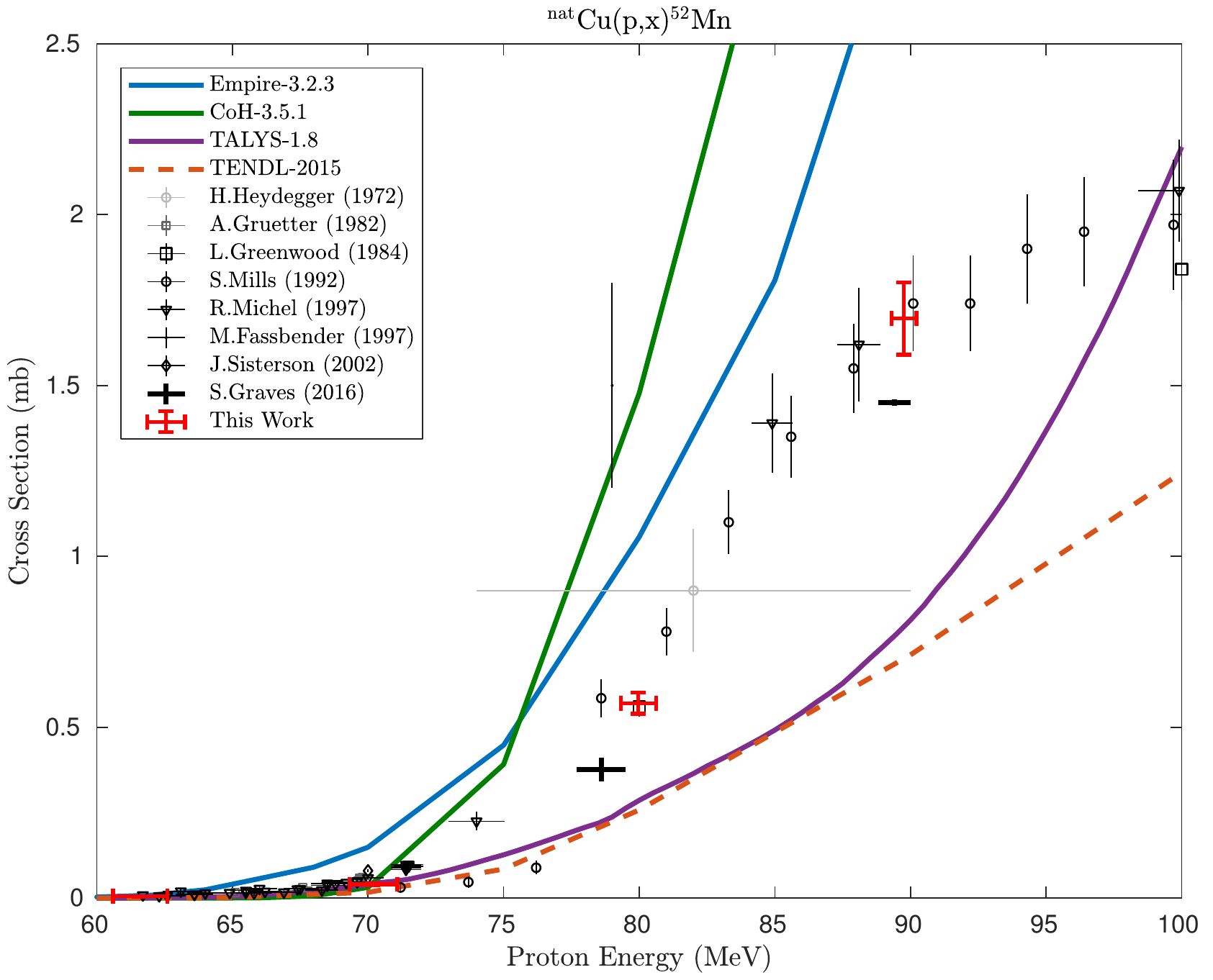}{50}
   \hspace{-10pt}}%
    \\
    \subfloat{
        \centering
        \subfigimg[width=0.496\textwidth]{}{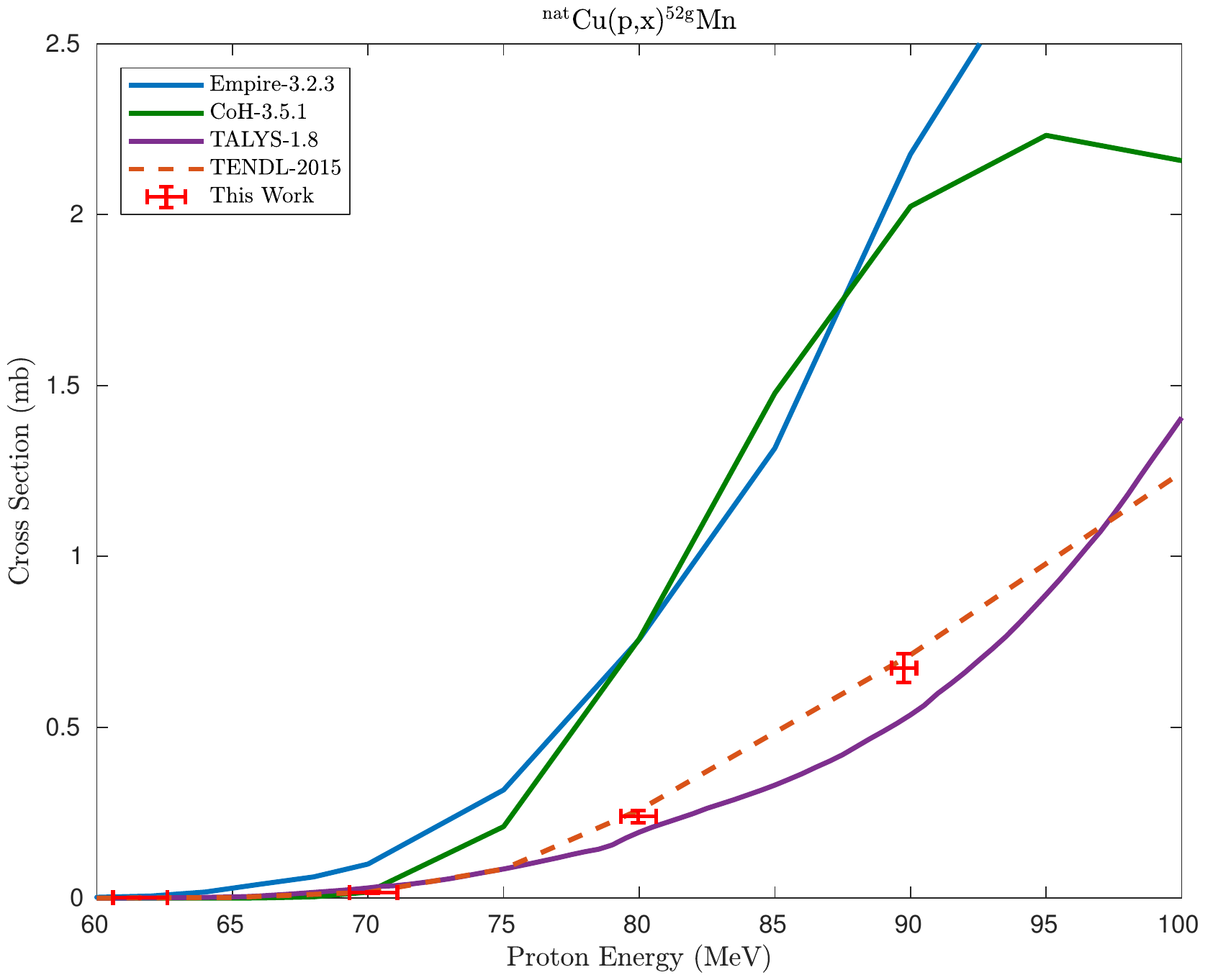}{50}
        \subfigimg[width=0.496\textwidth]{}{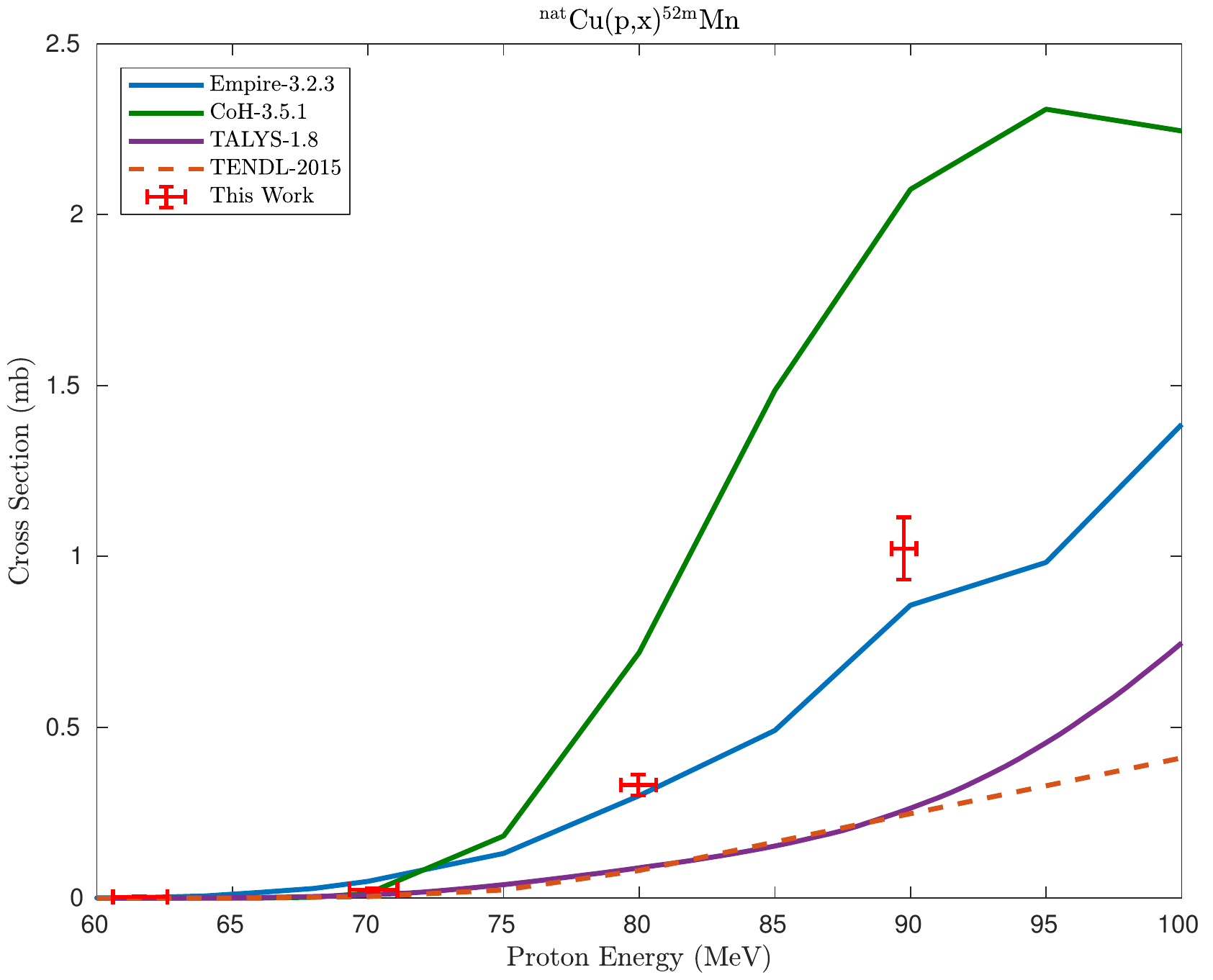}{50}
   \hspace{-10pt}}%
    \\
    \subfloat{
        \centering
        \subfigimg[width=0.496\textwidth]{}{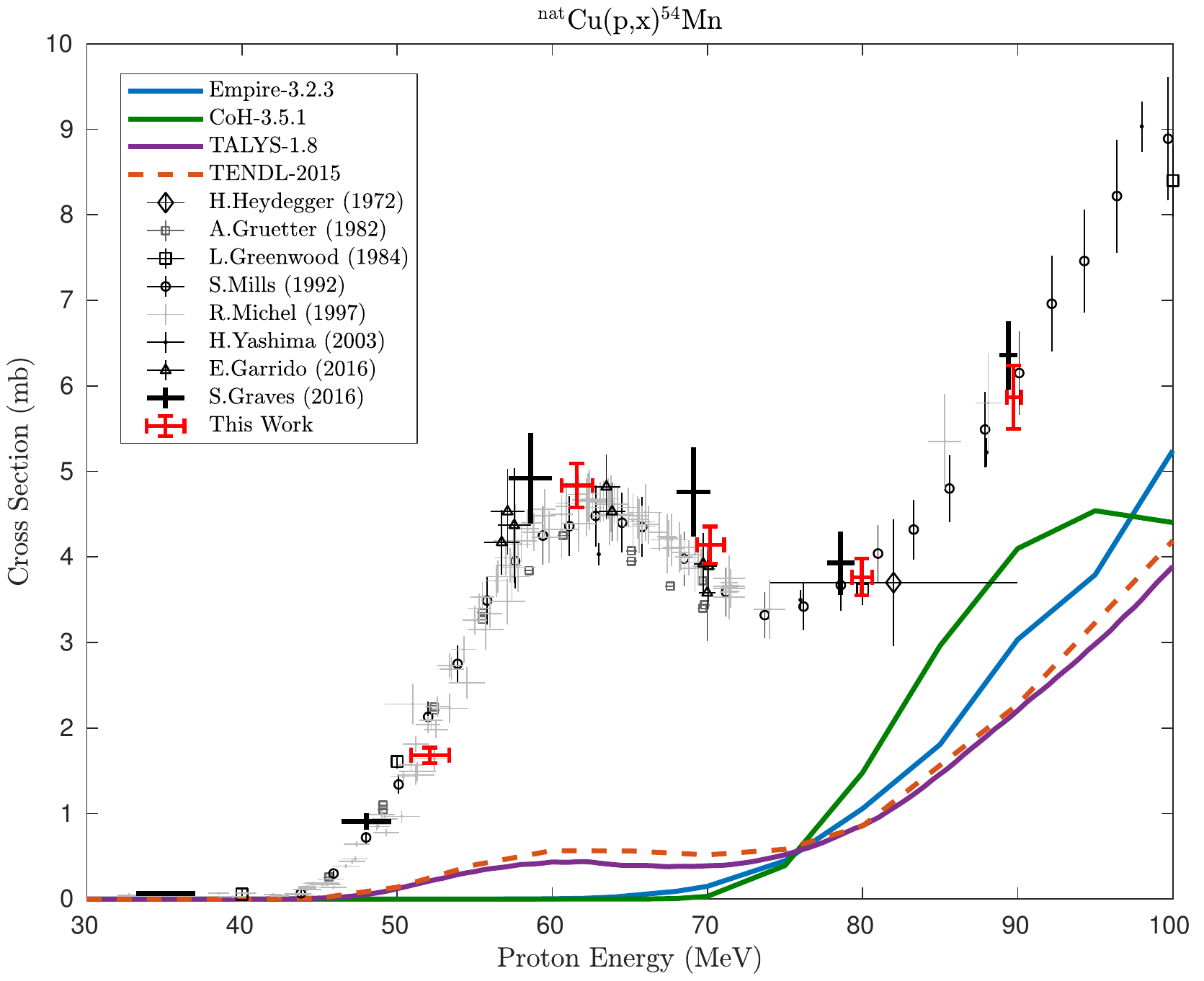}{50}
%
        \subfigimg[width=0.496\textwidth]{}{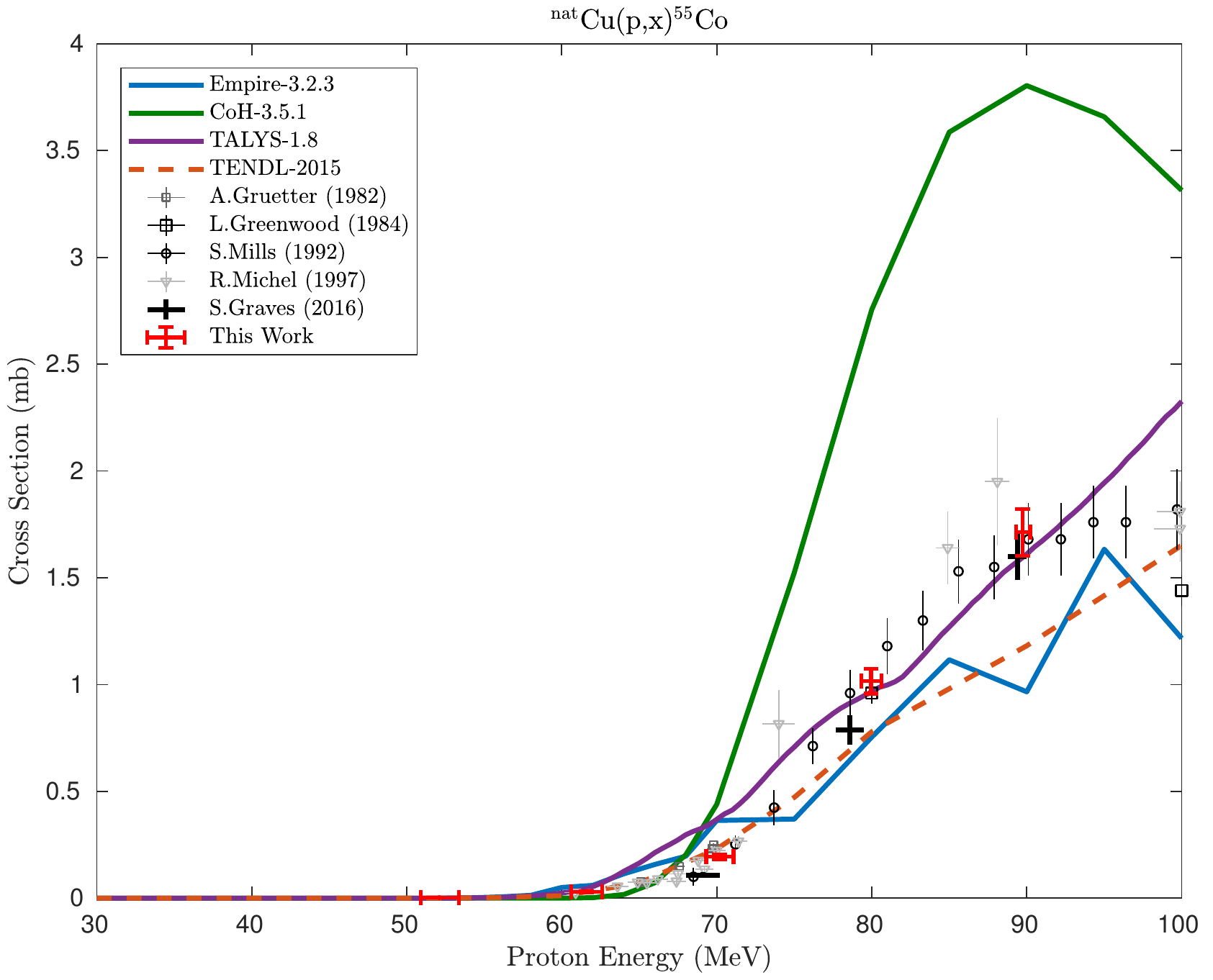}{50}
   \hspace{-10pt}}%
\end{figure*}

\begin{figure*}
    \centering
    \subfloat{
        \centering
        \subfigimg[width=0.496\textwidth]{}{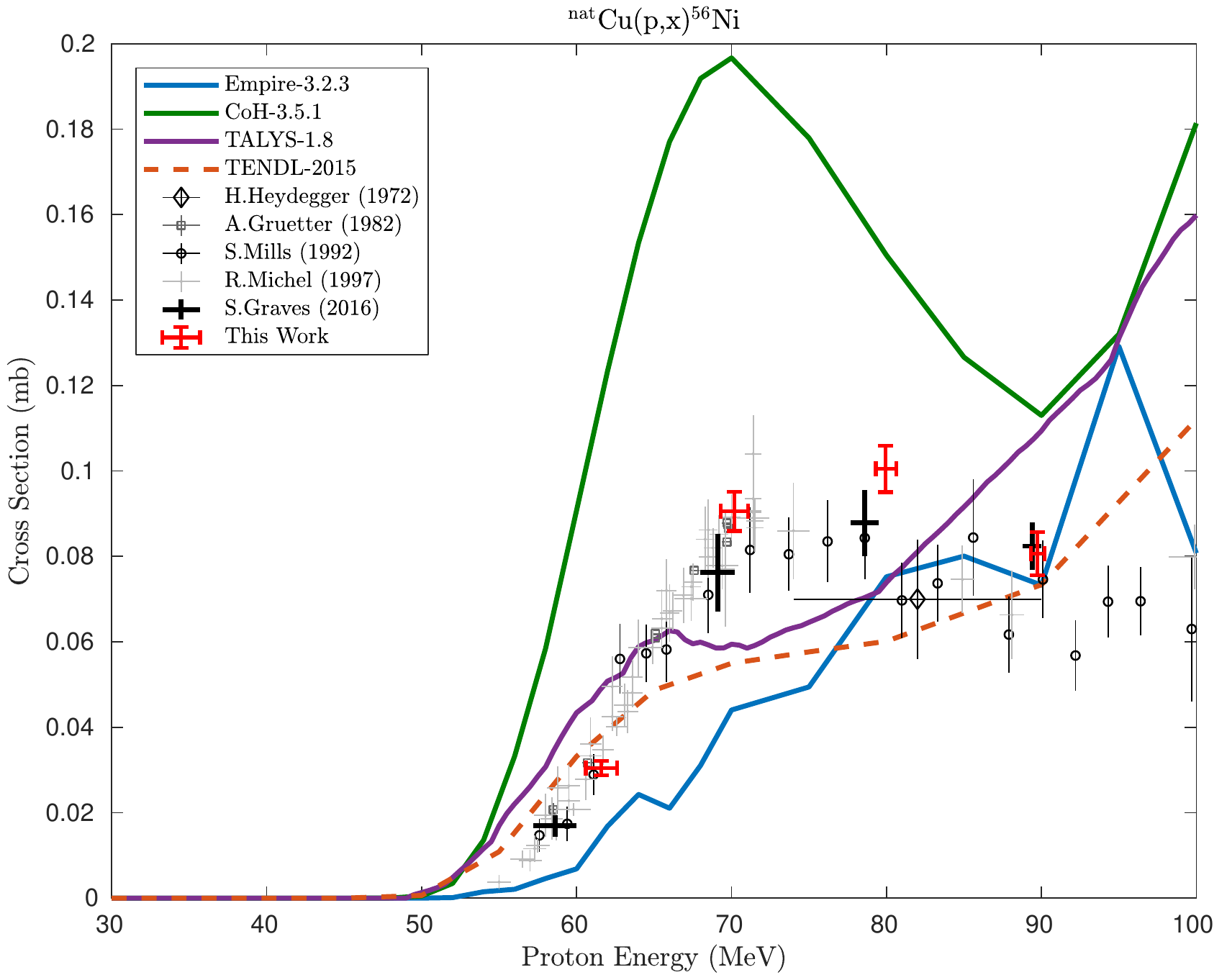}{50}
%
        \subfigimg[width=0.496\textwidth]{}{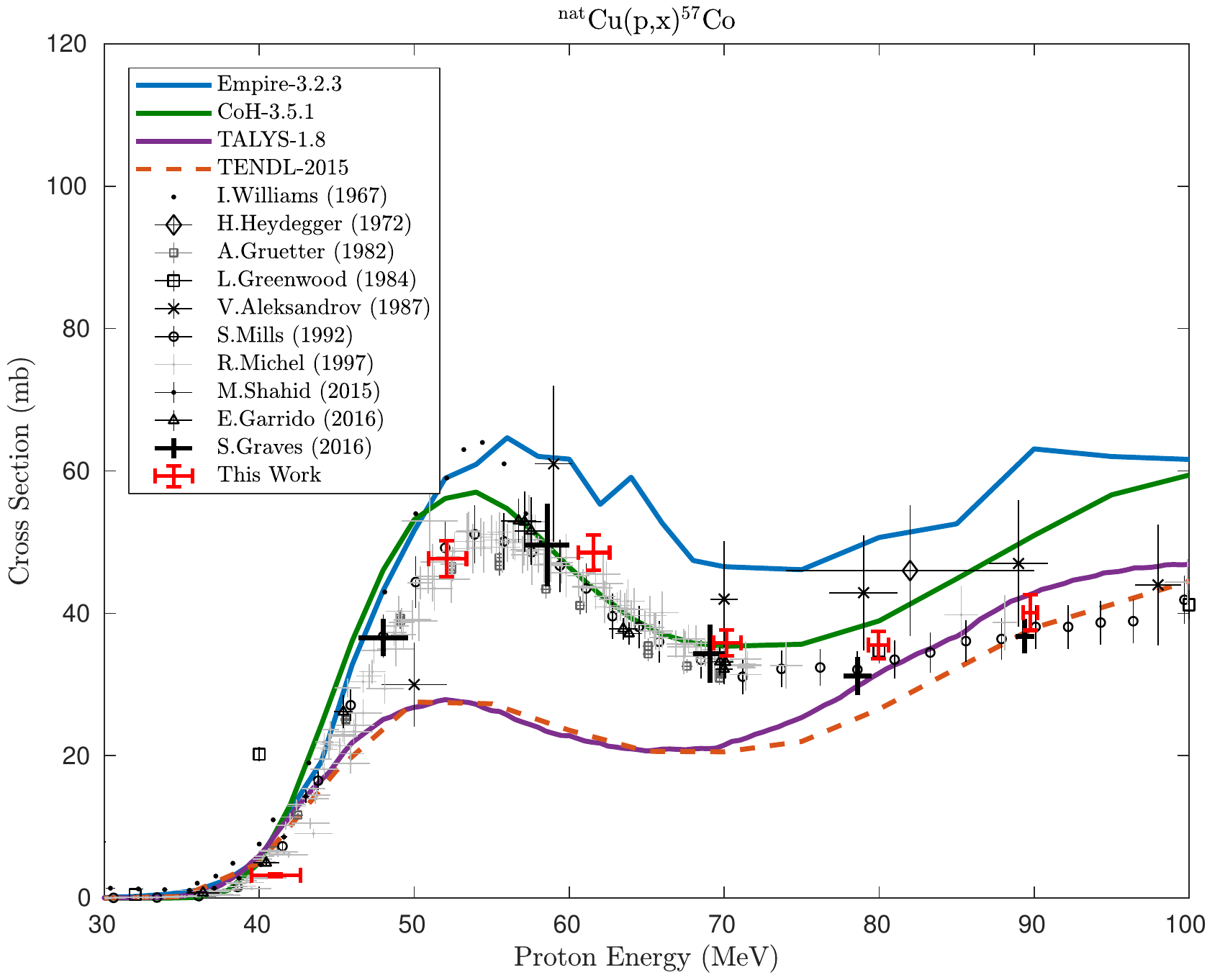}{50}
   \hspace{-10pt}}%
    \\
    \subfloat{
        \centering
        \subfigimg[width=0.496\textwidth]{}{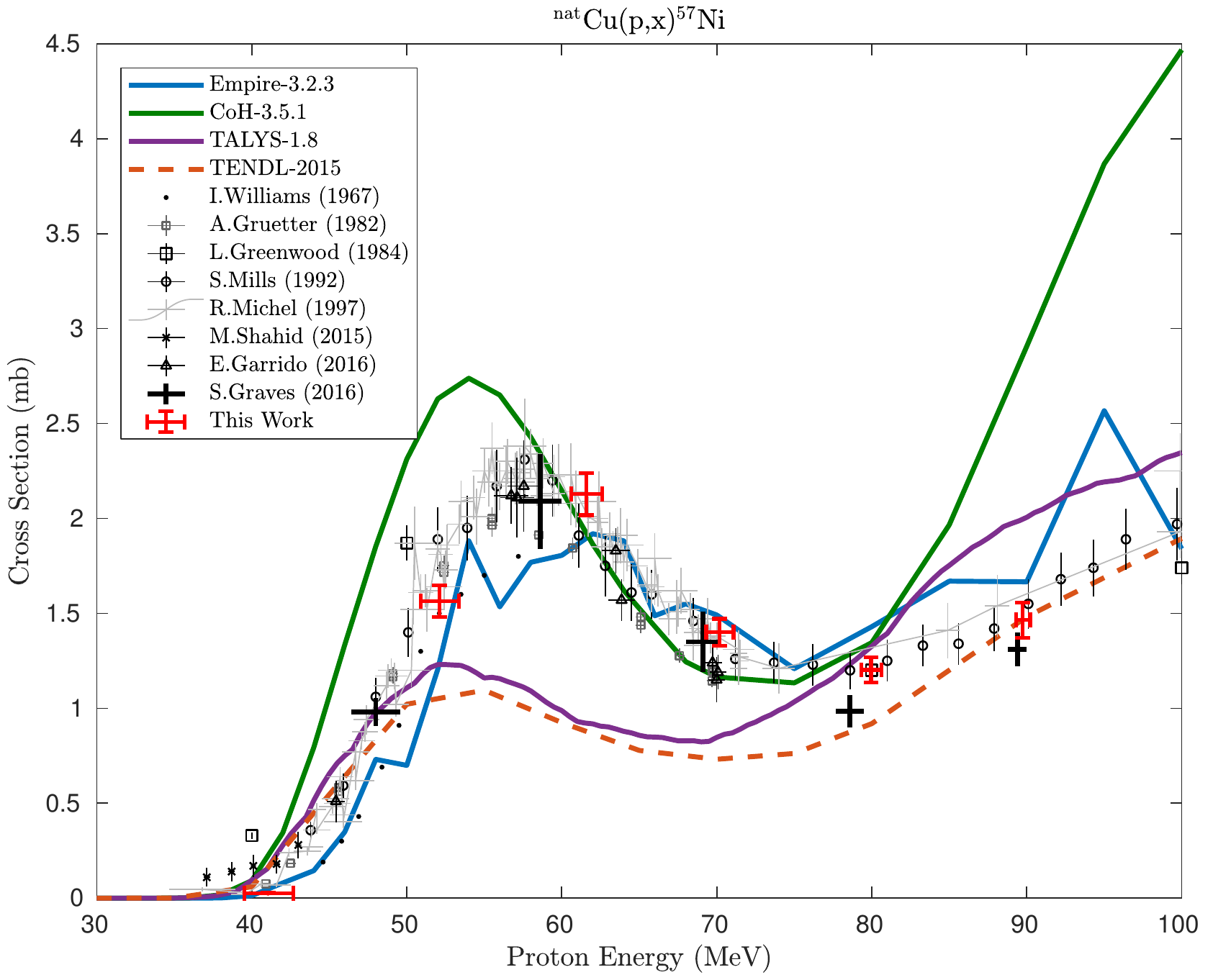}{50}
        \subfigimg[width=0.496\textwidth]{}{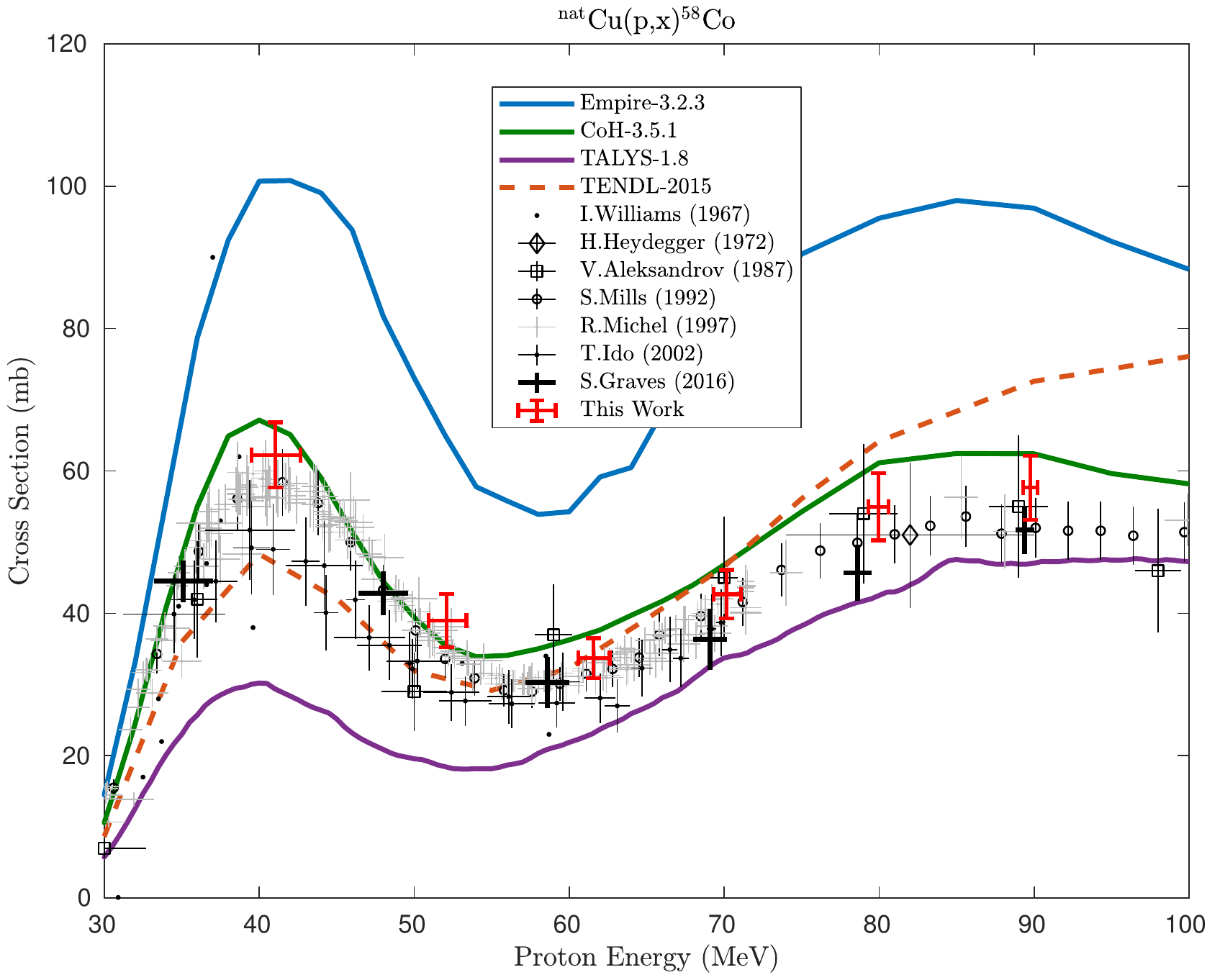}{50}
   \hspace{-10pt}}%
    \\
    \subfloat{
        \centering
        \subfigimg[width=0.496\textwidth]{}{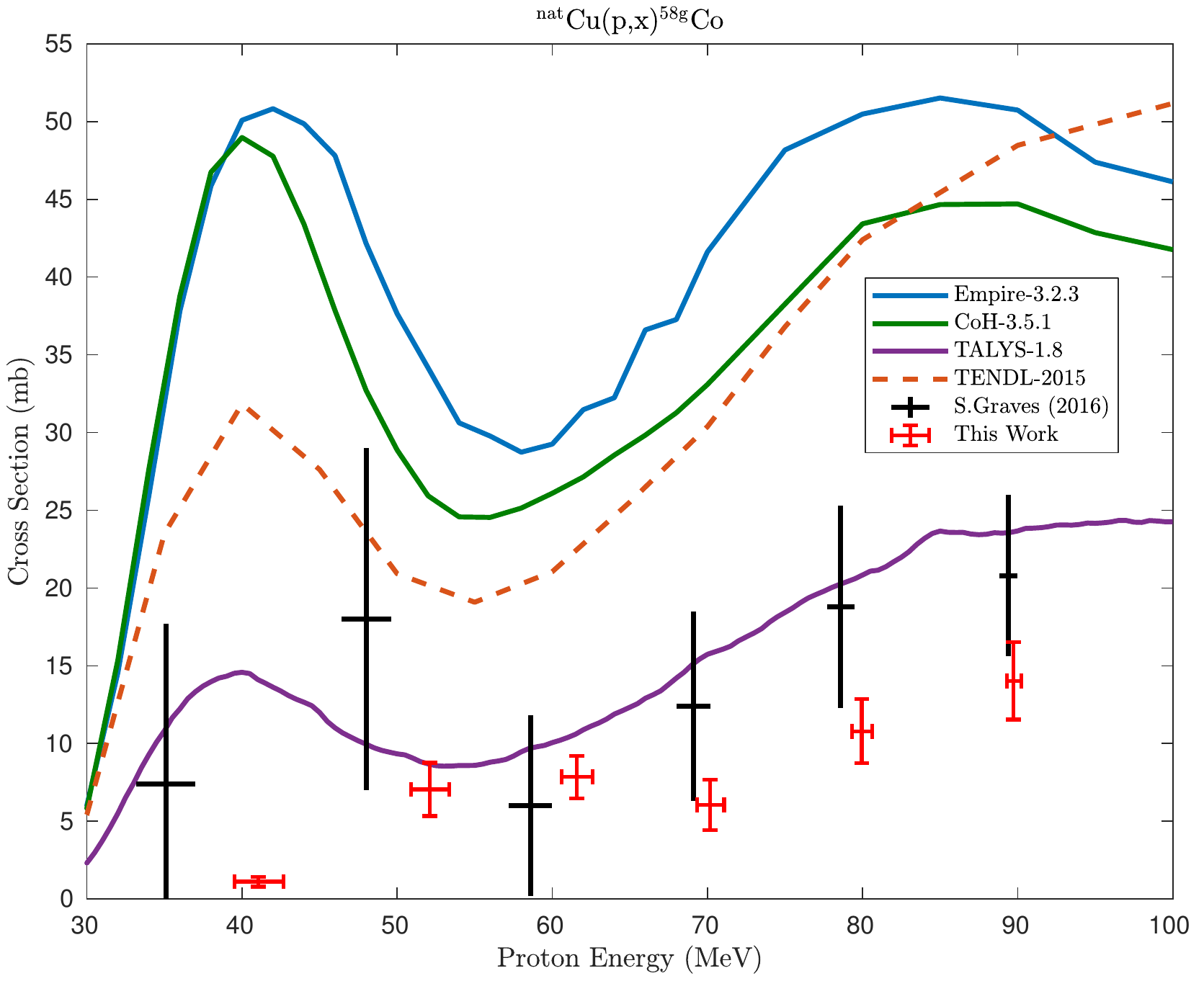}{50}
%
        \subfigimg[width=0.496\textwidth]{}{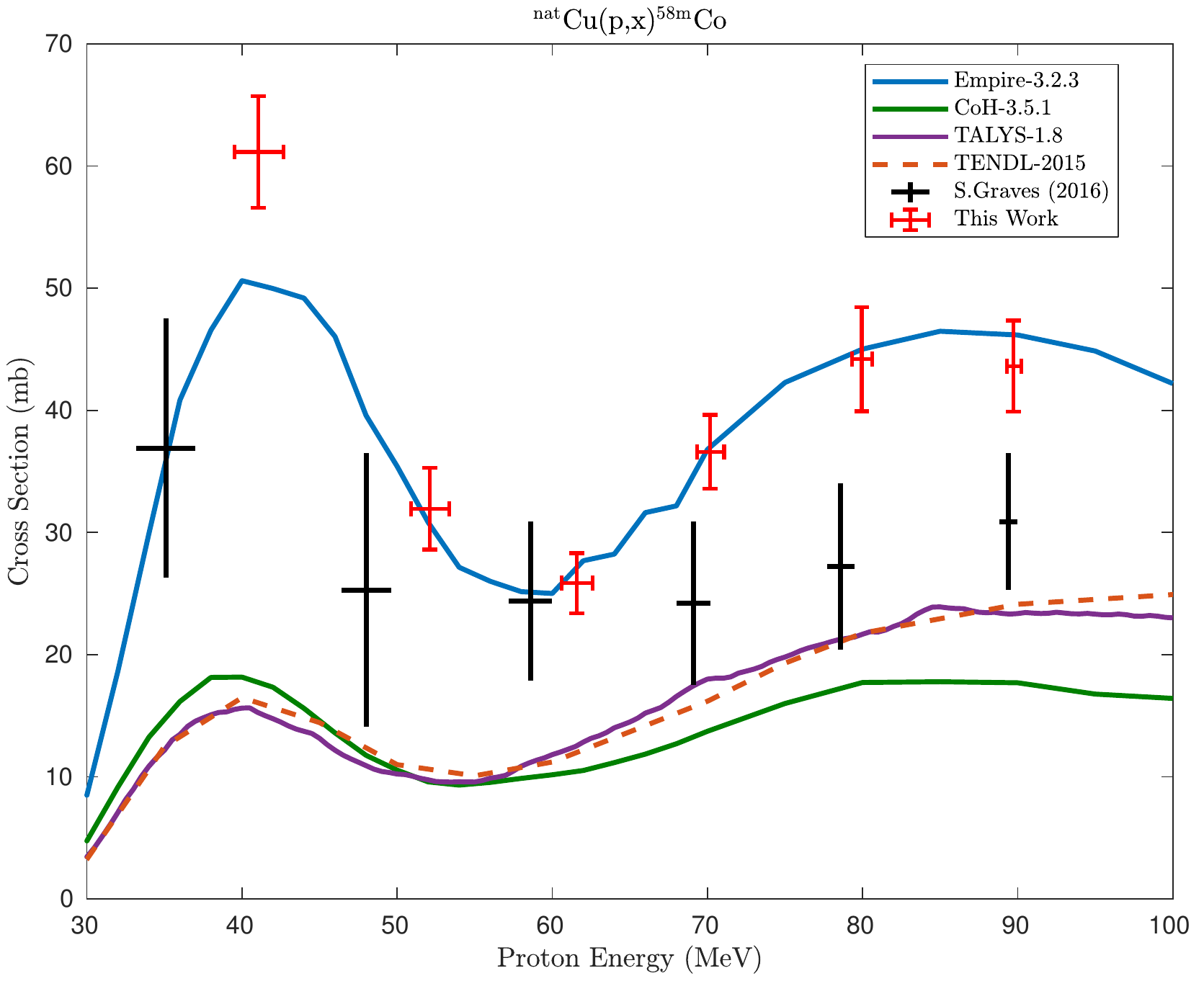}{50}
   \hspace{-10pt}}%
\end{figure*}

\begin{figure*}
    \centering
    \subfloat{
        \centering
        \subfigimg[width=0.496\textwidth]{}{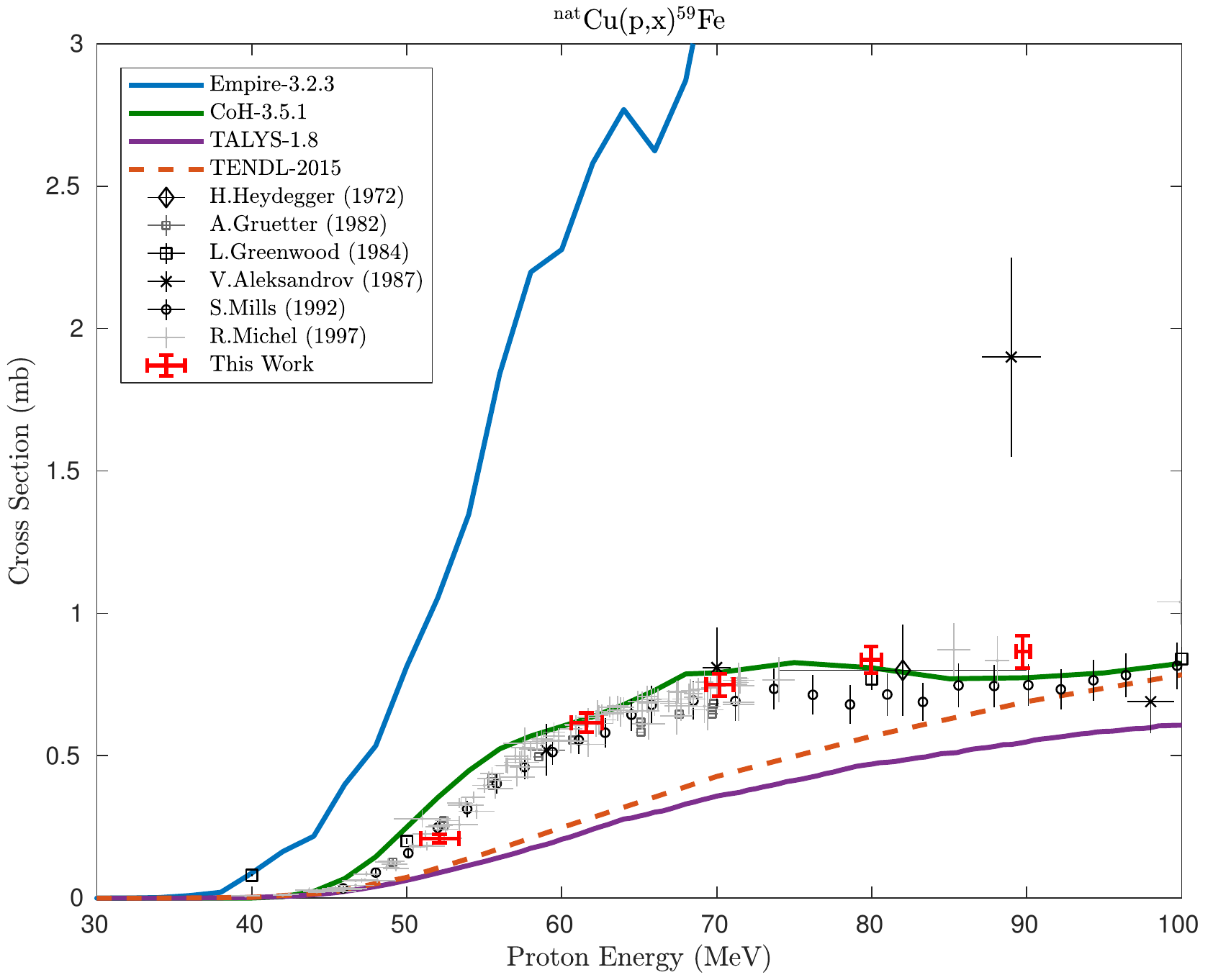}{50}
%
        \subfigimg[width=0.496\textwidth]{}{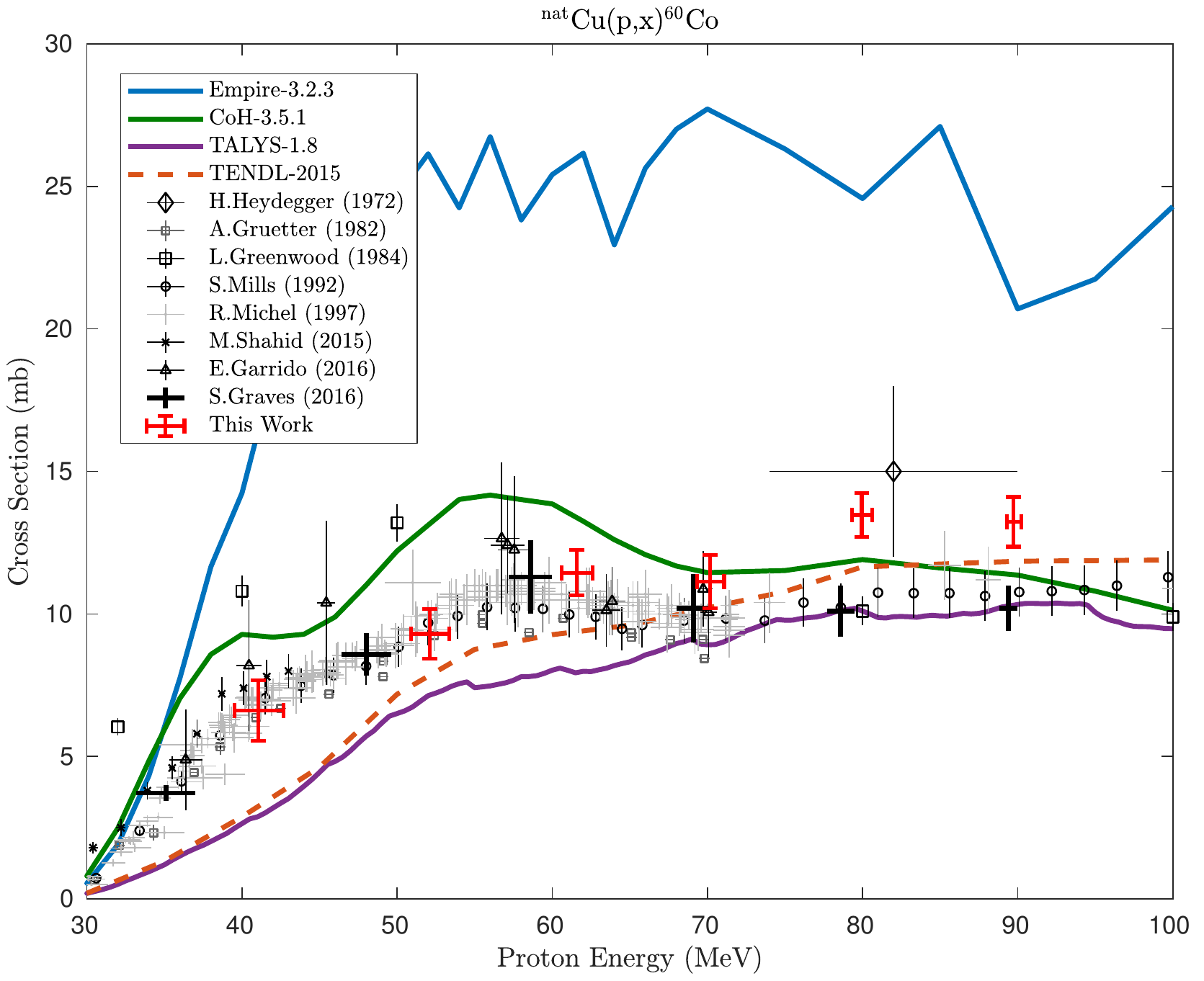}{50}
   \hspace{-10pt}}%
    \\
    \subfloat{
        \centering
        \subfigimg[width=0.496\textwidth]{}{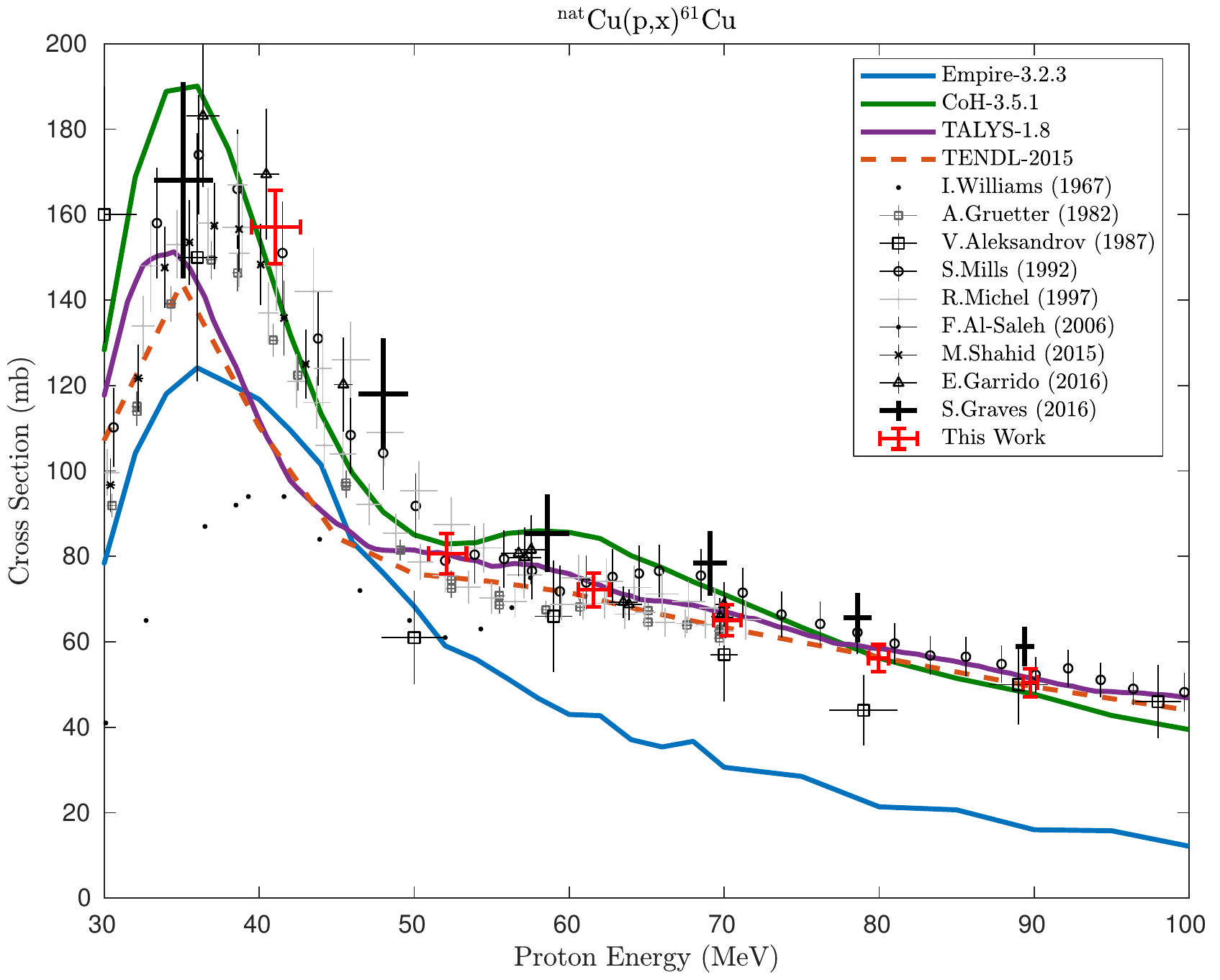}{50}
%
        \subfigimg[width=0.496\textwidth]{}{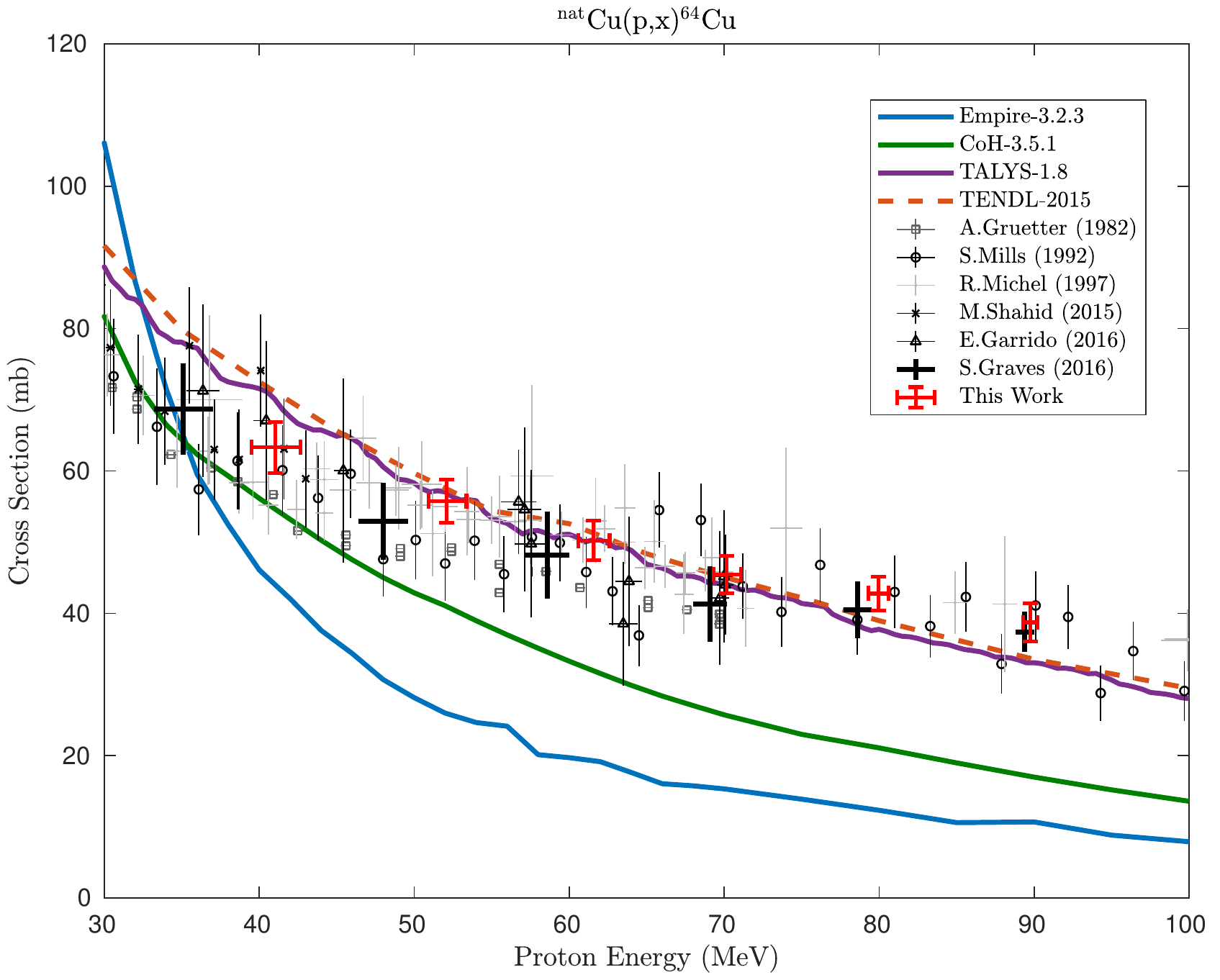}{50}
   \hspace{-10pt}}%
    \\
    \subfloat{
        \centering
        \subfigimg[width=0.496\textwidth]{}{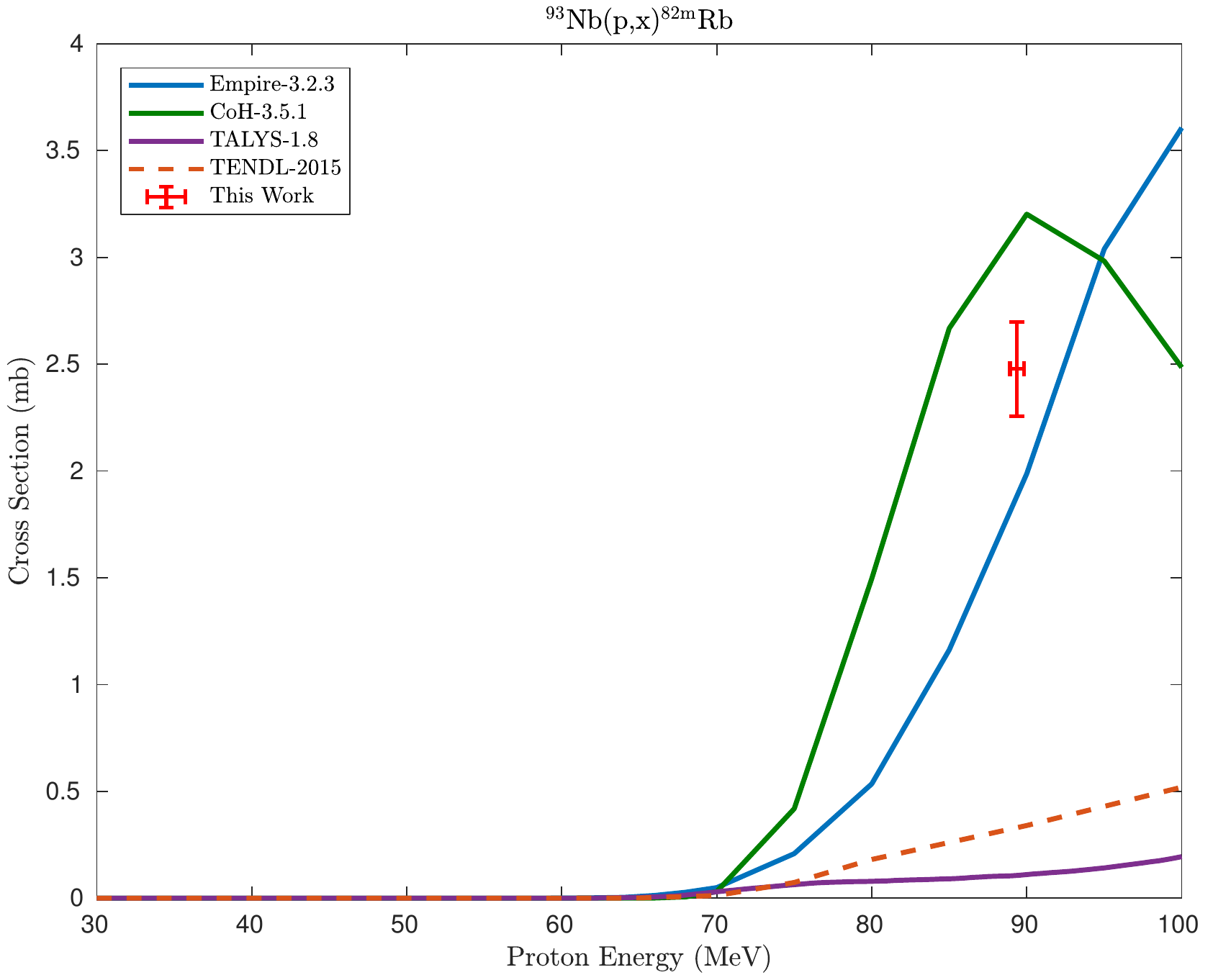}{50}
        \subfigimg[width=0.496\textwidth]{}{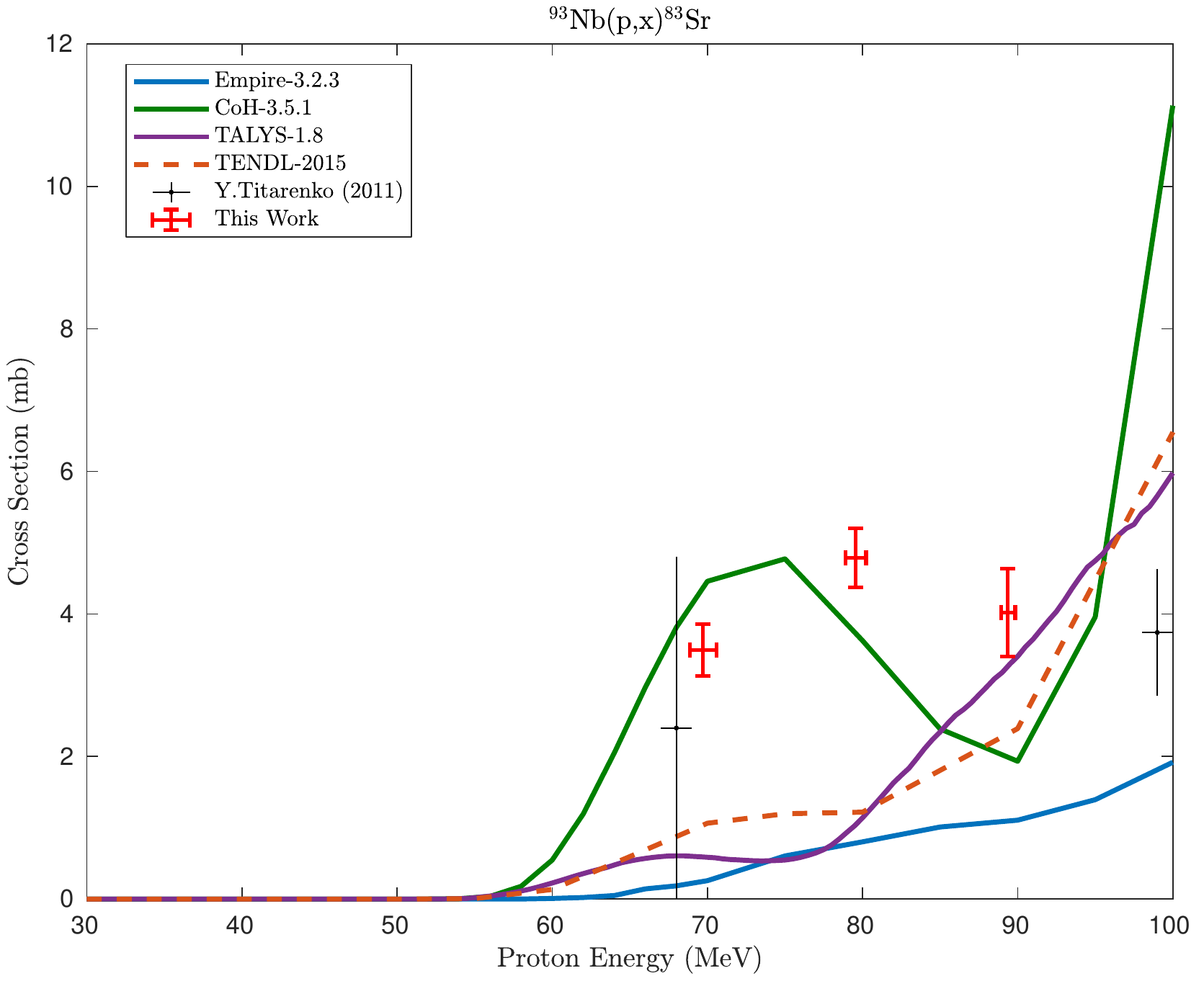}{50}
   \hspace{-10pt}}%
\end{figure*}

\begin{figure*}
    \centering
    \subfloat{
        \centering
        \subfigimg[width=0.496\textwidth]{}{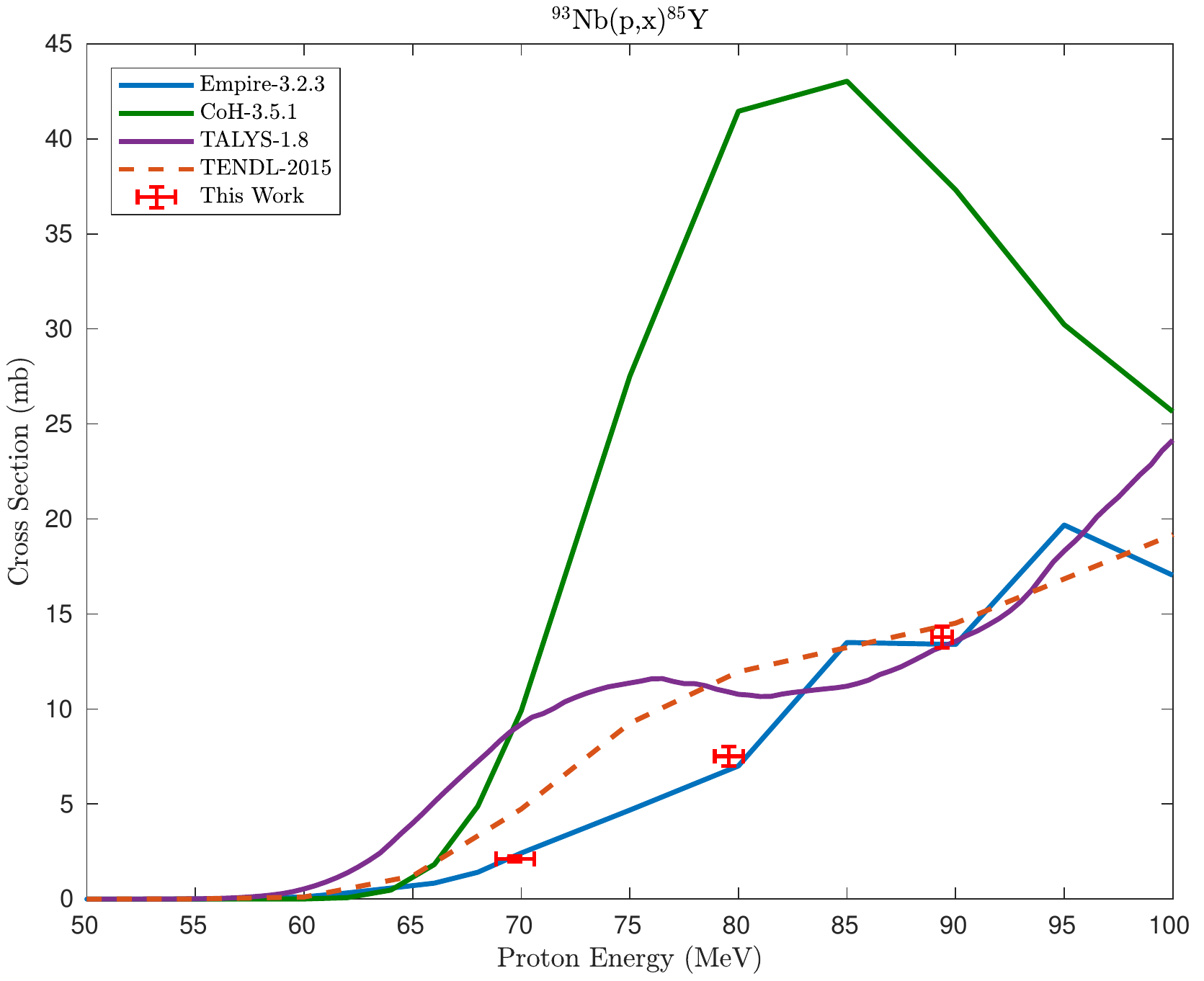}{50}
%
        \subfigimg[width=0.496\textwidth]{}{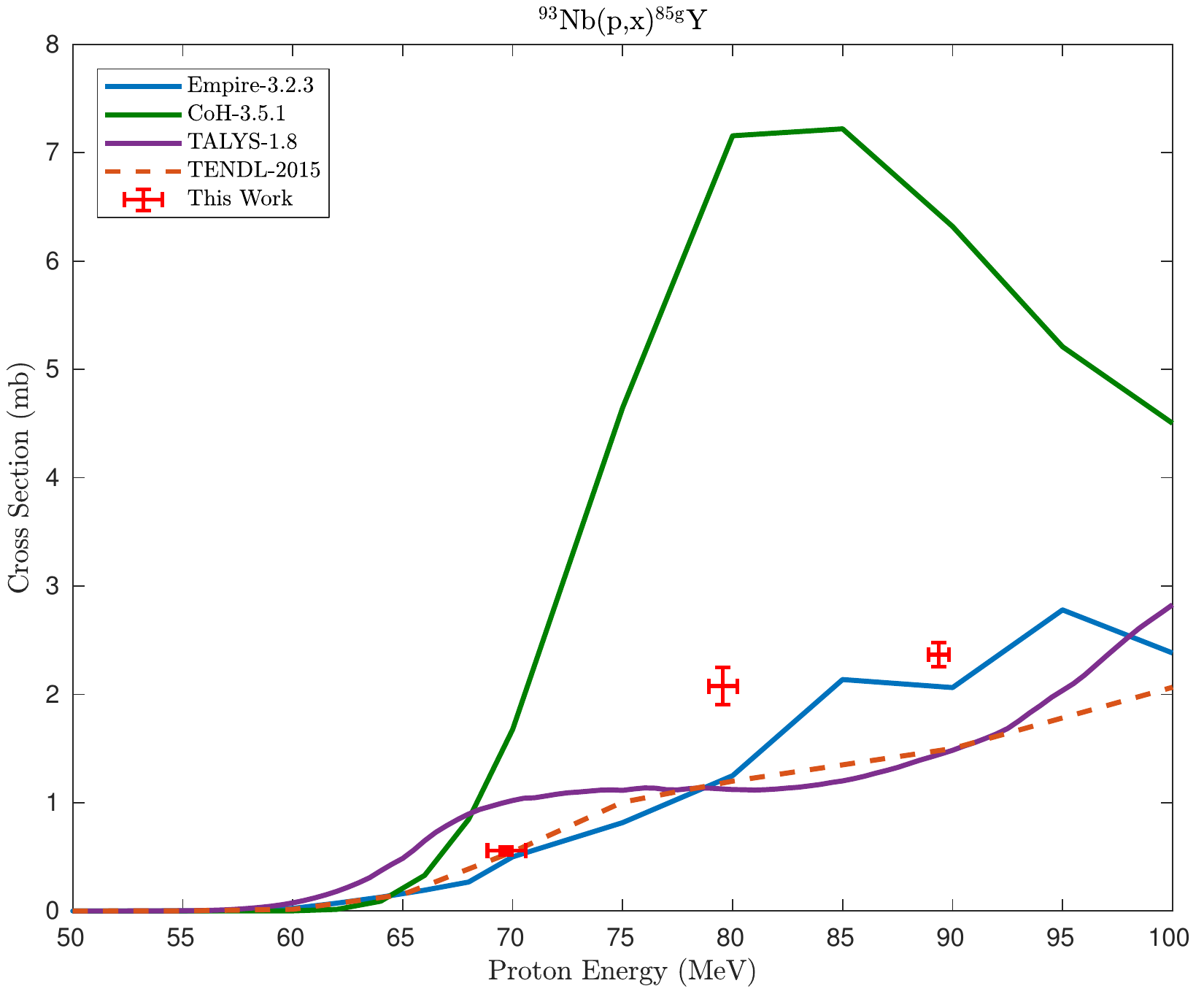}{50}
   \hspace{-10pt}}%
    \\
    \subfloat{
        \centering
        \subfigimg[width=0.496\textwidth]{}{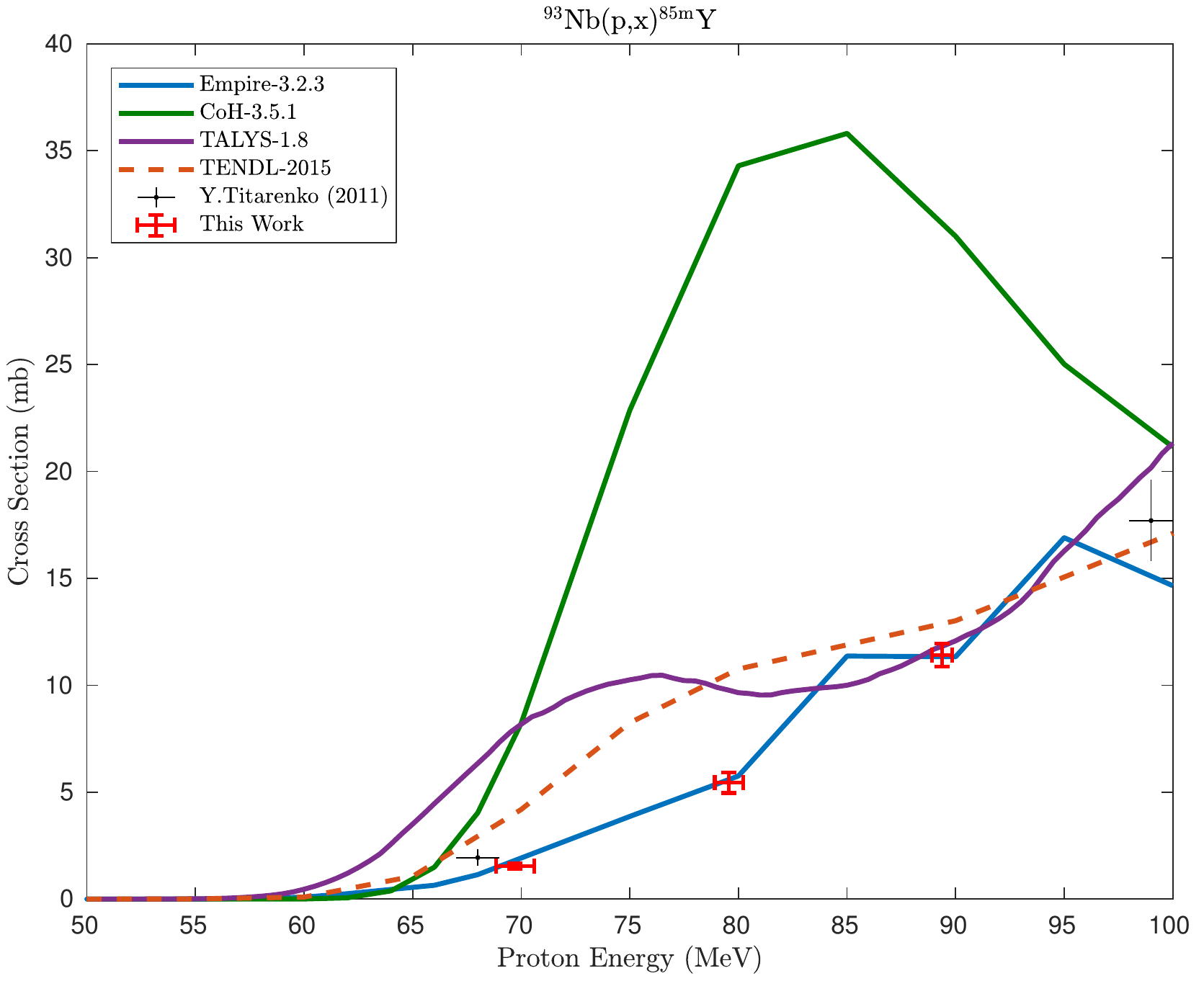}{50}
        \subfigimg[width=0.496\textwidth]{}{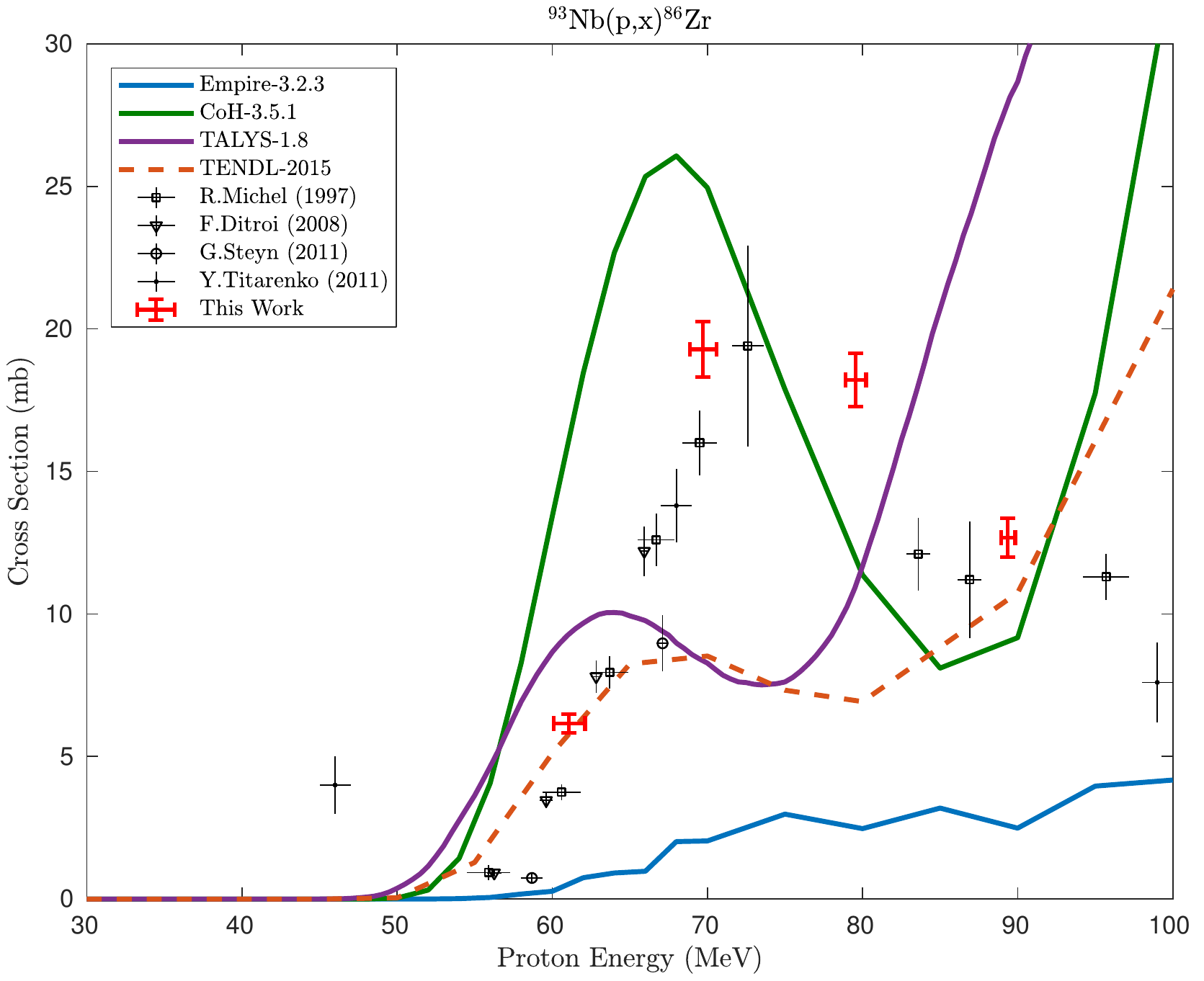}{50}
   \hspace{-10pt}}%
    \\
    \subfloat{
        \centering
        \subfigimg[width=0.496\textwidth]{}{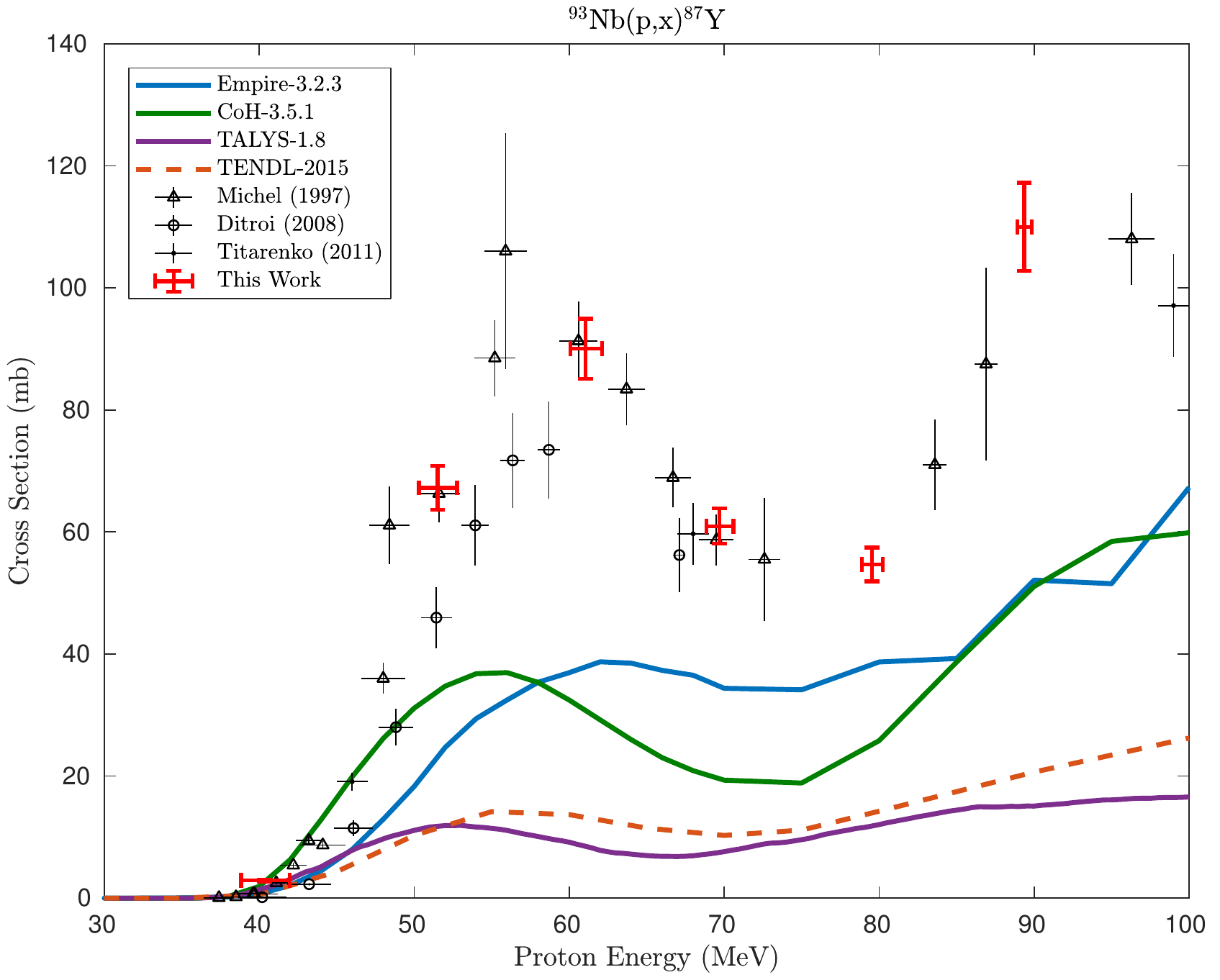}{50}
        \subfigimg[width=0.496\textwidth]{}{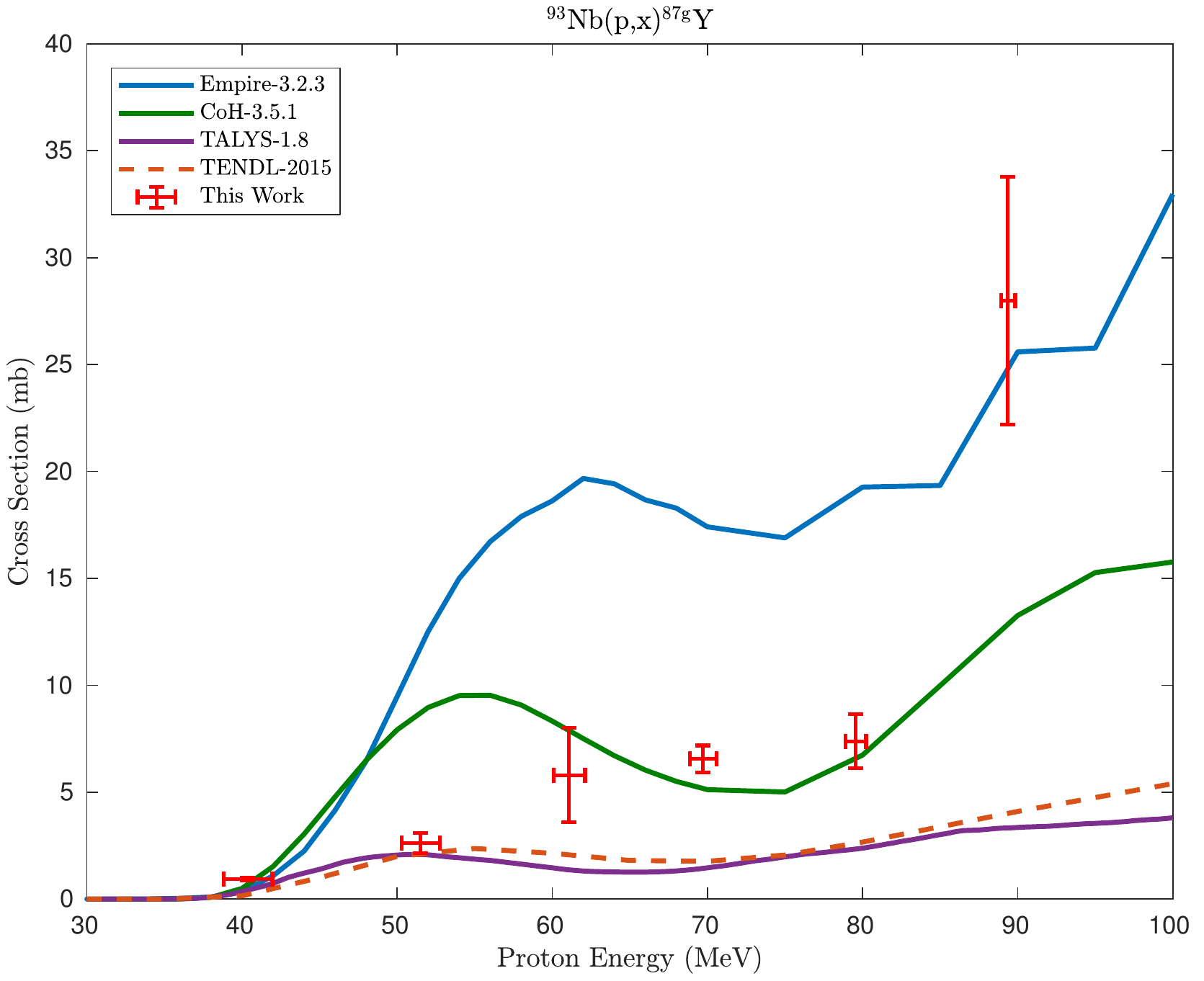}{50}
   \hspace{-10pt}}
\end{figure*}

\begin{figure*}
    \centering
     \subfloat{
        \centering
        \subfigimg[width=0.496\textwidth]{}{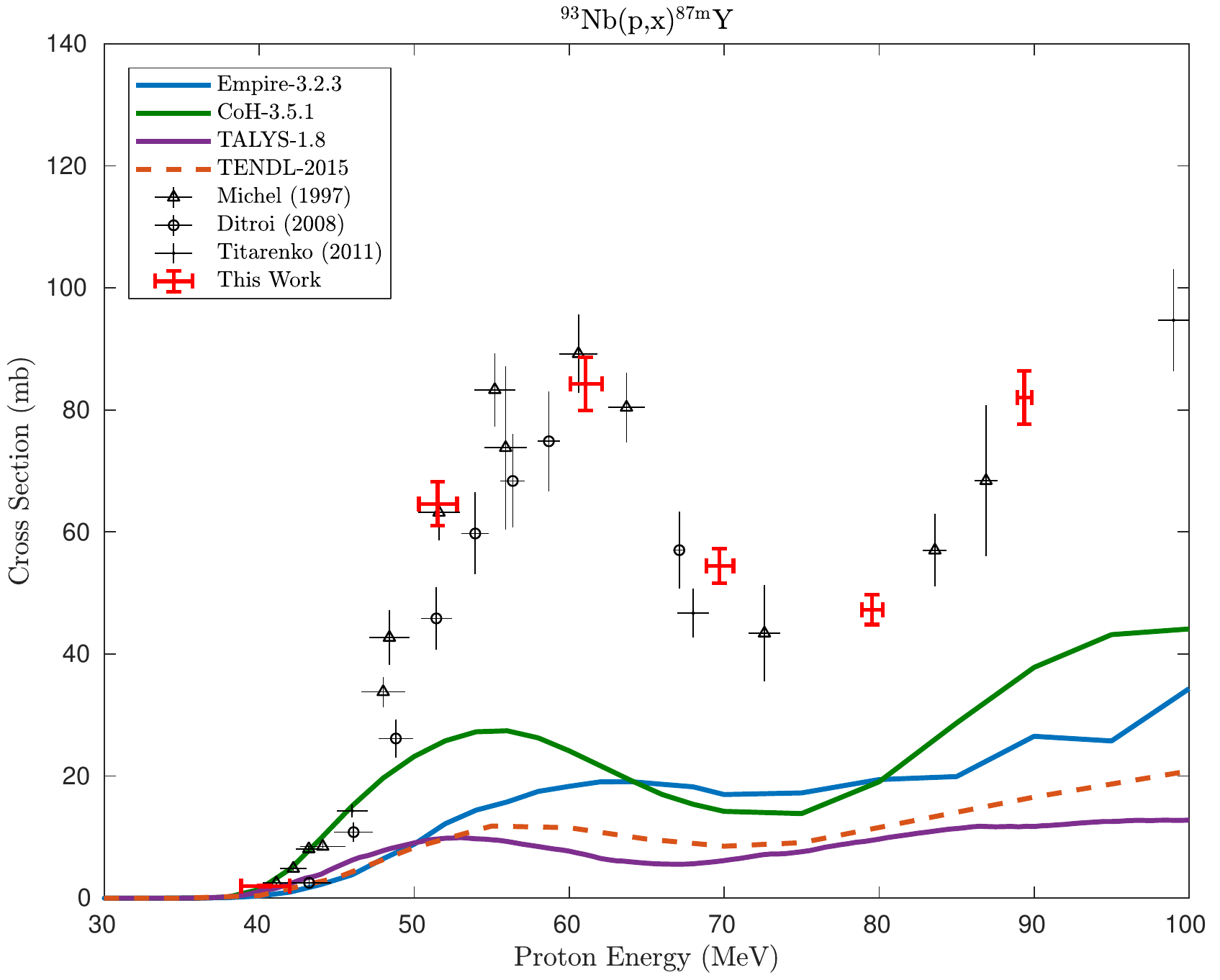}{50}
        \subfigimg[width=0.496\textwidth]{}{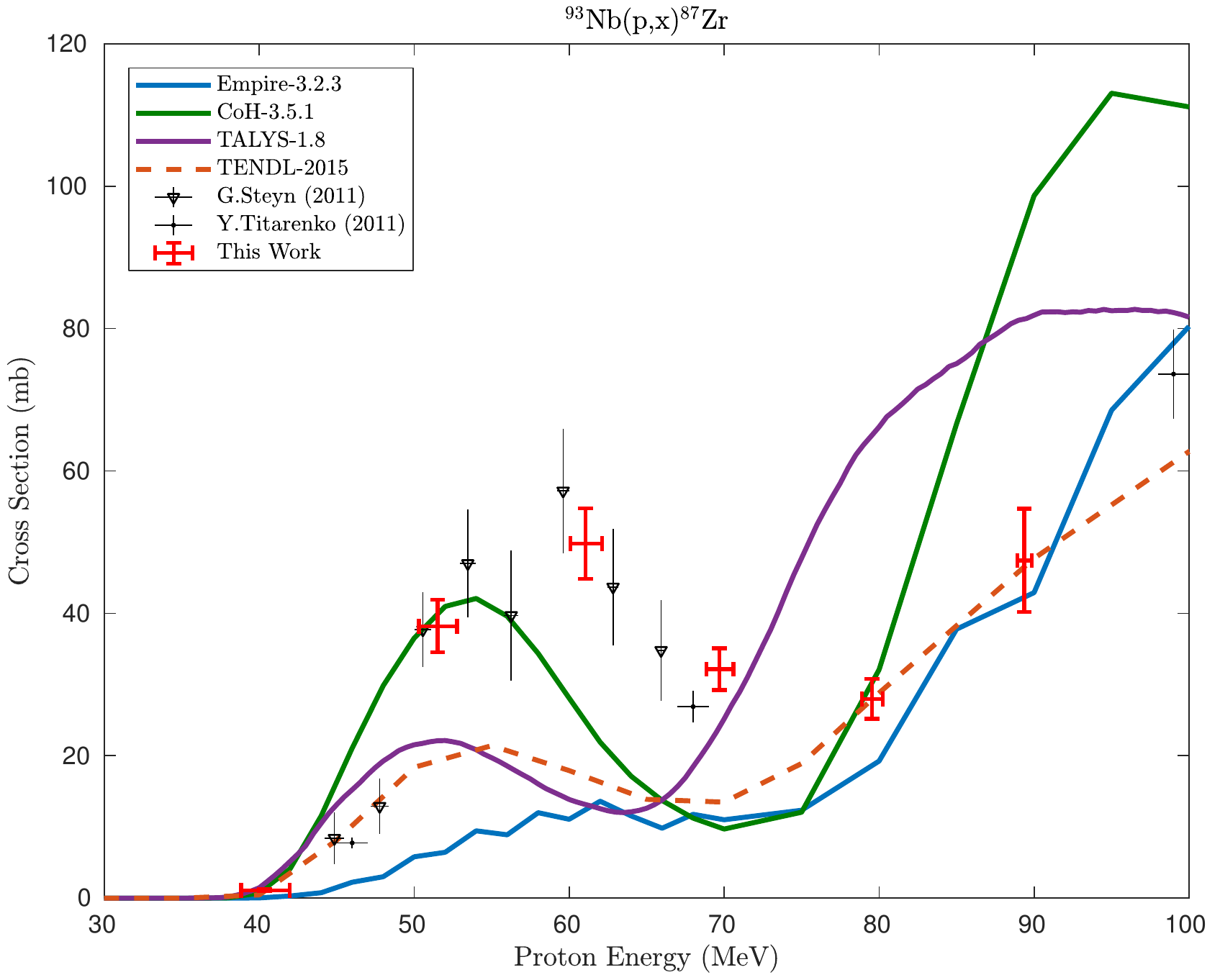}{50}
   \hspace{-10pt}}%
    \\
    \subfloat{
        \centering
        \subfigimg[width=0.496\textwidth]{}{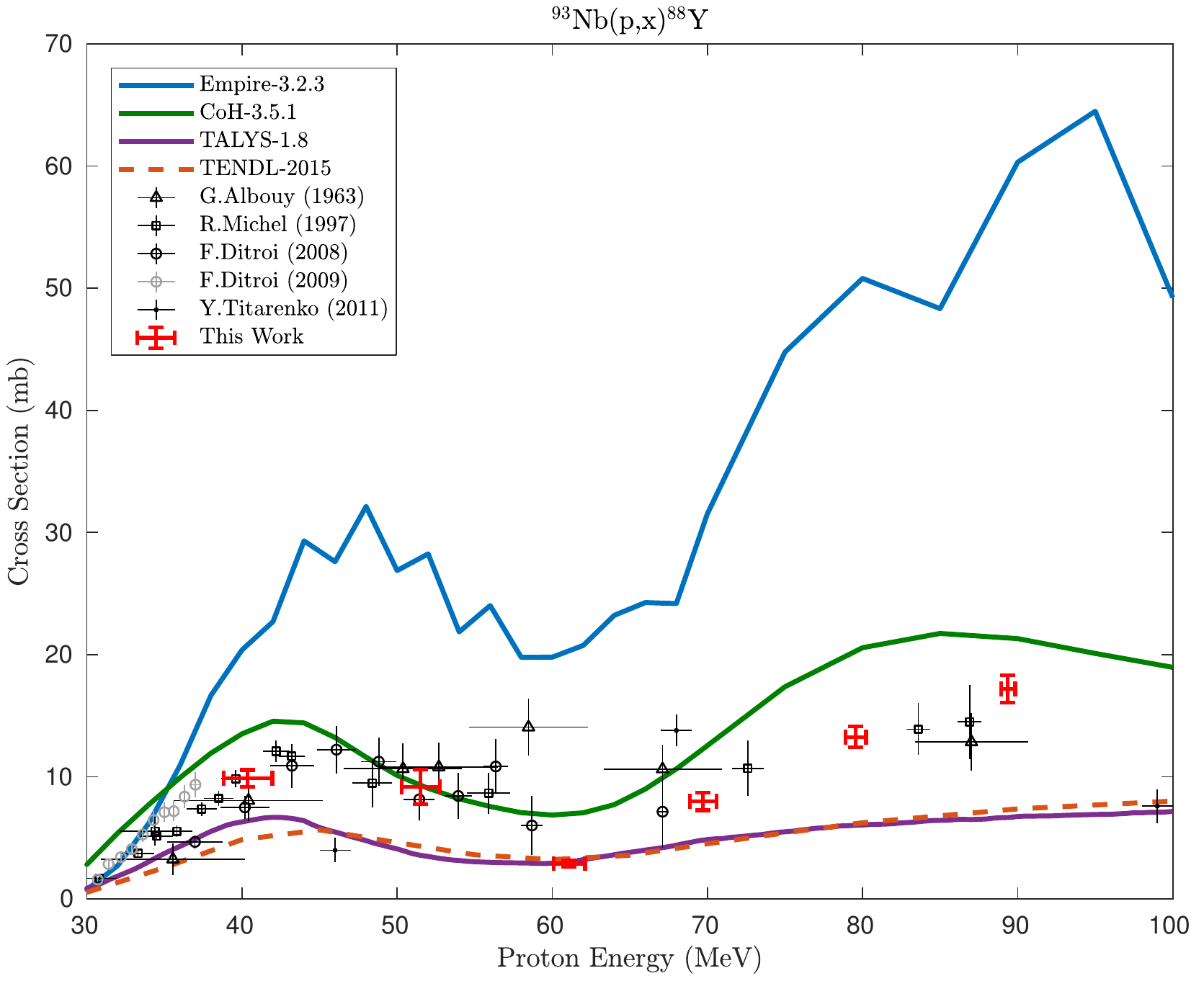}{50}
        \subfigimg[width=0.496\textwidth]{}{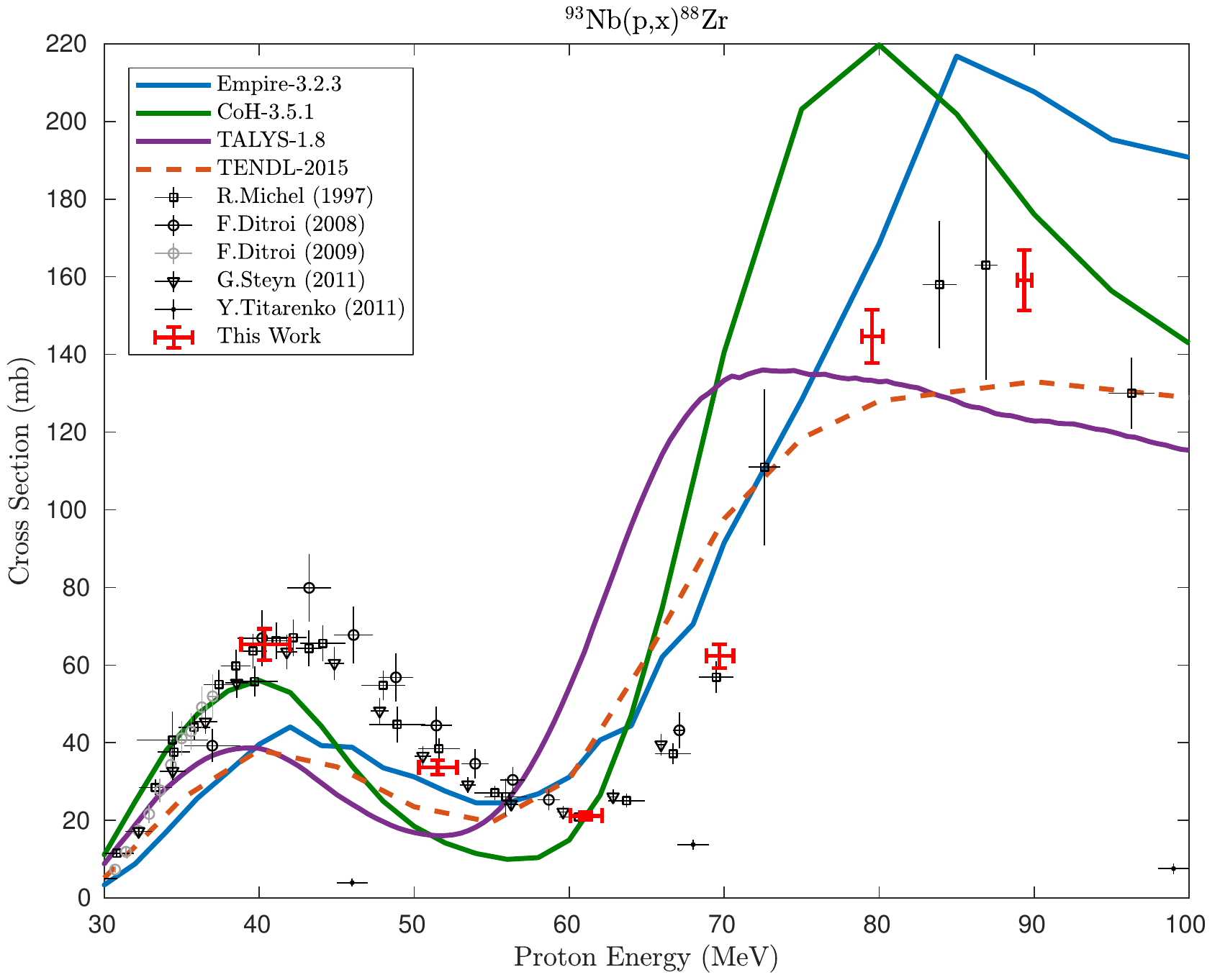}{50}
   \hspace{-10pt}}%
    \\
         \subfloat{
        \centering
        \subfigimg[width=0.496\textwidth]{}{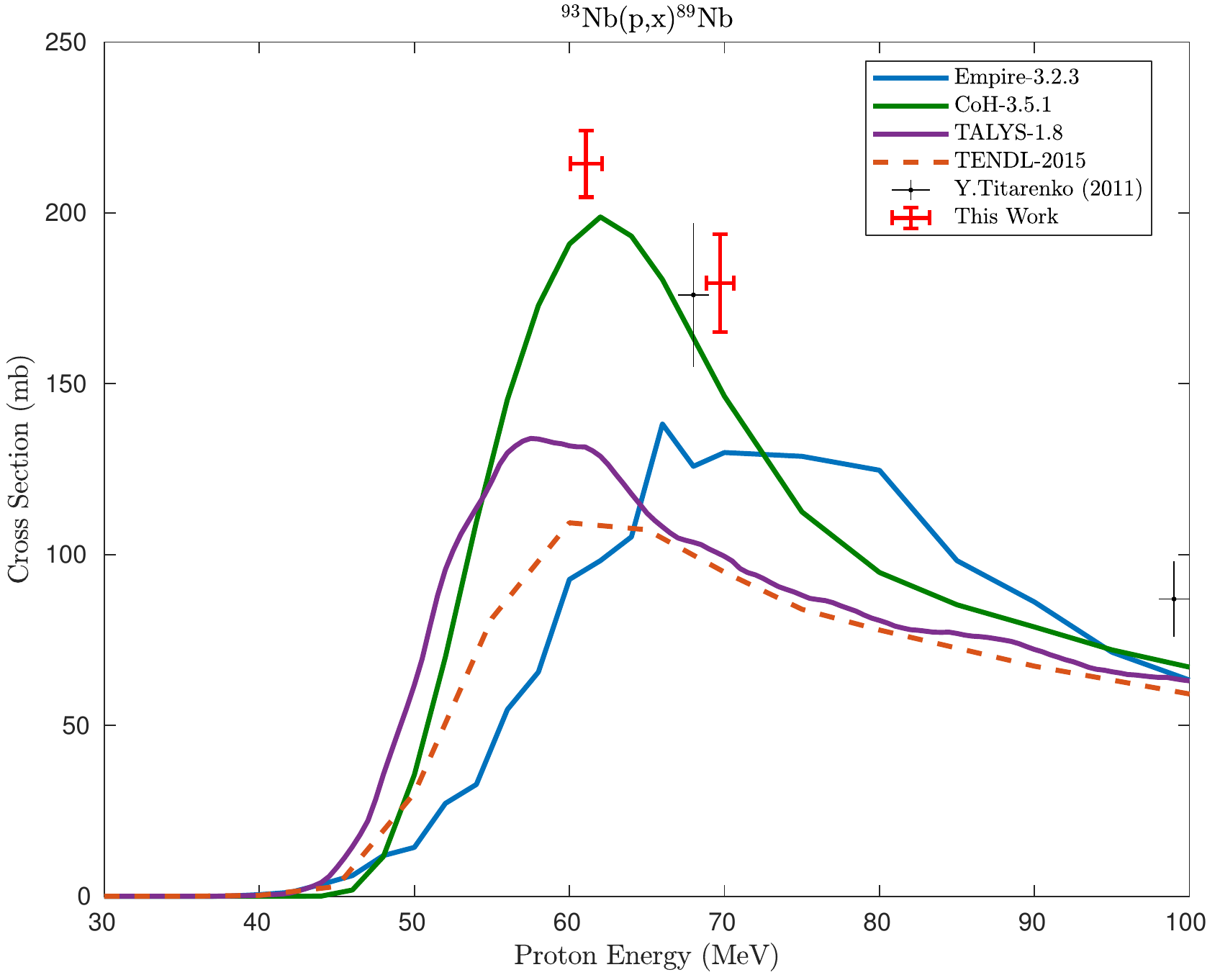}{50}
        \subfigimg[width=0.496\textwidth]{}{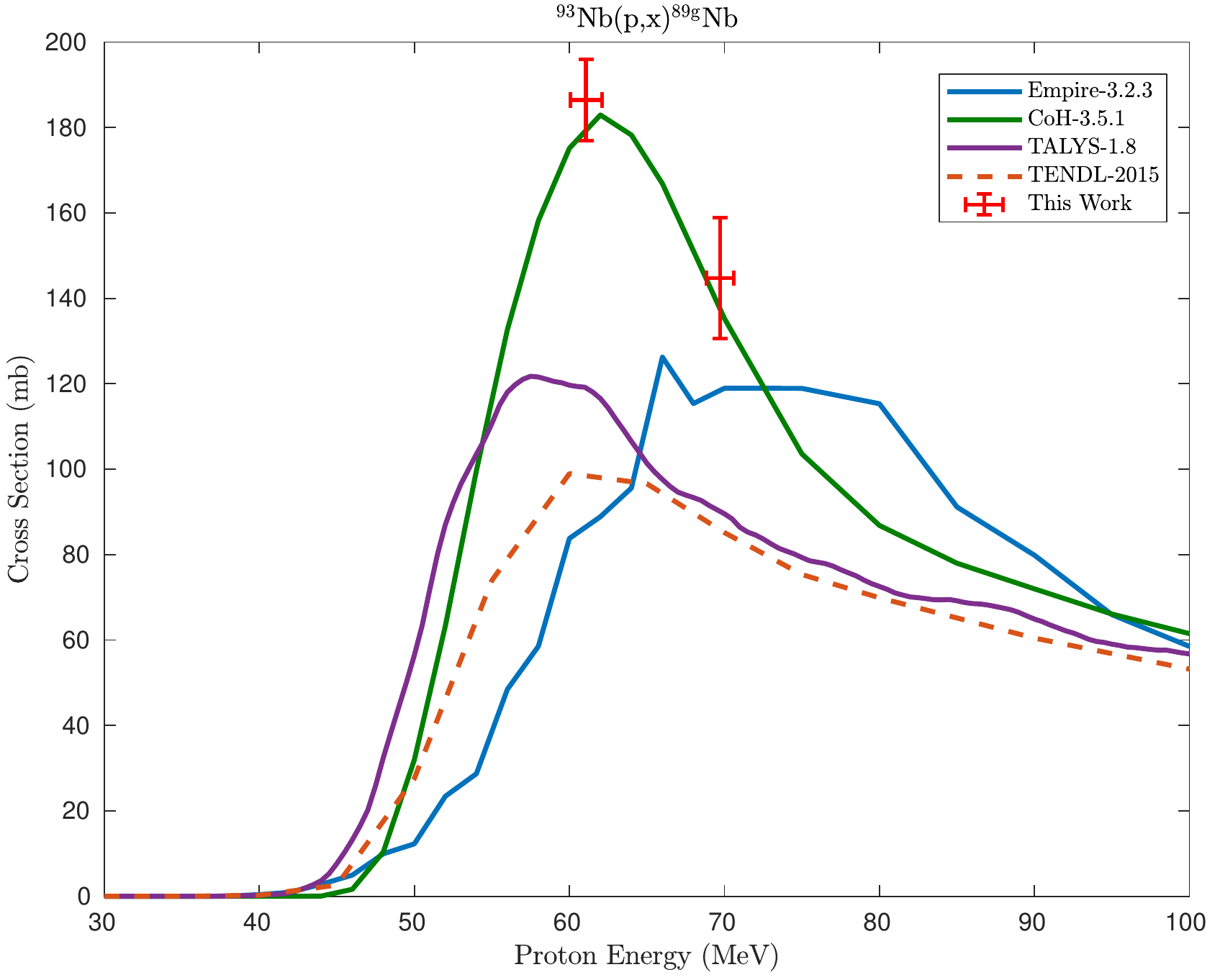}{50}
   \hspace{-10pt}}%
\end{figure*}

\begin{figure*}
    \centering    
         \subfloat{
        \centering
        \subfigimg[width=0.496\textwidth]{}{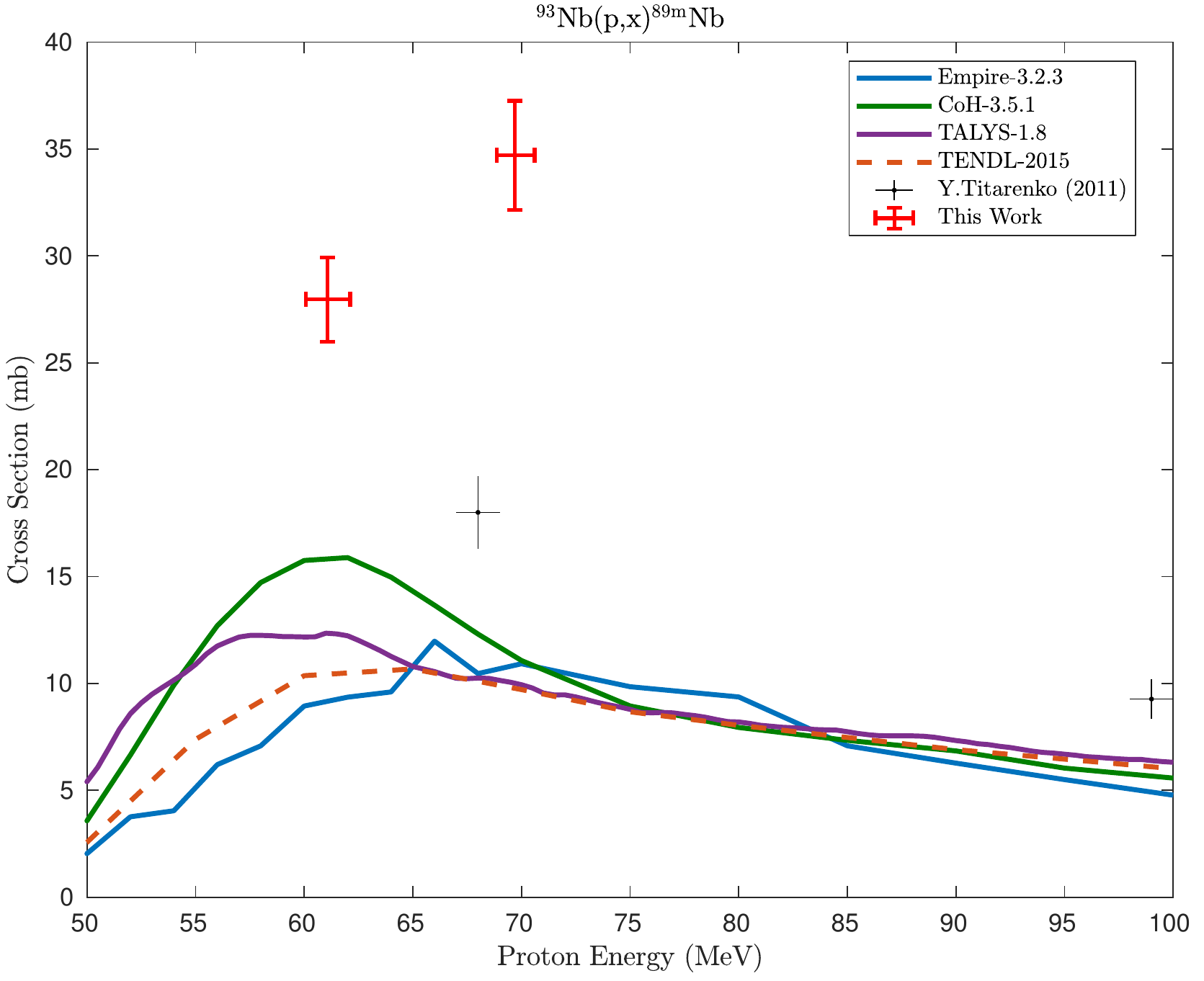}{50}
        \subfigimg[width=0.496\textwidth]{}{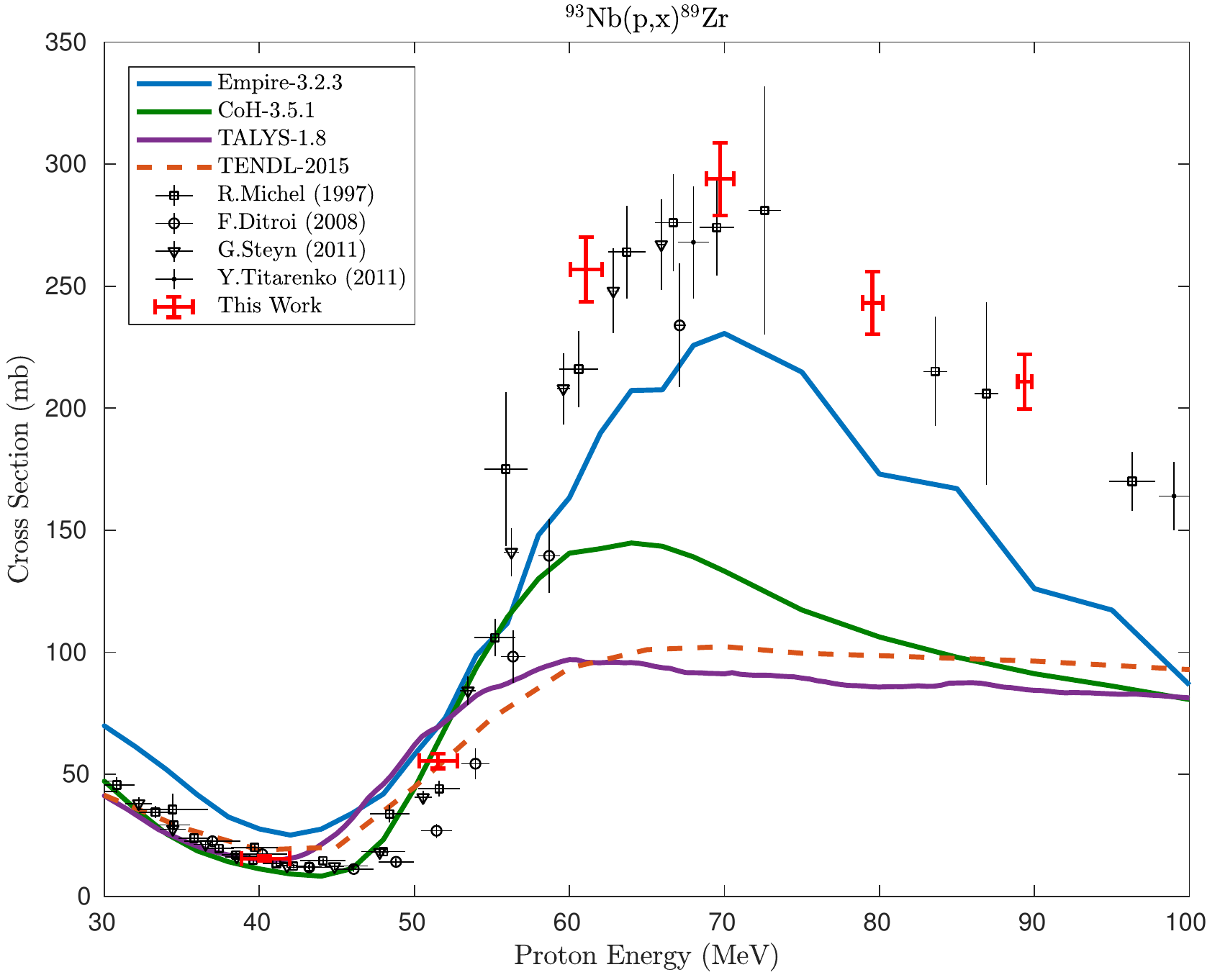}{50}
   \hspace{-10pt}}%
    \\
    \subfloat{
        \centering
        \subfigimg[width=0.496\textwidth]{}{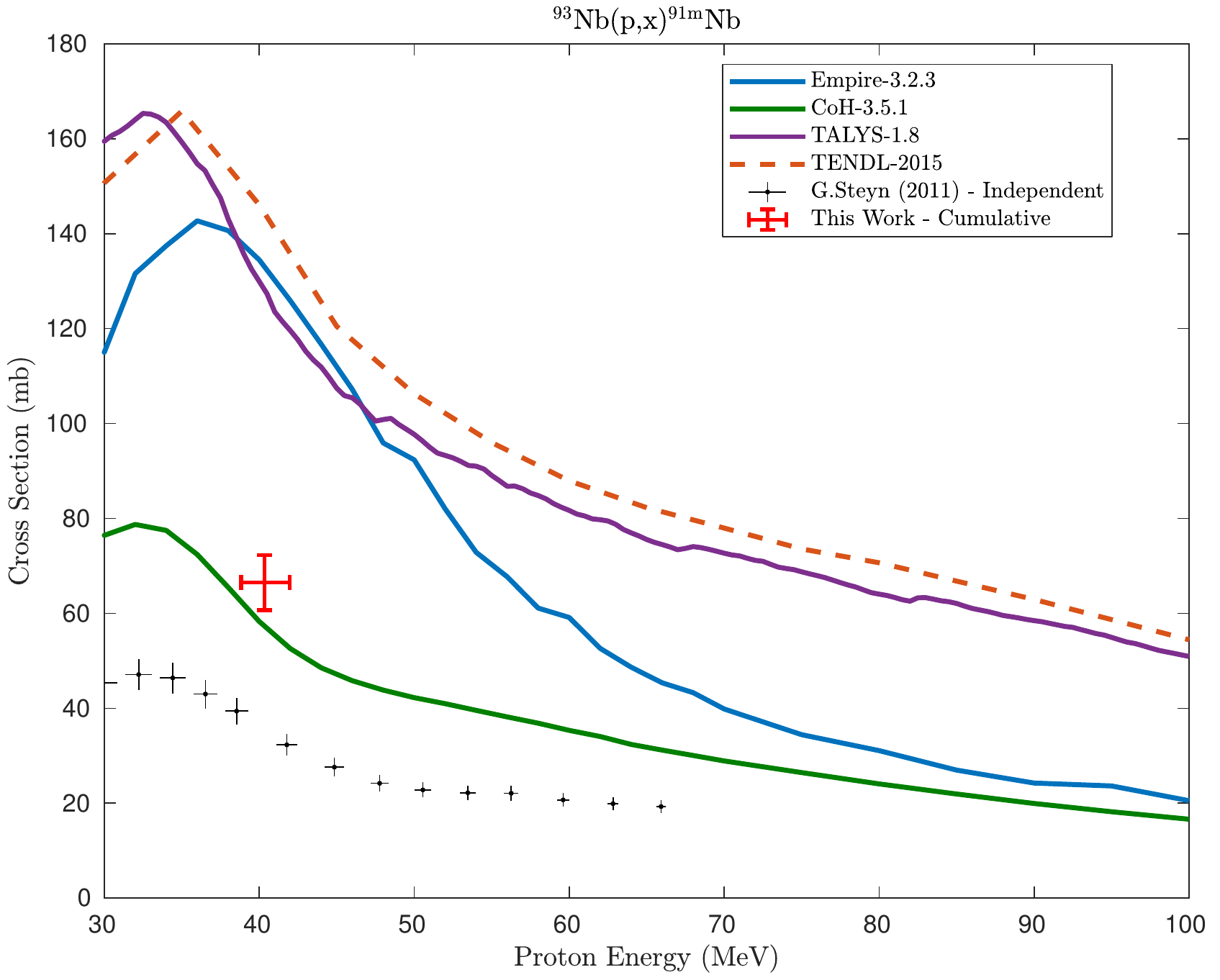}{50}
        \subfigimg[width=0.496\textwidth]{}{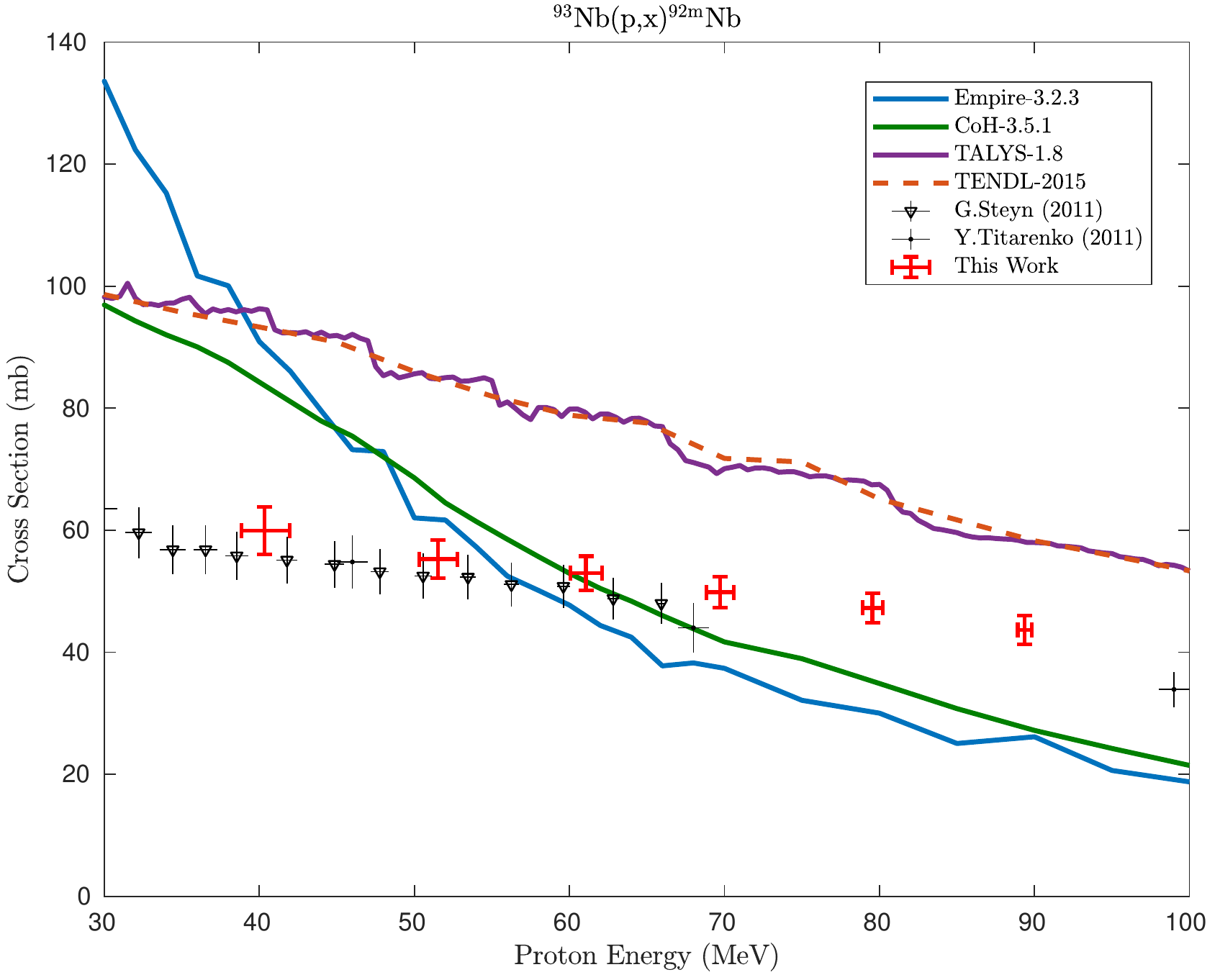}{50}
   \hspace{-10pt}}%
    \\
    \subfloat{
        \centering
        \subfigimg[width=0.496\textwidth]{}{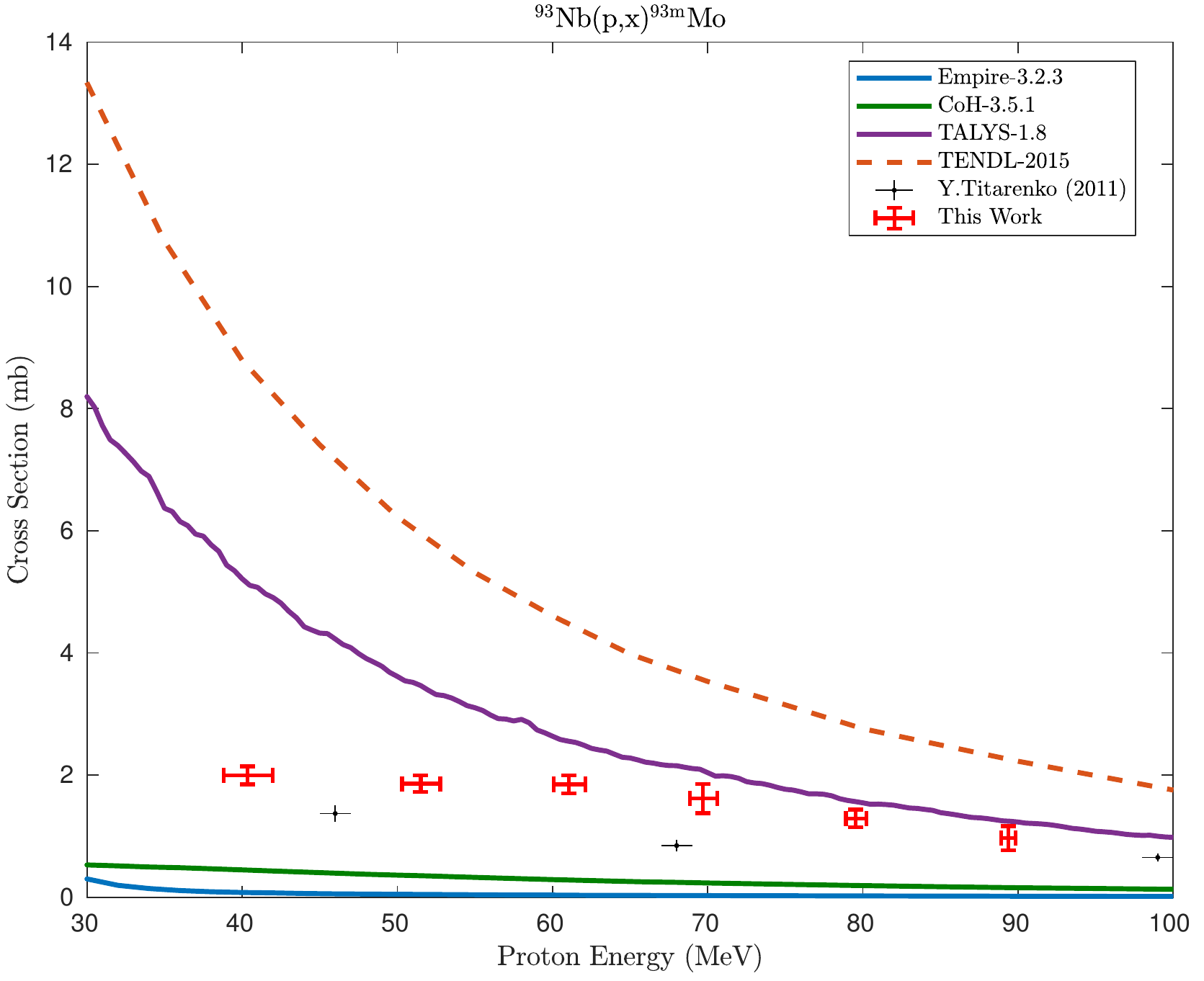}{50}
   }%
\end{figure*}

\section{Measured isomer-to-ground state branching ratios } \label{ibr_figures}

Plots of the isomer-to-ground state ratios measured in this work are presented here, in comparison with literature data and reaction modeling codes \cite{MICHEL1997153,Ditroi2008,Titarenko2011,Graves2016}.


\begin{figure*}
    \centering
    \subfloat{
        \centering
        \subfigimg[width=0.496\textwidth]{}{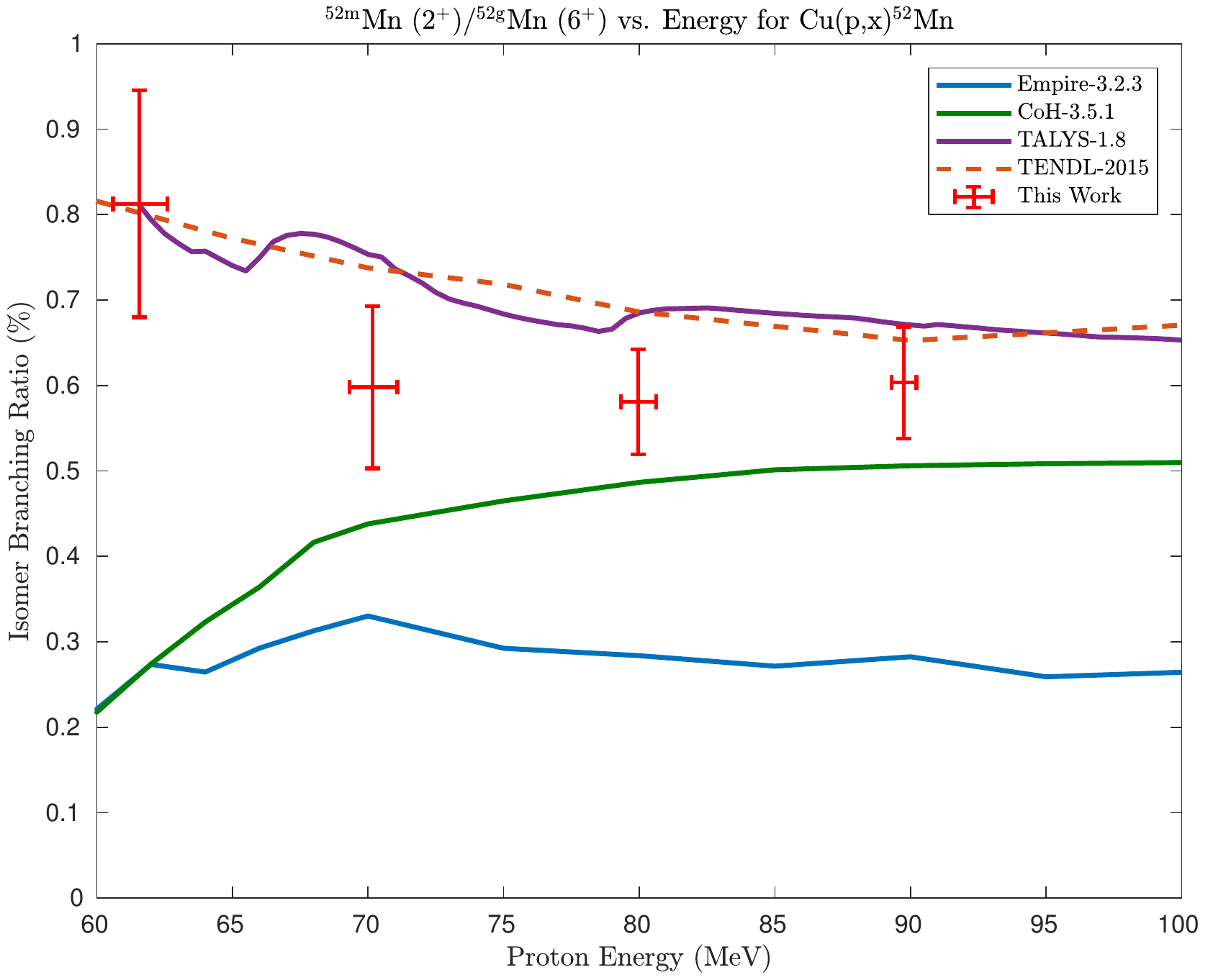}{50}
        \subfigimg[width=0.496\textwidth]{}{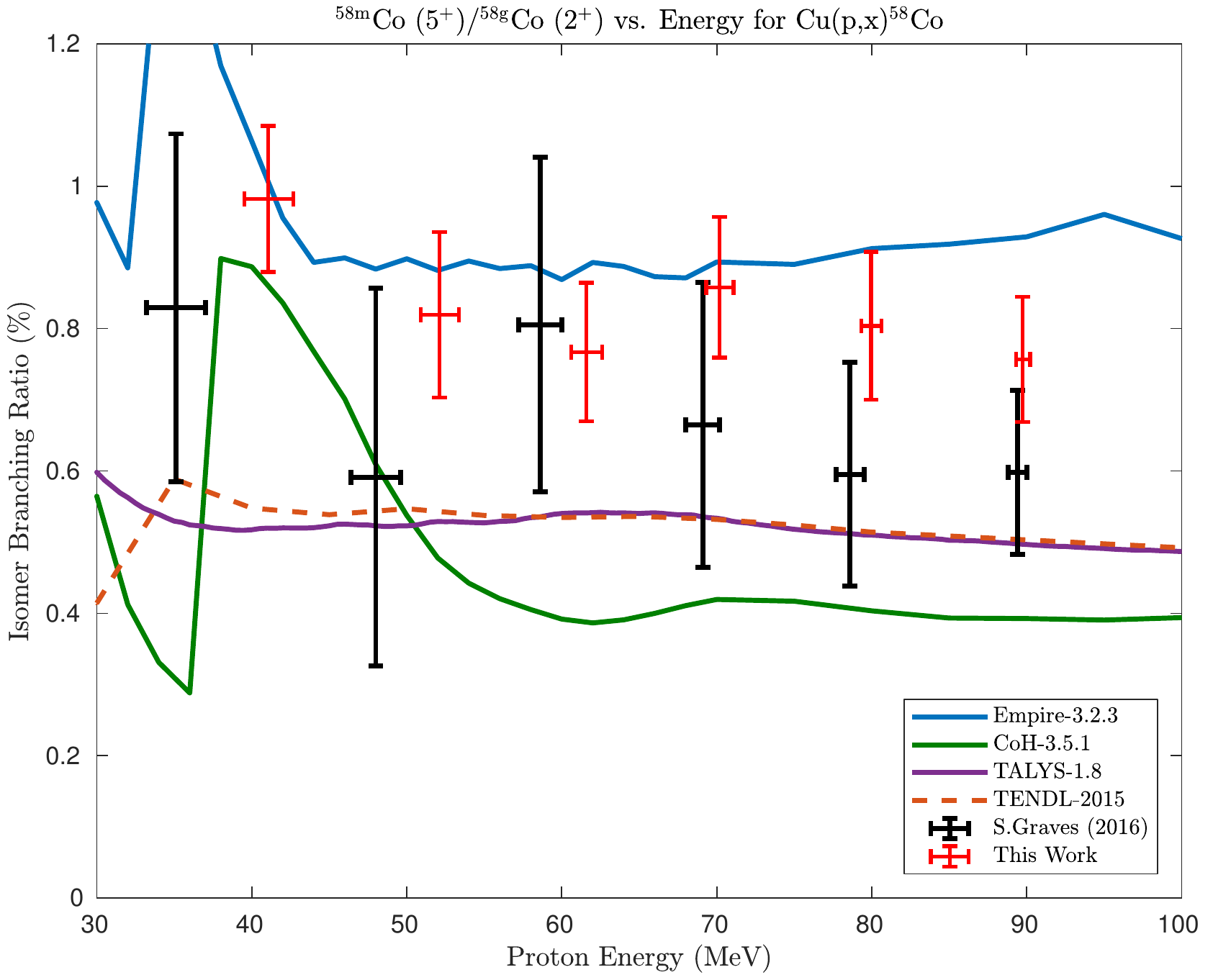}{50}
   \hspace{-10pt}}%
    \\
    \subfloat{
        \centering
        \subfigimg[width=0.496\textwidth]{}{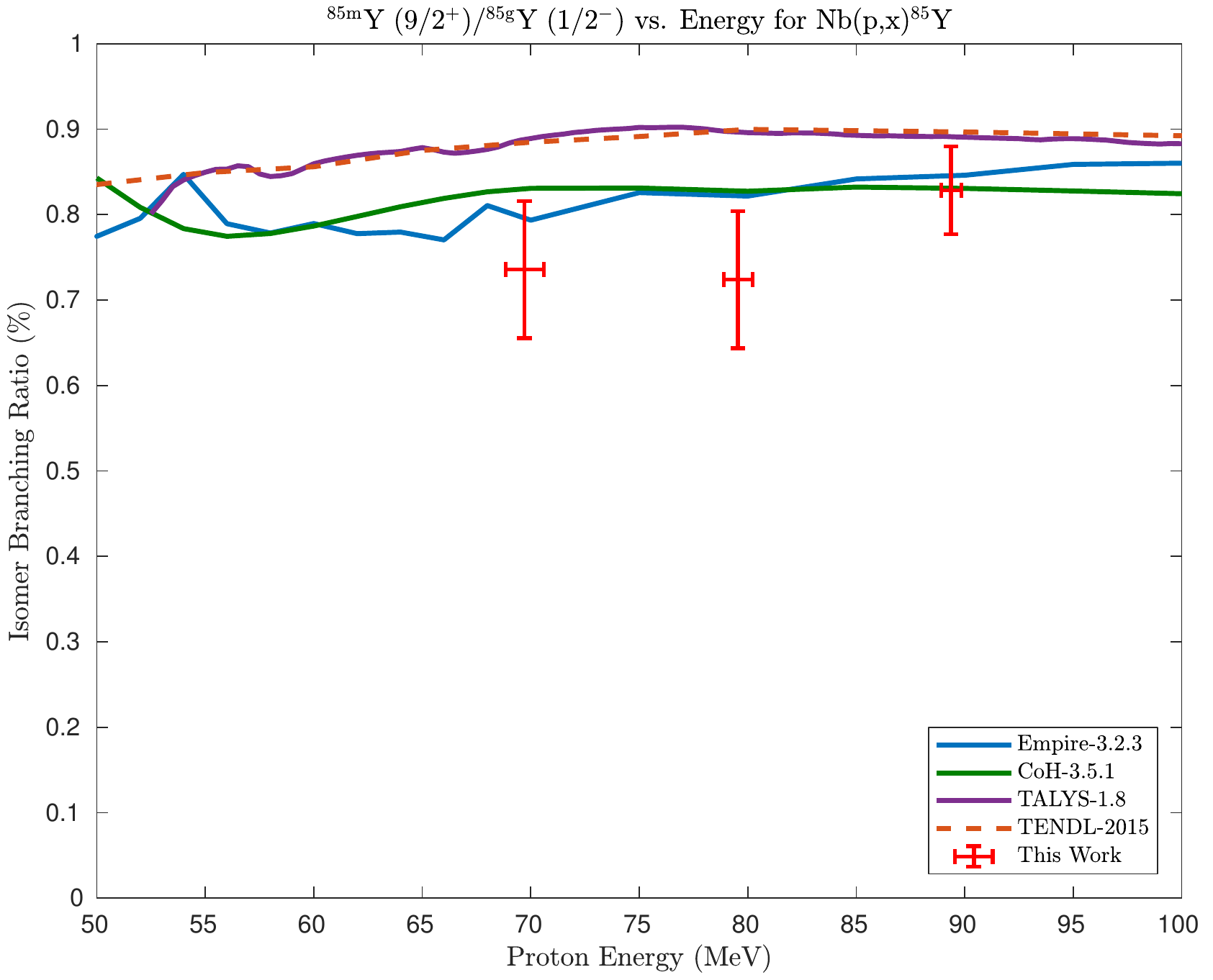}{50}
        \subfigimg[width=0.496\textwidth]{}{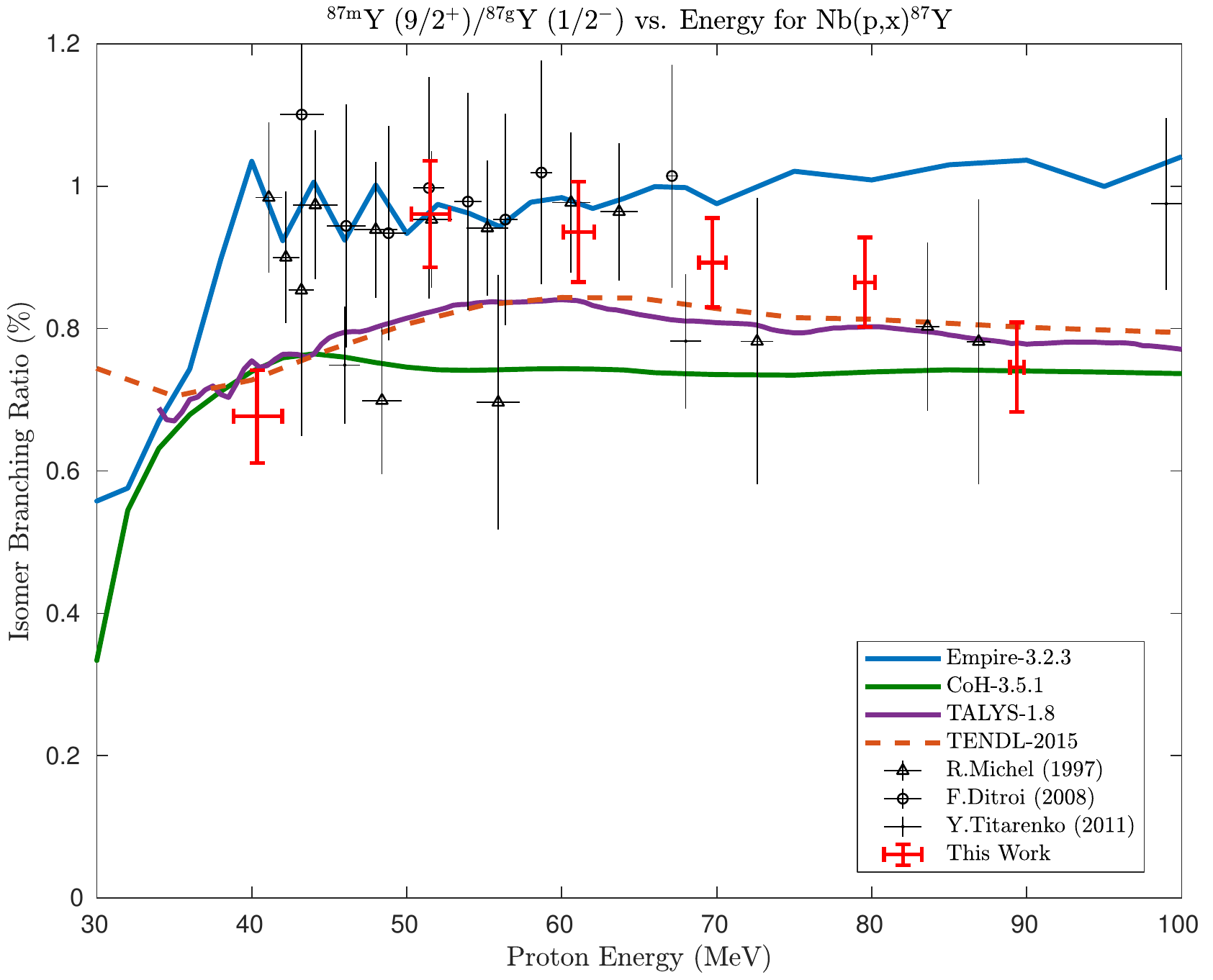}{50}
   \hspace{-10pt}}%
    \\
    \subfloat{
        \centering
        \subfigimg[width=0.496\textwidth]{}{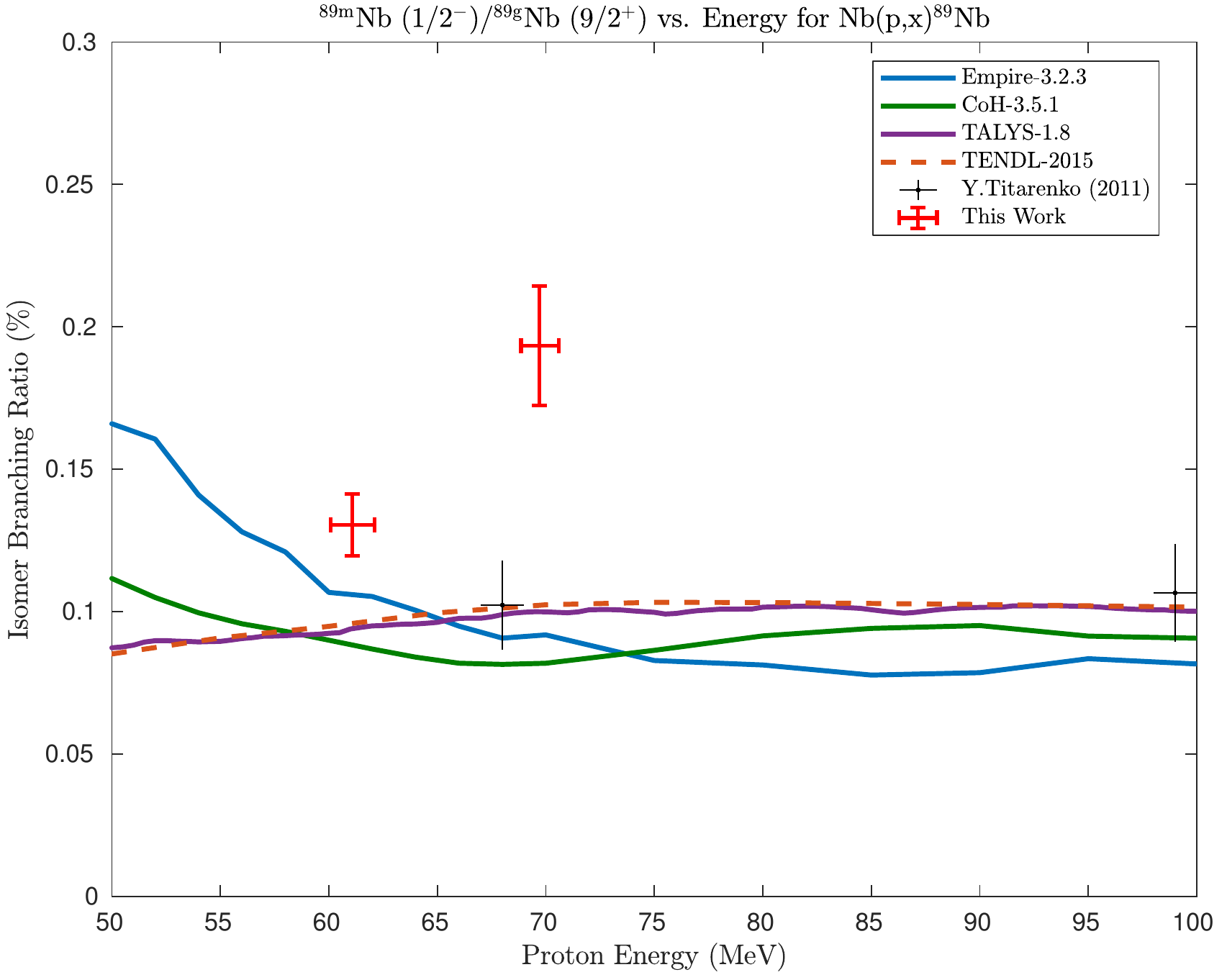}{50}
   }
\end{figure*}